
\input harvmac.tex
\input amssym.tex 


\def\a{\alpha}\def\b{\beta}\def\d{\delta}\def\e{\epsilon}
\def\g{\gamma}\def\k{\kappa}\def\l{\lambda}\def\m{\mu}\def\n{\nu}
\def\o{\omega}\def\r{\rho}\def\s{\sigma}\def\t{\tau}\def\z{\zeta}
\def\D{\Delta}\def\G{\Gamma}\def\L{\Lambda}\def\O{\Omega}
\def\CA{{\cal A}}\def\CB{{\cal B}}\def\CD{{\cal D}}
\def\CF{{\cal F}}\def\CI{{\cal I}}\def\CM{{\cal M}}\def\CN{{\cal N}}
\def\CS{{\cal S}}\def\CT{{\cal T}}\def\CW{{\cal W}}
\def\IC{{\Bbb C}} 
\def\IR{{\Bbb R}} 
\def\IZ{{\Bbb Z}} 
\def\IN{{\Bbb N}} 
\def\ibar{{\bar\imath}}
\def\jbar{{\bar\jmath}}
\def\det{\mathop{\rm det}\nolimits}
\def\diag{\mathop{\rm diag}\nolimits}
\def\Re{\mathop{\rm Re}\nolimits}
\def\Im{\mathop{\rm Im}\nolimits}
\def\Vol{\mathop{\rm Vol}\nolimits}
\def\sign{\mathop{\rm sign}\nolimits}
\def\pmod#1{\mkern10mu({\rm mod}\,\,#1)} 

\def\tilde{\widetilde}
\def\ul{\underline}
\def\w{\wedge}
\def\ha{{1\over2}}
\def\half{{\textstyle{1\over2}}}
\def\quarter{{\textstyle{1\over4}}}
\def\eighth{{\textstyle{1\over8}}}
\def\smallfrac#1#2{{\textstyle{{#1}\over{#2}}}}
\def\ssk{{\mkern1mu}} 
\def\msk{{\mkern2mu}} 
\def\bsk{{\mkern4mu}} 
\def\Fsl{{\mkern3.0mu\raise.15ex\hbox{/}\mkern-11.5mu F}}
\def\Hsl{{\mkern2.0mu\raise.15ex\hbox{/}\mkern-12.5mu H}}
\def\Gsl{{\mkern3.0mu\raise.15ex\hbox{/}\mkern-11.5mu G}}
\def\Asl{{\mkern2.5mu\raise.15ex\hbox{/}\mkern-12.0mu A}}
\def\Fslhat{{\mkern3.0mu\raise.15ex\hbox{/}\mkern-11.5mu\skew4\hat{\vphantom{/}F}}}

\def\dil{{\rm dil}}
\def\tdil{{\t_\dil}}
\def\fib{{\rm fib}}
\def\base{{\rm base}}
\def\Tfib{{T^2_{\rm fib}}}
\def\Tbase{{T^4_{\rm base}}}
\def\flux{{\rm flux}}
\def\bg{{\rm bg}}
\def\hatP{\hat P}
\def\footlabel#1{\xdef#1{\the\ftno}} 
\def\nonnumberedsubsec#1{\ifnum\lastpenalty>9000\else\bigbreak\fi
\noindent{\it{#1}}\par\nobreak\medskip\nobreak}


\def\preIIBfermvar{(D.7)}
\def\presignconv#1{(A.3{#1})}
\def\pretris{(C.1)}
\def\pretriF{(C.5)}
\def\pregtprod{(A.17)}


\lref\IIBsugra{
J.~H.~Schwarz and P.~C.~West, ``Symmetries And Transformations Of
Chiral $\CN=2$ $D=10$ Supergravity,'' Phys.\ Lett.\ B {\bf 126}, 301
(1983)\semi
J.~H.~Schwarz, ``Covariant Field Equations Of Chiral $\CN=2$ $D=10$
Supergravity,'' Nucl.\ Phys.\ B {\bf 226}, 269 (1983)\semi
P.~S.~Howe and P.~C.~West, ``The Complete $\CN=2$, $D=10$
Supergravity,'' Nucl.\ Phys.\ B {\bf 238}, 181 (1984).
}

\lref\BCJ{
K.~Behrndt and C.~Jeschek, ``Fluxes in M-theory on 7-manifolds and $G$
structures,'' JHEP {\bf 0304}, 002 (2003) [arXiv:hep-th/0302047]\semi
K.~Behrndt and M.~Cvetic, ``Supersymmetric intersecting D6-branes and
fluxes in massive type IIA string theory,'' Nucl.\ Phys.\ B {\bf 676},
149 (2004) [arXiv:hep-th/0308045]\semi
K.~Behrndt and C.~Jeschek, ``Fluxes in M-theory on 7-manifolds:
$G$-structures and superpotential,'' arXiv:hep-th/0311119\semi
K.~Behrndt and M.~Cvetic, ``General $\CN=1$ supersymmetric flux vacua
of (massive) type IIA string theory,'' arXiv:hep-th/0403049.
}

\lref\BeckerM{
K.~Becker and M.~Becker, ``M-Theory on Eight-Manifolds,'' Nucl.\
Phys.\ B {\bf 477}, 155 (1996) [arXiv:hep-th/9605053].
}

\lref\BeckerNew{
K.~Becker, M.~Becker and R.~Sriharsha, ``PP-waves, M-theory and
fluxes,'' Nucl.\ Phys.\ B {\bf 676}, 172 (2004)
[arXiv:hep-th/0308014].
}

\lref\BEM{
P.~Bouwknegt, J.~Evslin and V.~Mathai, ``T-duality: Topology change
from $H$-flux,'' arXiv:hep-th/0306062.
}

\lref\BEMmore{
P.~Bouwknegt, J.~Evslin and V.~Mathai, ``On the topology and $H$-flux of
T-dual manifolds,'' arXiv:hep-th/0312052\semi
P.~Bouwknegt and D.~Ridout, ``A note on the equality of algebraic and
geometric D-brane charges in WZW models,'' arXiv:hep-th/0312259\semi
P.~Bouwknegt, K.~Hannabuss and V.~Mathai, ``T-duality for principal
torus bundles,'' arXiv:hep-th/0312284\semi
V.~Mathai and J.~Rosenberg, ``T-duality for torus bundles via
noncommutative topology,'' arXiv:hep-th/0401168.
}

\lref\Bergman{
O.~Bergman, E.~G.~Gimon and S.~Sugimoto, ``Orientifolds, RR torsion,
and K-theory,'' JHEP {\bf 0105}, 047 (2001) [arXiv:hep-th/0103183].
}

\lref\Bergshoeff{
E.~Bergshoeff, C.~M.~Hull and T.~Ort\'\i n, ``Duality in the type II
superstring effective action,'' Nucl.\ Phys.\ B {\bf 451}, 547 (1995)
[arXiv:hep-th/9504081].
}

\lref\BF{
A.~Buchel and A.~R.~Frey, ``Comments on supergravity dual of pure
$\CN=1$ super Yang Mills theory with unbroken chiral symmetry,'' Phys.\
Rev.\ D {\bf 64}, 064007 (2001) [arXiv:hep-th/0103022].
}

\lref\bubbles{
S.~Kachru, X.~Liu, M.~B.~Schulz and S.~P.~Trivedi, ``Supersymmetry
changing bubbles in string theory,'' JHEP {\bf 0305}, 014 (2003)
[arXiv:hep-th/0205108].
}

\lref\Candelas{
P.~Candelas, G.~T.~Horowitz, A.~Strominger and E.~Witten,
``Vacuum Configurations For Superstrings,''
Nucl.\ Phys.\ B {\bf 258}, 46 (1985).
}

\lref\CardBPS{
G.~L.~Cardoso, G.~Curio, G.~Dall'Agata and D.~L\"ust, ``BPS action and
superpotential for heterotic string compactifications with fluxes,''
JHEP {\bf 0310}, 004 (2003) [arXiv:hep-th/0306088].
}

\lref\CardTor{
G.~L.~Cardoso, G.~Curio, G.~Dall'Agata, D.~L\"ust, P.~Manousselis and
G.~Zoupanos, ``Non-Kaehler string backgrounds and their five torsion
classes,'' Nucl.\ Phys.\ B {\bf 652}, 5 (2003) [arXiv:hep-th/0211118].
}

\lref\Castellani{
L.~Castellani, A.~Ceresole, S.~Ferrara, R.~D'Auria, P.~Fr\'e and
E.~Maina, ``The Complete $\CN=3$ Matter Coupled Supergravity,'' Nucl.\
Phys.\ B {\bf 268}, 317 (1986).
}

\lref\CYduals{M.~B.~Schulz, work in progress.}

\lref\DallAgataIIA{
G.~Dall'Agata and N.~Prezas, ``$\CN = 1$ geometries for M-theory and
type IIA strings with fluxes,'' Phys.\ Rev.\ D {\bf 69}, 066004 (2004)
[arXiv:hep-th/0311146].
}

\lref\DallAgataIIB{
G.~Dall'Agata, ``On supersymmetric solutions of type IIB supergravity
with general fluxes,'' arXiv:hep-th/0403220.
}

\lref\DeWolfe{
O.~DeWolfe and S.~B.~Giddings, ``Scales and hierarchies in warped
compactifications and brane worlds,'' Phys.\ Rev.\ D {\bf 67}, 066008
(2003) [arXiv:hep-th/0208123].
}

\lref\deSitterandInflation{
K.~Dasgupta, C.~Herdeiro, S.~Hirano and R.~Kallosh, ``D3/D7
inflationary model and M-theory,'' Phys.\ Rev.\ D {\bf 65}, 126002
(2002) [arXiv:hep-th/0203019]\semi
M.~Fabinger and E.~Silverstein, ``D-Sitter space: Causal structure,
thermodynamics, and entropy,'' arXiv:hep-th/0304220\semi
A.~R.~Frey, M.~Lippert and B.~Williams, ``The fall of stringy de
Sitter,'' Phys.\ Rev.\ D {\bf 68}, 046008 (2003)
[arXiv:hep-th/0305018]\semi
S.~Kachru, R.~Kallosh, A.~Linde, J.~Maldacena, L.~McAllister and
S.~P.~Trivedi, ``Towards inflation in string theory,'' JCAP {\bf
0310}, 013 (2003) [arXiv:hep-th/0308055]\semi
C.~P.~Burgess, R.~Kallosh and F.~Quevedo, ``de Sitter string vacua
from supersymmetric $D$-terms,'' JHEP {\bf 0310}, 056 (2003)
[arXiv:hep-th/0309187]\semi
E.~Silverstein and D.~Tong, ``Scalar speed limits and cosmology:
Acceleration from D-cceleration,'' arXiv:hep-th/0310221\semi
J.~P.~Hsu, R.~Kallosh and S.~Prokushkin, ``On brane inflation with
volume stabilization,'' JCAP {\bf 0312}, 009 (2003)
[arXiv:hep-th/0311077]\semi
A.~Buchel and R.~Roiban, ``Inflation in warped geometries,''
arXiv:hep-th/0311154\semi
F.~Koyama, Y.~Tachikawa and T.~Watari, ``Supergravity analysis of
hybrid inflation model from D3-D7 system,'' Phys.\ Rev.\ D {\bf 69},
106001 (2004) [arXiv:hep-th/0311191]\semi
P.~G.~Camara, L.~E.~Ibanez and A.~M.~Uranga, ``Flux-induced
SUSY-breaking soft terms,'' arXiv:hep-th/0311241\semi
H.~Firouzjahi and S.~H.~H.~Tye, ``Closer towards inflation in string
theory,'' Phys.\ Lett.\ B {\bf 584}, 147 (2004)
[arXiv:hep-th/0312020]\semi
A.~Buchel, ``On effective action of string theory flux
compactifications,'' Phys.\ Rev.\ D {\bf 69}, 106004 (2004)
[arXiv:hep-th/0312076]\semi
R.~Brustein and S.~P.~de Alwis, ``Moduli potentials in string
compactifications with fluxes: Mapping the discretuum,''
arXiv:hep-th/0402088\semi
L.~Kofman, A.~Linde, X.~Liu, A.~Maloney, L.~McAllister and
E.~Silverstein, ``Beauty is attractive: Moduli trapping at enhanced
symmetry points,'' arXiv:hep-th/0403001\semi
O.~DeWolfe, S.~Kachru and H.~Verlinde, ``The giant inflaton,''
arXiv:hep-th/0403123\semi
N.~Iizuka and S.~P.~Trivedi, ``An inflationary model in string
theory,'' arXiv:hep-th/0403203\semi
M.~Alishahiha, E.~Silverstein and D.~Tong, ``DBI in the sky,''
arXiv:hep-th/0404084\semi
M.~Berg, M.~Haack and B.~K\"ors, ``Loop corrections to volume moduli
and inflation in string theory,'' arXiv:hep-th/0404087\semi
A.~Buchel and A.~Ghodsi, ``Braneworld inflation,''
arXiv:hep-th/0404151.
}

\lref\Douglas{
L.~Susskind, ``The anthropic landscape of string theory,''
arXiv:hep-th/0302219\semi
M.~R.~Douglas, ``The statistics of string/M theory vacua,'' JHEP
{\bf 0305}, 046 (2003) [arXiv:hep-th/0303194]\semi
S.~Ashok and M.~R.~Douglas, ``Counting flux vacua,''
arXiv:hep-th/0307049\semi
T.~Banks, M.~Dine and E.~Gorbatov, ``Is there a string theory
landscape?,'' arXiv:hep-th/0309170\semi
M.~R.~Douglas, ``Statistics of string vacua,''
arXiv:hep-ph/0401004\semi
M.~Dine, ``Is there a string theory landscape: Some cautionary
notes,'' arXiv:hep-th/0402101\semi
M.~R.~Douglas, B.~Shiffman and S.~Zelditch, ``Critical points and
supersymmetric vacua,'' arXiv:math.cv/0402326\semi
F.~Denef and M.~R.~Douglas, ``Distributions of flux vacua,''
arXiv:hep-th/0404116\semi
A.~Giryavets, S.~Kachru and P.~K.~Tripathy, ``On the taxonomy of flux
vacua,'' arXiv:hep-th/0404243\semi
L.~Susskind, ``Supersymmetry breaking in the anthropic landscape,''
arXiv:hep-th/0405189\semi
M.~R.~Douglas, ``Statistical analysis of the supersymmetry breaking
scale,'' arXiv:hep-th/0405279.
}

\lref\DRS{
K.~Dasgupta, G.~Rajesh and S.~Sethi, ``M theory, orientifolds and
$G$-flux,'' JHEP {\bf 9908}, 023 (1999) [arXiv:hep-th/9908088].
}

\lref\DRSrefs{
C.~Vafa, ``Evidence for F-Theory,'' Nucl.\ Phys.\ B {\bf 469}, 403
(1996) [arXiv:hep-th/9602022]\semi
S.~Sethi, C.~Vafa and E.~Witten, ``Constraints on low-dimensional
string compactifications,'' Nucl.\ Phys.\ B {\bf 480}, 213 (1996)
[arXiv:hep-th/9606122]\semi
K.~Dasgupta and S.~Mukhi, ``A note on low-dimensional string
compactifications,'' Phys.\ Lett.\ B {\bf 398}, 285 (1997)
[arXiv:hep-th/9612188].
}
 
\lref\EvaScaling{
A.~Saltman and E.~Silverstein, ``The scaling of the no-scale potential
and de Sitter model building,'' arXiv:hep-th/0402135.
}

\lref\FG{
A.~R.~Frey and M.~Grana, ``Type IIB solutions with interpolating
supersymmetries,'' Phys.\ Rev.\ D {\bf 68}, 106002 (2003)
[arXiv:hep-th/0307142].
}

\lref\FP{
A.~R.~Frey and J.~Polchinski, ``$\CN=3$ warped compactifications,''
Phys.\ Rev.\ D {\bf 65}, 126009 (2002) [arXiv:hep-th/0201029].
}

\lref\Fre{
P.~Fr\'e, ``Lectures on Special Kahler Geometry and Electric--Magnetic
Duality Rotations,'' Nucl.\ Phys.\ Proc.\ Suppl.\ {\bf 45BC}, 59
(1996) [arXiv:hep-th/9512043].
}

\lref\FreySU{
A.~R.~Frey, ``Notes on $SU(3)$ structures in type IIB supergravity,''
arXiv:hep-th/0404107.
}

\lref\FreyThesis{
A.~R.~Frey, ``Warped strings: Self-dual flux and contemporary
compactifications,'' arXiv:hep-th/0308156.
}

\lref\Gates{
S.~J.~J.~Gates and V.~A.~Kostelecky, ``Supersymmetric Matter Gravitino
Multiplets,'' Nucl.\ Phys.\ B {\bf 248}, 570 (1984).
}

\lref\gaugedsugraI{
G.~Dall'Agata, ``Type IIB supergravity compactified on a Calabi-Yau
manifold with H-fluxes,'' JHEP {\bf 0111}, 005 (2001)
[arXiv:hep-th/0107264]\semi
L.~Andrianopoli, R.~D'Auria, S.~Ferrara and M.~A.~Lledo, ``Super Higgs
effect in extended supergravity,'' Nucl.\ Phys.\ B {\bf 640}, 46
(2002) [arXiv:hep-th/0202116]\semi
L.~Andrianopoli, R.~D'Auria, S.~Ferrara and M.~A.~Lledo, ``Gauging of
flat groups in four dimensional supergravity,'' JHEP {\bf 0207}, 010
(2002) [arXiv:hep-th/0203206]\semi
L.~Andrianopoli, R.~D'Auria, S.~Ferrara and M.~A.~Lledo, ``Duality and
spontaneously broken supergravity in flat backgrounds,'' Nucl.\ Phys.\
B {\bf 640}, 63 (2002) [arXiv:hep-th/0204145]\semi
R.~D'Auria, S.~Ferrara and S.~Vaula, ``$\CN=4$ gauged supergravity and a
IIB orientifold with fluxes,'' New J.\ Phys.\ {\bf 4}, 71 (2002)
[arXiv:hep-th/0206241]\semi
R.~D'Auria, S.~Ferrara, M.~A.~Lledo and S.~Vaula, ``No-scale $\CN=4$
supergravity coupled to Yang-Mills: The scalar potential and super
Higgs effect,'' Phys.\ Lett.\ B {\bf 557}, 278 (2003)
[arXiv:hep-th/0211027]\semi
S.~Ferrara, ``Duality, gauging and superHiggs effect in string and
M-theory,'' arXiv:hep-th/0211116\semi
L.~Andrianopoli, R.~D'Auria, S.~Ferrara and M.~A.~Lledo, ``$\CN=2$
super-Higgs, $\CN=1$ Poincar\'e vacua and quaternionic geometry,'' JHEP
{\bf 0301}, 045 (2003) [arXiv:hep-th/0212236]\semi
C.~Angelantonj, R.~D'Auria, S.~Ferrara and M.~Trigiante, ``$K3\times
T^2/\IZ_2$ orientifolds with fluxes, open string moduli and critical
points,'' Phys.\ Lett.\ B {\bf 583}, 331 (2004)
[arXiv:hep-th/0312019]\semi
M.~Grana, T.~W.~Grimm, H.~Jockers and J.~Louis, ``Soft supersymmetry
breaking in Calabi-Yau orientifolds with D-branes and fluxes,''
arXiv:hep-th/0312232\semi
T.~W.~Grimm and J.~Louis, ``The effective action of $\CN=1$ Calabi-Yau
orientifolds,'' arXiv:hep-th/0403067.
}

\lref\gaugedsugraII{
C.~Angelantonj, S.~Ferrara and M.~Trigiante, ``New $D=4$ gauged
supergravities from $\CN=4$ orientifolds with fluxes,'' JHEP {\bf 0310},
015 (2003) [arXiv:hep-th/0306185]\semi
C.~Angelantonj, S.~Ferrara and M.~Trigiante, ``Unusual gauged
supergravities from type IIA and type IIB orientifolds,'' Phys.\
Lett.\ B {\bf 582}, 263 (2004) [arXiv:hep-th/0310136]\semi
R.~D'Auria, S.~Ferrara and M.~Trigiante, ``Homogeneous special
manifolds, orientifolds and solvable coordinates,''
arXiv:hep-th/0403204.
}

\lref\GauntAdS{
J.~P.~Gauntlett, D.~Martelli, J.~Sparks and D.~Waldram,
``Supersymmetric $AdS_5$ solutions of M-theory,''
arXiv:hep-th/0402153.
}

\lref\GauntGP{
J.~P.~Gauntlett, J.~B.~Gutowski and S.~Pakis, ``The geometry of $D=11$
null Killing spinors,'' JHEP {\bf 0312}, 049 (2003)
[arXiv:hep-th/0311112].
}

\lref\GauntIntr{
J.~P.~Gauntlett, D.~Martelli and D.~Waldram, ``Superstrings with
intrinsic torsion,'' arXiv:hep-th/0302158.
}

\lref\GauntPakis{
J.~P.~Gauntlett and S.~Pakis, ``The geometry of $D=11$ Killing
spinors,'' JHEP {\bf 0304}, 039 (2003) [arXiv:hep-th/0212008].
}

\lref\GauntSLAG{
J.~P.~Gauntlett, N.~Kim, D.~Martelli and D.~Waldram, ``Fivebranes
wrapped on SLAG three-cycles and related geometry,'' JHEP {\bf 0111},
018 (2001) [arXiv:hep-th/0110034]\semi
J.~P.~Gauntlett, D.~Martelli, S.~Pakis and D.~Waldram, ``$G$-structures
and wrapped NS5-branes,'' arXiv:hep-th/0205050.
}

\lref\genholonomy{
M.~J.~Duff and J.~T.~Liu, ``Hidden spacetime symmetries and
generalized holonomy in M-theory,'' Nucl.\ Phys.\ B {\bf 674}, 217
(2003) [arXiv:hep-th/0303140]\semi
C.~Hull, ``Holonomy and symmetry in M-theory,''
arXiv:hep-th/0305039\semi 
G.~Papadopoulos and D.~Tsimpis, ``The holonomy of the supercovariant
connection and Killing spinors,'' JHEP {\bf 0307}, 018 (2003)
[arXiv:hep-th/0306117]\semi
G.~Papadopoulos and D.~Tsimpis, ``The holonomy of IIB supercovariant
connection,'' Class.\ Quant.\ Grav.\ {\bf 20}, L253 (2003)
[arXiv:hep-th/0307127]\semi
A.~Batrachenko, M.~J.~Duff, J.~T.~Liu and W.~Y.~Wen, ``Generalized
holonomy of M-theory vacua,'' arXiv:hep-th/0312165\semi
A.~Batrachenko and W.~Y.~Wen, ``Generalized holonomy of supergravities
with 8 real supercharges,'' arXiv:hep-th/0402141.
}

\lref\geotransitions{
R.~Gopakumar and C.~Vafa, ``On the gauge theory/geometry
correspondence,'' Adv.\ Theor.\ Math.\ Phys.\ {\bf 3}, 1415 (1999)
[arXiv:hep-th/9811131]\semi
R.~Dijkgraaf and C.~Vafa, ``Matrix models, topological strings, and
supersymmetric gauge theories,'' Nucl.\ Phys.\ B {\bf 644}, 3 (2002)
[arXiv:hep-th/0206255].
}

\lref\GKP{
S.~B.~Giddings, S.~Kachru and J.~Polchinski, ``Hierarchies from fluxes
in string compactifications,'' Phys.\ Rev.\ D {\bf 66}, 106006 (2002)
[arXiv:hep-th/0105097].
}

\lref\GM{S.~B.~Giddings and A.~Maharana, work in progress.}

\lref\GP{
M.~Gra\~na and J.~Polchinski, ``Gauge/gravity duals with holomorphic
dilaton,'' Phys.\ Rev.\ D {\bf 65}, 126005 (2002)
[arXiv:hep-th/0106014].
}

\lref\Gray{
A.~Gray and L.~M.~Hervella, ``The sixteen classes of almost Hermitian
manifolds and their linear invariants,'' Ann.\ Math.\ Pura Appl.\ {\bf
123} 35 (1980)\semi
S.~Chiossi and S.~Salamon, ``The Intrinsic Torsion of $SU(3)$ and
$G_2$ Structures,'' in {\it Differential Geometry, Valencia 2001\/},
World Sci.\ Publishing, River Edge, NJ, 115 (2002)
[arXiv:math.DG/0202282].  }

\lref\Gutowski{
J.~Gutowski, G.~Papadopoulos and P.~K.~Townsend, ``Supersymmetry and
generalized calibrations,'' Phys.\ Rev.\ D {\bf 60}, 106006 (1999)
[arXiv:hep-th/9905156]\semi
J.~Gutowski and G.~Papadopoulos, ``$AdS$ calibrations,'' Phys.\ Lett.\ B
{\bf 462}, 81 (1999) [arXiv:hep-th/9902034]\semi
J.~Gutowski, ``Generalized calibrations,'' arXiv:hep-th/9909096.
}

\lref\GVW{
S.~Gukov, C.~Vafa and E.~Witten, ``CFT's from Calabi-Yau four-folds,''
Nucl.\ Phys.\ B {\bf 584}, 69 (2000) [Erratum-ibid.\ B {\bf 608}, 477
(2001)] [arXiv:hep-th/9906070].
}

\lref\HananyKol{
A.~Hanany and B.~Kol, ``On orientifolds, discrete torsion, branes and
M theory,'' JHEP {\bf 0006}, 013 (2000) [arXiv:hep-th/0003025].
}

\lref\Hassan{
S.~F.~Hassan, ``T-duality, space-time spinors and R-R fields in curved
backgrounds,'' Nucl.\ Phys.\ B {\bf 568}, 145 (2000)
[arXiv:hep-th/9907152].
}

\lref\HullTor{
C.~M.~Hull, ``Superstring Compactifications With Torsion And
Space-Time Supersymmetry,'' Print-86-0251 (CAMBRIDGE).
}

\lref\HumboldtPSflow{
G.~L.~Cardoso, G.~Curio, G.~Dall'Agata and D.~L\"ust, work in
progress.
}

\lref\Husain{
T.~Z.~Husain, ``M2-branes wrapped on holomorphic curves,'' JHEP {\bf
0312}, 037 (2003) [arXiv:hep-th/0211030]\semi
T.~Z.~Husain, ``That's a wrap!,'' JHEP {\bf 0304}, 053 (2003)
[arXiv:hep-th/0302071].
}

\lref\Johnson{
C.~V.~Johnson, ``D-brane primer,'' arXiv:hep-th/0007170.
}

\lref\Keshav{
K.~Becker and K.~Dasgupta, ``Heterotic strings with torsion,'' JHEP
{\bf 0211}, 006 (2002) [arXiv:hep-th/0209077]\semi
K.~Becker, M.~Becker, K.~Dasgupta and P.~S.~Green, ``Compactifications
of heterotic theory on non-Kaehler complex manifolds.  I,'' JHEP {\bf
0304}, 007 (2003) [arXiv:hep-th/0301161]\semi
K.~Becker, M.~Becker, K.~Dasgupta and S.~Prokushkin, ``Properties of
heterotic vacua from superpotentials,'' Nucl.\ Phys.\ B {\bf 666}, 144
(2003) [arXiv:hep-th/0304001].
}

\lref\Keshavgeo{
M.~Becker, K.~Dasgupta, A.~Knauf and R.~Tatar, ``Geometric
transitions, flops and non-Kaehler manifolds. I,''
arXiv:hep-th/0403288.
}

\lref\KeshavHetII{
K.~Becker, M.~Becker, P.~S.~Green, K.~Dasgupta and E.~Sharpe,
``Compactifications of heterotic strings on non-Kaehler complex
manifolds. II,'' arXiv:hep-th/0310058.
}

\lref\KKLT{
S.~Kachru, R.~Kallosh, A.~Linde and S.~P.~Trivedi, ``De Sitter vacua
in string theory,'' Phys.\ Rev.\ D {\bf 68}, 046005 (2003)
[arXiv:hep-th/0301240].
}

\lref\KreuzerSkarke{
M.~Kreuzer and H.~Skarke,
``Reflexive polyhedra, weights and toric Calabi-Yau fibrations,''
Rev.\ Math.\ Phys.\  {\bf 14}, 343 (2002)
[arXiv:math.ag/0001106]\semi
M.~Kreuzer and H.~Skarke, ``Complete classification of reflexive
polyhedra in four dimensions,'' Adv.\ Theor.\ Math.\ Phys.\ {\bf 4},
1209 (2002) [arXiv:hep-th/0002240].
}

\lref\KS{
I.~R.~Klebanov and M.~J.~Strassler, ``Supergravity and a confining
gauge theory: Duality cascades and $\chi$SB-resolution of naked
singularities,'' JHEP {\bf 0008}, 052 (2000) [arXiv:hep-th/0007191].
}

\lref\KST{
S.~Kachru, M.~B.~Schulz and S.~Trivedi, ``Moduli stabilization from
fluxes in a simple IIB orientifold,'' JHEP {\bf 0310}, 007 (2003)
[arXiv:hep-th/0201028].
}

\lref\KSTT{
S.~Kachru, M.~B.~Schulz, P.~K.~Tripathy and S.~P.~Trivedi, ``New
supersymmetric string compactifications,'' JHEP {\bf 0303}, 061 (2003)
[arXiv:hep-th/0211182].
}

\lref\Kthreelift{
N.~Seiberg, ``IR dynamics on branes and space-time geometry,'' Phys.\
Lett.\ B {\bf 384}, 81 (1996) [arXiv:hep-th/9606017]\semi
N.~Seiberg and E.~Witten, ``Gauge dynamics and compactification to
three dimensions,'' arXiv:hep-th/9607163.
}

\lref\Louis{
S.~Gurrieri, J.~Louis, A.~Micu and D.~Waldram, ``Mirror symmetry in
generalized Calabi-Yau compactifications,'' Nucl.\ Phys.\ B {\bf 654},
61 (2003) [arXiv:hep-th/0211102].
}

\lref\LM{
A.~Lawrence and J.~McGreevy, ``Local string models of soft
supersymmetry breaking,'' arXiv:hep-th/0401034.
}

\lref\MaharanaSchwarz{
J.~Maharana and J.~H.~Schwarz, ``Noncompact Symmetries In String
Theory,'' Nucl.\ Phys.\ B {\bf 390}, 3 (1993) [arXiv:hep-th/9207016].
}

\lref\Marino{
P.~Candelas, ``Lectures On Complex Manifolds,'' in {\it Superstrings
87}, proceedings of the Trieste Spring School (1987) 1-88\semi
M.~Mari\~no, R.~Minasian, G.~W.~Moore and A.~Strominger, ``Nonlinear
instantons from supersymmetric $p$-branes,'' JHEP {\bf 0001}, 005
(2000) [arXiv:hep-th/9911206].
}

\lref\Marolf{
D.~Marolf, ``Chern-Simons terms and the three notions of charge,''
arXiv:hep-th/0006117.
}

\lref\MartelliSparks{
D.~Martelli and J.~Sparks, ``$G$-structures, fluxes and calibrations
in M-theory,'' Phys.\ Rev.\ D {\bf 68}, 085014 (2003)
[arXiv:hep-th/0306225].
}

\lref\Minetal{
P.~Kaste, R.~Minasian, M.~Petrini and A.~Tomasiello, ``Kaluza-Klein
bundles and manifolds of exceptional holonomy,'' JHEP {\bf 0209}, 033
(2002) [arXiv:hep-th/0206213]\semi
P.~Kaste, R.~Minasian, M.~Petrini and A.~Tomasiello, ``Nontrivial RR
two-form field strength and $SU(3)$-structure,'' Fortsch.\ Phys.\ {\bf
51}, 764 (2003) [arXiv:hep-th/0301063]\semi
P.~Kaste, R.~Minasian and A.~Tomasiello, ``Supersymmetric M-theory
compactifications with fluxes on seven-manifolds and $G$-structures,''
JHEP {\bf 0307}, 004 (2003) [arXiv:hep-th/0303127].
}

\lref\MinMirror{
S.~Fidanza, R.~Minasian and A.~Tomasiello, ``Mirror symmetric
$SU(3)$-structure manifolds with NS fluxes,'' arXiv:hep-th/0311122.
}

\lref\modstab{
J.~Polchinski and A.~Strominger, ``New Vacua for Type II String
Theory,'' Phys.\ Lett.\ B {\bf 388}, 736 (1996)
[arXiv:hep-th/9510227]\semi
T.~R.~Taylor and C.~Vafa, ``RR flux on Calabi-Yau and partial
supersymmetry breaking,'' Phys.\ Lett.\ B {\bf 474}, 130 (2000)
[arXiv:hep-th/9912152]\semi
P.~Mayr, ``On supersymmetry breaking in string theory and its
realization in brane worlds,'' Nucl.\ Phys.\ B {\bf 593}, 99 (2001)
[arXiv:hep-th/0003198]\semi
B.~R.~Greene, K.~Schalm and G.~Shiu, ``Warped compactifications in M
and F theory,'' Nucl.\ Phys.\ B {\bf 584}, 480 (2000)
[arXiv:hep-th/0004103]\semi
G.~Curio, A.~Klemm, D.~L\"ust and S.~Theisen, ``On the vacuum structure
of type II string compactifications on Calabi-Yau spaces with
$H$-fluxes,'' Nucl.\ Phys.\ B {\bf 609}, 3 (2001)
[arXiv:hep-th/0012213]\semi
G.~Curio, A.~Klemm, B.~K\"ors and D.~L\"ust, ``Fluxes in heterotic and
type II string compactifications,'' Nucl.\ Phys.\ B {\bf 620}, 237
(2002) [arXiv:hep-th/0106155]\semi
G.~Curio, B.~K\"ors and D.~L\"ust, ``Fluxes and branes in type II vacua
and M-theory geometry with $G_2$ and Spin(7) holonomy,'' Nucl.\ Phys.\
B {\bf 636}, 197 (2002) [arXiv:hep-th/0111165]\semi
J.~Louis and A.~Micu, ``Type II theories compactified on Calabi-Yau
threefolds in the presence of background fluxes,'' Nucl.\ Phys.\ B
{\bf 635}, 395 (2002) [arXiv:hep-th/0202168]\semi
K.~Becker, M.~Becker, M.~Haack and J.~Louis, ``Supersymmetry breaking
and $\alpha'$-corrections to flux induced potentials,'' JHEP {\bf
0206}, 060 (2002) [arXiv:hep-th/0204254]\semi
A.~R.~Frey and A.~Mazumdar, ``3-form induced potentials, dilaton
stabilization, and running moduli,'' Phys.\ Rev.\ D {\bf 67}, 046006
(2003) [arXiv:hep-th/0210254]\semi
R.~Blumenhagen, D.~L\"ust and T.~R.~Taylor, ``Moduli stabilization in
chiral type IIB orientifold models with fluxes,'' Nucl.\ Phys.\ B {\bf
663}, 319 (2003) [arXiv:hep-th/0303016]\semi
M.~Berg, M.~Haack and B.~K\"ors, ``An orientifold with fluxes and
branes via T-duality,'' Nucl.\ Phys.\ B {\bf 669}, 3 (2003)
[arXiv:hep-th/0305183]\semi
S.~P.~de Alwis, ``On potentials from fluxes,'' Phys.\ Rev.\ D {\bf
68}, 126001 (2003) [arXiv:hep-th/0307084]\semi
J.~F.~G.~Cascales and A.~M.~Uranga, ``Chiral 4d string vacua with
D-branes and moduli stabilization,'' arXiv:hep-th/0311250\semi
J.~F.~G.~Cascales, M.~P.~Garcia del Moral, F.~Quevedo and
A.~M.~Uranga, ``Realistic D-brane models on warped throats: Fluxes,
hierarchies and moduli stabilization,'' JHEP {\bf 0402}, 031 (2004)
[arXiv:hep-th/0312051]\semi
D.~L\"ust, S.~Reffert and S.~Stieberger, ``Flux-induced Soft
Supersymmetry Breaking in Chiral Type IIB Orientifolds with
D3/D7-Branes,'' arXiv:hep-th/0406092.
}

\lref\moreKKLT{
F.~Denef, M.~R.~Douglas and B.~Florea, ``Building a better
racetrack,'' arXiv:hep-th/0404257.
}

\lref\MTW{
C.~W.~Misner, K.~S.~Thorne and J.~A.~Wheeler, ``Gravitation,''
W.~H.~Freeman and Company, New York (2000).}

\lref\Myers{
R.~C.~Myers, ``Dielectric-branes,'' JHEP {\bf 9912}, 022 (1999)
[arXiv:hep-th/9910053].
}

\lref\Nakahara{
M.~Nakahara, ``Geometry, Topology and Physics,'' (Graduate student
series in physics), Institute of Physics Publishing, Bristol, UK
(1990).
}

\lref\NoGo{
B.~de Wit, D.~J.~Smit and N.~D.~Hari Dass, ``Residual Supersymmetry Of
Compactified $D=10$ Supergravity,'' Nucl.\ Phys.\ B {\bf 283}, 165
(1987)\semi
J.~M.~Maldacena and C.~Nu\~nez, ``Supergravity description of field
theories on curved manifolds and a no go theorem,'' Int.\ J.\ Mod.\
Phys.\ A {\bf 16}, 822 (2001) [arXiv:hep-th/0007018]\semi
S.~Ivanov and G.~Papadopoulos, ``A no-go theorem for string warped
compactifications,'' Phys.\ Lett.\ B {\bf 497}, 309 (2001)
[arXiv:hep-th/0008232].
}

\lref\nongeo{
S.~Hellerman, J.~McGreevy and B.~Williams, ``Geometric constructions
of nongeometric string theories,'' arXiv:hep-th/0208174\semi
A.~Dabholkar and C.~Hull, ``Duality twists, orbifolds, and fluxes,''
JHEP {\bf 0309}, 054 (2003) [arXiv:hep-th/0210209]\semi
A.~Flournoy, B.~Wecht and B.~Williams, ``Constructing nongeometric
vacua in string theory,'' arXiv:hep-th/0404217.
}

\lref\noscale{
S.~Ferrara and M.~Porrati, ``$\CN=1$ no-scale supergravity from IIB
orientifolds,'' Phys.\ Lett.\ B {\bf 545}, 411 (2002)
[arXiv:hep-th/0207135]\semi
L.~Andrianopoli, R.~D'Auria, S.~Ferrara and M.~A.~Lledo, ``Gauged
extended supergravity without cosmological constant: No-scale
structure and supersymmetry breaking,'' Mod.\ Phys.\ Lett.\ A {\bf
18}, 1001 (2003) [arXiv:hep-th/0212141].
}

\lref\oldnoscale{
E.~Cremmer, S.~Ferrara, C.~Kounnas and D.~V.~Nanopoulos,
``Naturally Vanishing Cosmological Constant In $\CN=1$ Supergravity,''
Phys.\ Lett.\ B {\bf 133}, 61 (1983)\semi
J.~R.~Ellis, A.~B.~Lahanas, D.~V.~Nanopoulos and K.~Tamvakis,
``No-Scale Supersymmetric Standard Model,''
Phys.\ Lett.\ B {\bf 134}, 429 (1984).
}

\lref\otherTdual{
J.~F.~G.~Cascales and A.~M.~Uranga, ``M5-brane geometries, T-duality
and fluxes,'' JHEP {\bf 0401}, 021 (2004) [arXiv:hep-th/0307156]\semi
M.~Serone and M.~Trapletti, ``String vacua with flux from
freely-acting obifolds,'' JHEP {\bf 0401}, 012 (2004)
[arXiv:hep-th/0310245].
}

\lref\Prokushkin{
E.~Goldstein and S.~Prokushkin, ``Geometric model for complex
non-Kaehler manifolds with $SU(3)$ structure,'' arXiv:hep-th/0212307.
}

\lref\PS{
J.~Polchinski and M.~J.~Strassler, ``The string dual of a confining
four-dimensional gauge theory,'' arXiv:hep-th/0003136.
}

\lref\RSIandII{
L.~Randall and R.~Sundrum, `A large mass hierarchy from a small extra
dimension,'' Phys.\ Rev.\ Lett.\ {\bf 83}, 3370 (1999)
[arXiv:hep-ph/9905221]\semi
L.~Randall and R.~Sundrum, ``An alternative to compactification,''
Phys.\ Rev.\ Lett.\ {\bf 83}, 4690 (1999) [arXiv:hep-th/9906064].
}

\lref\Salamon{
S.~Salamon, ``Complex structures on nilpotent Lie algebras,''
arXiv:math.DG/9808025\semi
E.~Abbena, S.~Garbiero, S.~Salamon, ``Almost Hermitian Geometry on Six
Dimensional Nilmanifolds,'' arXiv:math.DG/0007066.
}

\lref\ScherkSchwarz{
J.~Scherk and J.~H.~Schwarz, ``How To Get Masses From Extra
Dimensions,'' Nucl.\ Phys.\ B {\bf 153}, 61 (1979).
}

\lref\Sen{
A.~Sen, ``A note on enhanced gauge symmetries in M- and string
theory,'' JHEP {\bf 9709}, 001 (1997) [arXiv:hep-th/9707123].
}

\lref\SethiBarren{
D.~Robbins and S.~Sethi, ``A barren landscape,'' arXiv:hep-th/0405011.
}

\lref\Strom{
A.~Strominger, ``Superstrings With Torsion,'' Nucl.\ Phys.\ B {\bf
274}, 253 (1986).
}

\lref\SUSYhiggs{
R.~D'Auria, S.~Ferrara, F.~Gargiulo, M.~Trigiante and S.~Vaula,
``$\CN=4$ supergravity Lagrangian for type IIB on $T^6/\IZ_2$ in
presence of fluxes and D3-branes,'' JHEP {\bf 0306}, 045 (2003)
[arXiv:hep-th/0303049].
}

\lref\Tripathy{
P.~K.~Tripathy and S.~P.~Trivedi, ``Compactification with flux on $K3$
and tori,'' JHEP {\bf 0303}, 028 (2003) [arXiv:hep-th/0301139]\semi
A.~Giryavets, S.~Kachru, P.~K.~Tripathy and S.~P.~Trivedi, ``Flux
compactifications on Calabi-Yau threefolds,'' JHEP {\bf 0404}, 003
(2004) [arXiv:hep-th/0312104].
}

\lref\Verlinde{
H.~Verlinde, ``Holography and compactification,'' Nucl.\ Phys.\ B {\bf
580}, 264 (2000) [arXiv:hep-th/9906182]\semi
C.~S.~Chan, P.~L.~Paul and H.~Verlinde, ``A note on warped string
compactification,'' Nucl.\ Phys.\ B {\bf 581}, 156 (2000)
[arXiv:hep-th/0003236].
}

\lref\Warneretal{
C.~N.~Gowdigere and N.~P.~Warner, ``Flowing with eight supersymmetries
in M-theory and F-theory,'' JHEP {\bf 0312}, 048 (2003)
[arXiv:hep-th/0212190]\semi
C.~N.~Pope and N.~P.~Warner, ``A dielectric flow solution with maximal
supersymmetry,'' arXiv:hep-th/0304132\semi
C.~N.~Gowdigere, D.~Nemeschansky and N.~P.~Warner, ``Supersymmetric
solutions with fluxes from algebraic Killing spinors,''
arXiv:hep-th/0306097\semi
K.~Pilch and N.~P.~Warner, ``Generalizing the $\CN=2$ supersymmetric RG
flow solution of IIB supergravity,'' Nucl.\ Phys.\ B {\bf 675}, 99
(2003) [arXiv:hep-th/0306098]\semi
K.~Pilch and N.~P.~Warner, ``$\CN=1$ supersymmetric solutions of IIB
supergravity from Killing spinors,'' arXiv:hep-th/0403005\semi
D.~Nemeschansky and N.~P.~Warner, ``A family of M-theory flows with
four supersymmetries,'' arXiv:hep-th/0403006.
}

\lref\WittenNPSup{
E.~Witten, ``Non-Perturbative Superpotentials In String Theory,''
Nucl.\ Phys.\ B {\bf 474}, 343 (1996) [arXiv:hep-th/9604030].
}


\Title{\vbox{\hbox{hep-th/0406001}\hbox{CALT-68-2471}
\hbox{NSF-ITP-03-108}}}
{\vbox{\baselineskip=22pt
\centerline{Superstring Orientifolds with Torsion:}
\centerline{O5 Orientifolds of Torus Fibrations}
\centerline{and their Massless Spectra}}}
\centerline{Michael B.~Schulz}
\bigskip\centerline{\it {California Institute of Technology 452-48}}
\centerline{{\it Pasadena, CA 91125 USA}}

\vskip .3in 

Using a ``Superstrings with Torsion'' type description, we study a
class of IIB orientifolds in which spacefilling O5 planes and D5
branes wrap the $T^2$ fiber in a warped modification of the product of
4D Minkowski space and a $T^2$ fibration.  For the case that the base
is $T^4$, we provide examples that preserve 4D $\CN=1$, 2, and 3
supersymmetry, both with internal RR flux, and with a combination of
internal RR and NS flux.  In these examples, the internal geometries
admit integrable complex structure; however, the almost complex
structure selected by the supersymmetry conditions is nonintegrable in
the case that there is NS flux.  We indicate explicitly the massless
spectrum of gauge fields and moduli in each example.  In a previous
investigation, this class of orientifolds was studied using T-duality.
Here, we extend the previous analysis, first by providing an intrinsic
description that does not rely on duality, and then by elaborating on
details of the T-duality map, which we use to check our results.

\Date{1 June 2004}


\def\TOCchapter#1#2#3{\noindent {\hyperref{}{section}{#1}{#1.}} {#2} 
\leaderfill{\hyperref{}{page}{#3}{#3}} \par}
\centerline{\bf Contents}\nobreak\medskip%
{\baselineskip=12pt\footnotefont\parskip=0pt
\TOCchapter{1}{Introduction}{1}
\TOCchapter{2}{{O5 Orientifolds of $T^2$ Fibrations}}{9}
\TOCchapter{3}{Equations of Motion and Integrability Constraints}{14}
\TOCchapter{4}{Supersymmetry Conditions}{27}
\TOCchapter{5}{Relation to T-Dual O3 Orientifolds}{41}
\TOCchapter{6}{$\CN=2$ Examples}{48}
\TOCchapter{7}{$\CN=3$ Example: The Warped Iwasawa Manifold}{62}
\TOCchapter{8}{$\CN=1$ Example}{71}
\TOCchapter{9}{Preview: $\CN=2$ Calabi-Yau Duals without Flux}{78}
\TOCchapter{10}{Conclusions and Outlook}{81}
\noindent {\hyperref{}{appendix}{A}{Appendices}} {}
\leaderfill{\hyperref{}{page}{84}{84}} \par
\bigbreak\bigskip}


\newsec{Introduction}

String theory admits an enormous set of seemingly consistent Minkowski
vacua, yet the number of classes of compact geometries that have
appeared in these vacua is small.  With a few noteworthy exceptions
\refs{\HullTor,\Strom,\DRS}, all supersymmetric compactifications of
string theory to 4D Minkowski space studied until the past two years
have involved only five types of special holonomy manifolds.  The
canonical building blocks include Calabi-Yau threefolds, $K3$
surfaces, and tori.  Beyond these, the list also includes $G_2$
manifolds for M~theory compactifications and Calabi-Yau fourfolds for
F~theory.  It is natural to ask to what extent there exist sensible 4D
Minkowski vacua based on other compactification geometries.  The goal
of this paper is to make contact between certain exotic orientifold
vacua deduced via string dualities \KSTT\ and other work based on a
``Superstrings with Torsion'' type description \Strom.

In the absence of flux, when the geometry is a direct product of
$\IR^{3,1}$ and a compact manifold, 4D $\CN\ge1$ supersymmetry of the
low-energy supergravity action demands that the compact manifold be a
product of the special holonomy manifolds listed above \Candelas.  In
the case of type IIB string theory, a similar statement can also be
made for a more general class of warped compactifications with
internal flux and chiral spinor constraints.  Here, the size of
spacetime is allowed to vary over the internal manifold, and the data
defining the compactification is enlarged to include a discrete choice
of branes and internal NS and RR flux.  For the class that has
received the most attention, chiral (Becker-type) constraints are also
imposed on the spinors generating the supersymmetries; these are the
the constraints associated with D3 branes and D7 branes \GP.

Such D3/D7 warped compactifications are attractive for a number of
reasons.  They provide a stringy embedding \refs{\Verlinde,\KS,\GKP}\
of the Randall-Sundrum approach to the hierarchy problem \RSIandII, as
well as a mechanism for perturbative stabilization of moduli
\refs{\DRS,\GKP,\KST,\Tripathy,\DeWolfe,\modstab,\FreyThesis}.  (See
Refs.~\refs{\gaugedsugraI,\SUSYhiggs,\noscale} for a gauged
supergravity perspective; also, see Ref.~\LM, for a discussion of
local models that includes a worldsheet perspective.)  In addition,
they are amenable to a statistical analysis as an ensemble of vacua
\refs{\Douglas}, and with a few new ingredients, have lead to a
construction of metastable de~Sitter vacua of string theory
\KKLT.\foot{An explicit model of this type with all K\"ahler moduli
stabilized was recently exhibited in Ref.~\moreKKLT.  See also
Ref.~\SethiBarren\ for a nonexistence proof for one K\"ahler modulus,
and Ref.~\EvaScaling\ for a related discussion in the $T^6/\IZ_2$
orientifold.  For work on inflation and de~Sitter space in models with
D3 branes and D7 branes, see Ref.~\deSitterandInflation.}

However, for compactification to 4D Minkowski space they require the
same special holonomy manifolds as do conventional compactifications,
up to an overall conformal rescaling by the warp factor \GP.
Schematically, the reason is as follows.  For supersymmetric IIB
vacua, the gravitino variation is
$$\d\psi_M = (D_M + {\rm flux})\e + ({\rm flux})\e^* = 0,$$
where $\e$ is a 10D Weyl spinor.  In the case of Becker-type
constraints, the $\e$ and $\e^*$ terms separately vanish.  The $\e$
term decomposes into the usual covariant constancy condition plus a
relation between the RR flux $\tilde F_{(5)}$ and the warp factor.
The $\e^*$ term gives conditions on the complex three-form flux
$G_{(3)}$.

For other spinor constraints, or for vacua other than type IIB warped
compactifications, the fluxes can mix with covariant derivatives.  One
obtains generalized covariant constancy conditions based on torsionful
connections, whose solutions, if any, involve internal manifolds not
appearing in the list above.  These exotic compactification manifolds
can be non-K\"ahler and even non-complex.  However, while it is easy
write down the equations of motions and supersymmetry conditions, it
is not so easy to find solutions.  Given certain assumptions, one can
prove no-go theorems on their existence
\refs{\NoGo,\GKP,\GauntIntr}.\foot{The no-go theorems state that in the
absence of negative-tension sources, the leading $\a'$-order
supergravity action has no solutions with internal flux and compact
internal manifold.  These theorems are evaded in string theory by the
existence of negative-tension orientifold planes, by $\a' R\w R$ D7
brane worldvolume curvature couplings (which give D7 branes negative
D3 brane charge and tension), and $\a'R\w R$ corrections to the
heterotic Bianchi identities.}

\subsec{Older Progress}

Most of the progress toward understanding torsionful supersymmetric
vacua builds on work performed by Strominger nearly two decades ago,
in the context of the heterotic string with NS flux
\Strom.  In the heterotic theories, the gravitino and dilatino
variations are
\eqn\stromeq{\eqalign{\d\psi_M
& = \bigl(\nabla_M + \quarter\Hsl_{(3)M}\bigr)\e,\cr
\d\l & = \bigl(\dsl\phi+\half\Hsl_{(3)}\bigr)\e,}}
where $\e$ is a 10D Majorana-Weyl spinor, and $\e = (u\otimes\chi +
u^*\otimes\chi^*)/\sqrt{2}$ in terms of a 4D Weyl spinor $u$ and 6D
Weyl spinor $\chi$.  (Strominger did not specialize to
compactifications to 4D, but we will do so here).  For supersymmetric
vacua, the gravitino variation directly implies that $\chi$ and
$\chi^*$ are both covariantly constant with respect to the same
connection of torsion $T^a_{bc} = H_{(3)}{}^a{}_{bc}$.  If we define
an almost complex structure (ACS)
\eqn\acs{J_a{}^b = i\chi^\dagger\gamma_a{}^b\chi,}
then Eqs.~\stromeq\ together imply that the ACS is integrable.
Lowering one index on $J_a{}^b$ gives the fundamental form\foot{In the
special case that $J$ is closed, the internal manifold is K\"ahler and
$J$ is the K\"ahler form.} $J$, which is related to the flux via
\eqna\dJH
$$\eqalignno{-e^{2\phi}d\bigl(e^{-2\phi}J\bigr) & = *_6 H_{(3)}, &\dJH
a\cr d\bigl(e^{-2\phi} *_6 J\bigr) & = 0. &\dJH b}$$
Eqs.~\dJH{}\ are not quite the relations that appeared in Strominger's
work \Strom, but are equivalent to them
\refs{\GauntSLAG,\GauntIntr}.

Strominger went on to obtain a complete set of geometrical conditions
for supersymmetry, as well as a set of equations expressing the
dilaton and flux in terms the geometry.  In addition, he provided a
worldsheet description, and for compactifications to 6D was able to
give exact solutions.  For compactifications to 4D on a compact
manifold other than a Calabi-Yau, he argued that the solutions could
not extend to large volume, since his equations reduced to Calabi-Yau
conditions in this limit.  He did not provide any solutions, but did
mention the Iwasawa manifold as an example of a complex non-K\"ahler
manifold satisfying the topological condition $h^{3,0}=1$, that could
conceivably be used as the basis for a consistent torsionful
supersymmetric solution.\foot{Heterotic compactifications on the
Iwasawa manifold have been re-explored using the tools discussed in
the first part of Sec.~1.2 \CardTor.  In Sec.~7, we discuss
orientifolds based on a warped Iwasawa manifold.}

This is the way things stood for thirteen years until the first
non-K\"ahler heterotic compactifications to 4D were constructed using
string dualities \DRS.  The starting point was a IIB warped
compactification, obtained as the orientifold limit of a warped
F-theory compactification on $K3\times K3$ with internal flux
\refs{\BeckerM,\DRSrefs}.   After two T-dualities,\foot{The geometry
of the IIB orientifold is a warped version of $\IR^{3,1} \times
K3\times T^2/\bigl(\O(-1)^{F_L}\CI_2\bigr) $, where $\CI_2$ inverts
the $T^2$.  The T-dualities are performed in the $T^2$ directions.}
this theory becomes type I on a non-K\"ahler manifold with RR
three-form flux.  The heterotic theory is then obtained via S-duality.
The resulting geometry takes the form of a warped $T^2$ fibration over
$K3$, with the noteworthy property that the fiber volume is fixed in
terms of the fiber complex structure modulus.  So, lengths cannot
simply be scaled by an overall factor to obtain a large volume
solution, in agreement with Strominger's earlier observation
forbidding an overall volume modulus.

\subsec{More Recent Progress}

Two important organizing principles that have emerged more recently
are the notions of $G$-structures and generalized calibrations.  These
tools were first applied to studying supergravity solutions with
background flux in Ref.~\GauntSLAG. Since then, the subject has
flourished
\refs{\Louis,\CardTor,\GauntPakis,\GauntIntr,\MartelliSparks,%
\GauntGP,\GauntAdS,\Minetal,\MinMirror,\Husain,\genholonomy,%
\DallAgataIIA,\DallAgataIIB,\BCJ,\FreySU}.  In the context of the
heterotic string with 4D $\CN=1$ supersymmetry, the appropriate group
$G$ is $SU(3)$, and the idea is as follows
\refs{\Louis,\CardTor,\GauntIntr}.  The existence of the covariantly
constant spinor $\chi$ discussed above is equivalent to the statement
that the compactification manifold $X_6$ has $SU(3)$ holonomy with
respect to the torsionful connection.  As a consequence of the
existence of a privileged 6D spinor, the usual $SU(3)$ structures $J$
and $\O$ can be canonically defined in terms of this spinor, just as
for a Calabi-Yau.  However, these objects are no longer closed.  A
useful mathematical characterization of the precise sense in which
$X_6$ differs from a Calabi-Yau comes from the fact that the $SU(3)$
structures induce a natural $SU(3)$ decomposition of the torsion into
five torsion classes \Gray:
\eqn\Tclasses{\matrix{T& = & \CW_1 & + & \CW_2 & + & \CW_3
& + &\CW_4 & + & \CW_5.\cr
&& (1+1) && (8+8) && (6+\bar6) && (3+\bar3) && (3+\bar3)}}
In terms of this decomposition,
\eqn\JOW{\eqalign{dJ &= {3\over4}i(\CW_1\bar\O - {\bar\CW}_1\O)
+\CW_3 + J\w\CW_4,\cr
d\O &= \CW_1 J\w J + J\w\CW_2 + \O\w\CW_5.}}
The supersymmetry conditions can then be translated into constraints
on the $\CW_i$.  Conversely, once these constraints are known, we can
instead start with a manifold satisfying the constraints, and then
$H_{(3)}$ and $e^{\phi}$ follow from Eqs.~\dJH{}.  

The connection to generalized calibrations arises when, following
Ref.~\refs{\GauntSLAG,\GauntIntr}, we interpret Eq.~\dJH{a} as a
consequence of the fact that (i) NS5 branes are a source of $H_{(3)}$
flux, and (ii) $J$ is a generalized calibration \Gutowski\ for the
two-cycles on which we can wrap a supersymmetric NS5 brane probe.  In
some sense, we can think of the manifold $X_6$ as a fully backreacted
geometry involving one or more wrapped NS5 branes.  The appeal of this
formalism is that it readily generalizes.  Instead of considering the
heterotic string on a 6D manifold with a torsionful connection of
$SU(3)$ holonomy, we can consider type I, II, or the heterotic
theories on a manifold of some other dimension, with torsionful
connection of holonomy group $G$.  The $SU(3)$ structures $J$ and $\O$
just get replaced by the appropriate $G$-structures.  The
generalization of Eq.~\dJH{a} is an analogous relation between some NS
or RR flux and the $G$-structure that calibrates the cycles on which
we can wrap a corresponding NS or D brane probe.

Beyond $G$-structures, another important advance has been the
reformulation of the heterotic moduli constraints in a way analogous
to the usual formulation of the constraints for IIB Becker-type warped
compactifications.  For the latter, one defines a complex flux
$G_{(3)} = F_{(3)}-\tdil H_{(3)}$.  The supersymmetry conditions
constrain the moduli so that $G_{(3)}$ is primitive\foot{Primitivity
means that $J\w G_{(3)}=0$.}  and of type (2,1).  The (2,1) condition
follows from a superpotential $W = \int G_{(3)}\w\O$.  Finally, the
scalar potential comes from Kaluza-Klein reduction of the kinetic term
for $G_{(3)}$ in the 10D supergravity action.  A similar structure has
been exhibited in the heterotic theories, except that the scalar
potential now comes from a combination of the $H_{(3)}$,
Einstein-Hilbert, and dilaton kinetic terms in the 10D supergravity
action \CardBPS.  The complex flux $H_{(3)} -
ie^{2\phi}d\bigl(e^{-2\phi}J\bigr)$ is required to be (2,1) and
primitive.  It has been argued that the (2,1) condition follows from
an analogous superpotential \refs{\Keshav,\CardBPS,\KeshavHetII},
although this is more subtle than in the IIB case.

The duality chains, which proved so useful in the first non-K\"ahler
heterotic compactifications to four dimensions, have also been
revisited.  In the original construction \DRS, all analysis was
performed in the dual IIB/F-theory description.  Much effort has been
devoted to providing an intrinsically heterotic description of these
vacua and their generalizations, as well as to the mathematical
properties of the compact manifolds involved
\refs{\Keshav,\Prokushkin,\KeshavHetII}.  From the results of
Ref.~\KeshavHetII: the previously known class of admissible compact
manifolds obtained via duality has been generalized to include
manifolds with nonzero Euler characteristic, and a better
understanding now exists of phenomenologically relevant details
concerning the vector bundles on these manifolds and the numbers of
fermion generations; moreover, through a different duality, there is
now a relation to brane-box constructions.  A particular duality
symmetry that has proven extremely useful in the past is mirror
symmetry, and steps have been taken toward generalizing Calabi-Yau
mirror symmetry to a mirror symmetry of non-K\"ahler manifolds with
nonvanishing NS flux \refs{\KSTT,\Louis,\MinMirror}.  Even for
geometric transitions, the cornerstones in our modern understanding of
topological string theory \geotransitions, there now exists a sequence
of supergravity descriptions for a complete duality cycle connecting
the various IIB, IIA, and M~theory descriptions before and after the
transition \Keshavgeo.  This duality cycle relates D branes wrapped on
cycles of non-K\"ahler manifolds to fluxes in other non-K\"ahler
manifolds.  For other interesting results involving dualities and
torsion, see Ref.~\otherTdual.

\subsec{Work Reported Here}

A final way in which duality symmetries have been applied is in the
construction of a new class of non-K\"ahler orientifold vacua \KSTT.
These vacua will be our focus here.  They are relatively simple to
describe---the geometry is a warped torus fibration over a torus base,
with O planes and D branes wrapping the fiber and filling spacetime.
In addition, there is internal flux.  These vacua were constructed
using a duality argument similar to the one used to obtain
non-K\"ahler heterotic compactifications.  However, the orientifold
vacua that we discuss here do not have obvious geometrical heterotic
duals.  The starting point is a $T^6$ orientifold of type IIB with
internal flux, where the orientifold inverts all of the torus
directions.  So, the initial theory contains O3 planes and possibly D3
branes, but no O7 planes or D7 branes.  The non-K\"ahler orientifolds
are obtained via T-duality.  For supersymmetric flux, the maximum
number of T-dualities that can be performed is either two or
three,\foot{After this number of T-dualities, the metric has no
further isometries that can be used to perform additional standard
T-dualities.  There does exist a generalized sense in which we can
perform additional T-dualities. However, the result is a
nongeometrical string compactification \refs{\nongeo,\KSTT}, involving
duality twists that mix the metric and NS $B$-field, so that neither
of these quantities is globally well-defined by itself.  In contrast,
standard T-dualities only result in geometric fibrations
(Scherk-Schwarz \ScherkSchwarz\ twists of the metric) and NS flux
(Scherk-Schwarz twists of the NS $B$-field).}{\footlabel\nongeofootI}
depending on the choice of flux.  Therefore, we cannot relate these
string vacua to type I with O9 planes, or subsequently to the
heterotic theory by S-duality.

The main motivation for the present investigation is that the
orientifolds just described have been analyzed in a way that relies
almost exclusively on the original pre-duality theory, with little or
no intrinsic description in the final non-K\"ahler orientifold.  On
the other hand, the geometry of these orientifolds is very similar to
that of certain noncompact manifolds of $SU(2)$ or $SU(3)$ structure
that have been described elsewhere, and used in consistent
supergravity solutions with a ``Superstrings with Torsion'' type
description \GauntIntr.  The primary goal below is to provide such a
description for torsionful orientifolds of type IIB string theory in
which O5 planes are wrapped on the fiber of a $T^2$ fibration.  A
secondary goal is to study moduli stabilization in these orientifolds.

An outline of the paper is as follows.  We begin in Sec.~2 by defining
the class of O5 orientifolds that we consider.  Sec.~2.1 contains a
description of the 10D supergravity background, including the 10D
metric, the orientifold operation, and fluxes. In Secs.~2.2 and 2.3 we
discuss the fields and energy scales of the 4D effective field theory.

Sec.~3 is devoted to analyzing the model to the extent possible
without any assumptions regarding supersymmetry.  Here, and also in
Sec.~4, we relax the ansatz that the base of the internal manifold is
$T^4$, and assume only that it is some compact manifold $\CB$.
However, order to proceed systematically, we find it necessary to
impose one restriction on the NS flux.  This restriction is discussed
at the end of Sec.~3.1.  The main results of Sec.~3 are the Gauss's
law constraint in Sec.~3.2, a pseudo-BPS constraint on the fluxes in
Sec.~3.3, expressions for the dilaton and warp factor in Secs.~3.4 and
3.5, and the requirement that the base be Ricci flat.  The pseudo-BPS
constraint follows from an equality between an exact four-form on the
base and sum of positive semidefinite squares of fluxes.  Integration
of this relation imposes a set of Hodge duality conditions on the
fluxes.  There is a naive paradox concerning the Gauss's law
constraint.  Due to the fibration, one can have $d\tilde F_{(3)}$
supported on local sources, the sum of whose charges does not vanish.
This paradox is resolved by correctly pushing forward the $\tilde
F_{(3)}$ Bianchi indentity to the base.  We conclude the section with
a precise definition of the moduli, and a decomposition of the fluxes
into a quantized discrete part and moduli dependent deformations.

Sec.~4 contains a discussion of the supersymmetry conditions.  After a
brief discussion in Sec.~4.1 of the contraints on the spinors from the
orientifold projection, in Sec.~4.2 we analyze the case in which only
the RR three-form flux is nonzero.  Here, the discussion closely
parallels that given by Strominger for the heterotic string with NS
flux.  The vacua that we describe are related by S-duality to dual
vacua with NS-flux, NS5 branes, and ON5 planes, that are very similar
to those discussed by Gauntlett, Martelli, and Waldram \GauntIntr\ in
the case of 6D internal manifold.  The S-duality map is discussed in
Sec.~4.3.  We then consider the case of more general flux in Sec.~4.4,
and derive the corresponding generalization of the supersymmetry
conditions.  These conditions are stated in terms of the flux, the
right-$SU(3)$ structures $J$ and $\O$ constructed from the
right-moving part of the Killing spinors only (or alternatively,
left-$SU(3)$ structures), and the volume form $\Vol_\fib$ on the $T^2$
fiber.  Note that while we employ $SU(3)$ structures in analyzing the
supersymmetry conditions, we do not find it useful here to work in
terms of $SU(3)$ (or $SU(2)$) torsion classes in order to satisfy
these conditions.  Therefore, we will not compute torsion classes in
this paper, except for a fleeting instance in Sec.~6.3 to verify the
nonintegrability of the almost complex structure.  See, however, the
discussion in Sec.~10.

In Sec.~5, we relate O5 orientifolds with $T^4$ base to T-dual O3
orientifolds with internal $T^6$.  Then, in Secs.~6--8 we study a
number of specific classes of O5 backgrounds preserving various
amounts of supersymmetry, and their O3 duals.  Sec~6.3 is devoted to a
class with nonvanishing $F_{(1)}$ and $H_{(3)}$ flux, and
nonintegrable complex structure, but the examples in Secs.~6--8
otherwise contain only $\tilde F_{(3)}$ flux and integrable complex
structure.  Our discussion of the O3 duals extends the known results
for the $T^6/\IZ_2$ orientifold.

Sec.~9 contains a preview of work to appear elsewhere \CYduals.  In
the $\CN=2$ case, our examples are dual to M~theory on a Calabi-Yau
threefold $Y_6$ times a circle. That is, they can be viewed as
F~theory on $Y_6\times T^2$, where the F~theory fiber is taken to be
the product of an $S^1$ in the $T^2$ and an $S^1$ in $Y_6$ (with $Y_6$
viewed as a fibration).  Here, we provide a few preliminary
observations concerning this duality.

Finally, in Sec.~10, we conclude and discuss possibilities for future
work.

The appendices contain auxilliary details not found in the body of the
paper.  App.~A contains a summary of our conventions, mostly with
regard to Dirac matrices.  In App.~B, we describe the twisted
coordinate identifications that generalize the $T^6$ identifications
$x^a\cong x^a+1$ to a $T^2$ fibration over $T^4$.  In App.~C, we
derive the first equation of Sec.~3.3, from which the pseudo-BPS
constraints follow.  App.~D is a reference for the IIB supersymmetry
variations of fermions in various forms (string frame, Einstein frame,
in terms of $\e_{L,R}$ and in terms of $\e =
\e_L+i\e_R)$.  In App.~E, we discuss subtleties in applying the
superpotential of Gukov, Vafa, and Witten \GVW\ to the $T^6/\IZ_2$ O3
orientifold in which the fluxes (partially) break $\CN=4$ extended
supersymmetry.  The T-duality map of RR moduli, as defined in
Secs.~3.9 and 5.1, is worked out in App.~F\null.  App.~G contains a
discussion of moduli space metrics for the $T^6/\IZ_2$ orientifold;
here, we first review the results of Frey and Polchinski for the case
of $\CN=3$ supersymmetry, and then consider the $\CN=2,1$ cases
relevant to Secs.~6 and 8.  Lastly, in App.~H we prove that a
particular class of fluxes discussed in Sec.~8.2 reduces to a unique
choice of flux modulo the discrete identifications of the
axion-dilaton and $T^6$ complex structure.


\newsec{{O5 Orientifolds of $T^2$ Fibrations}}

\subsec{Supergravity Background}

As a starting point, recall that the supergravity background for $N$
coincident D5 branes in flat noncompact spacetime is \Johnson
\eqna\Dpsoln
$$\eqalignno{ds^2_{\rm string} 
&= Z^{-1/2}\eta_{\mu\nu}dx^\mu dx^\nu
+ Z^{1/2}\delta_{mn}dx^m dx^n, &\Dpsoln a\cr 
e^{\phi} &= g_s Z^{-1/2}, &\Dpsoln b\cr 
F_{(7)} &= -g_s^{-1} d(Z^{-1})\wedge dx^0\wedge\ldots\wedge
dx^5, &\Dpsoln c}$$
where $\mu,\nu = 0,\ldots,5$ and $m,n = 6,\ldots,9$.  The subscript
``string'' indicates that Eq.~\Dpsoln{a} gives the string frame
metric. The warp factor is
\eqn\ZDp{Z = 1 + N {\a'g_s\over 2r^2}.}
This function is harmonic away from $r=0$ and is a solution to Poisson's
equation
\eqn\PoissonDfive{\delta^{mn}\partial_m\partial_n Z = 
N (2\pi)^2\a' g_s\delta^4(x)}
in the non-warped metric $\delta_{mn}$ on the $\IR^4$ transverse to
the D5 branes.

We would like to study the analogous supergravity backgrounds for
orientifolds of type IIB string theory in which 4D spacetime-filling
O5 planes and D5 branes wrap the $T^2$ fiber over a $T^4$
base.\foot{Although we focus on the case of $T^4$ base, most of the
analysis in Secs.~3 and~4 is performed for arbitrary compact base
$\CB$.} The new metric ansatz is
\eqna\dsstring
$$\eqalignno{ds_{\rm string}^2 &= Z^{-1/2}(\eta_{\m\n}dx^\m dx^\n +
ds_\Tfib^2) + Z^{1/2}ds^2_\Tbase, &\dsstring a}$$
where $\eta_{\m\n}$ is the 4D Minkowski metric, and the fiber and base
metrics are
$$\eqalignno{ds_\Tfib^2 &= g^{(\Tfib)}_{\a\b}\eta^\a\eta^\b,
\quad\eta^\a = dx^\a +\CA^\a,\quad \a,\b = 4,5,&\dsstring b\cr
ds_\Tbase^2 &= g^{(\Tbase)}_{mn} dx^m dx^n,\quad
m,n=5,\ldots,9. &\dsstring c}$$
Here, we assume that $g^{(\Tfib)}_{\a\b}$ and $g^{(\Tbase)}_{mn}$ are
flat metrics, and that the warp factor depends only on the base
coordinates.  The fiber and base coordinates are identified via a
twisted generalization of $x^\a\cong x^\a+1$, $x^m\cong x^m+1$ that is
discussed in App.~B\null.  The quantity $\CA^\a
=\CA^\a{}_m\bigl(\{x^n\}\bigr)dx^m$ is the fiber connection.  The
one-forms $\eta^\alpha$ are globally defined and are the
generalizations of $dx^\alpha$ consistent with the twisting of the
fibration.
Nontrivial twisting is measured by nonvanishing cohomology classes
\eqn\Chern{[\CF^\alpha] \in H^2(\Tbase,2\IZ),}
where
\eqn\Falpha{\CF^\alpha = d\CA^\alpha.}
These are the Chern classes of the $x^\alpha$ circle fibrations (or
equivalently $U(1)$ fibrations) over $\Tbase$.

For later convenience, we define a 6D metric
\eqn\sixDg{ds_6{}^2 = ds^2_\Tfib + Z ds^2_\Tbase,}
so that
\eqn\tensix{ds^2_{\rm string} = Z^{-1/2}(\eta_{\m\n}dx^\m dx^\n +
ds_6{}^2).}
We denote by $X_6$ the 6D space whose metric is \sixDg.

In addition to this geometry, the model includes 16 O5 planes and $2M$
D5 branes.  The $\IZ_2$ orientifold operation that defines the O5 planes
is $\O\CI_4$, where $\O$ is worldsheet parity and $\CI_4$ inverts the
base $T^4$:
\eqn\Ibase{\CI_4\colon\quad x^m \to -x^m.}
The orientifold planes wrap the fiber and are located at the $2^4$
$\IZ_2$ fixed points on the base where $x^m=0,1/2$.\foot{The
orientifold planes are assumed to be standard O5${}^-$ planes in the
terminology of \Bergman, as opposed to the exotic O5${}^+$ or
$\widetilde{\rm O5}^\pm$ planes that lead to shifted flux quantization
conditions.  Similarly, we assume that there is no localized $\CF^\a$
curvature at the $\CI_4$ fixed points.}  The $2M$ D5 branes also wrap
the fiber, and are located at $M$ arbitrary points on the base
together with their $M$ $\IZ_2$ images.  The $2\IZ$ rather than $\IZ$
valued cohomology in Eq.~\Chern\ ensures that $[\CF^\a]\in
H^2(T^4/\CI_4,\IZ)$, which is needed to define the orientifold.  (See
Sec.~3.7).

Finally, the orientifold projection preserves the following even
fluxes together with their 10D Hodge duals:\foot{Here, all components
of fluxes and potentials are given in the basis $dx^\m,\eta^\a,dx^m$.
These components should not be confused with those in the basis
$dx^m,dx^\a,dx^m$.}
\eqn\evenfluxes{\hbox{Even fluxes:}\quad\tilde F_{(1)m}, \quad\tilde
F_{(3)\a mn},\quad\tilde F_{(5)\a\b mnp},\quad H_{(3)\a\b m},\quad
H_{(3)mnp}.}
By even, we mean that these components must be even functions of the
transverse coordinates $x^m$.  The orientifold projection also
preserves the complementary set of components, provided these
components are odd functions of the transverse coordinates.  With the
exception of the special case that $2M=32$ and each O5 plane is
coincident with exactly one D5 brane and its image (so that charges
and tensions cancel locally), there do not exist string vacua with all
odd fluxes set to zero.  In order to satisfy the low energy equations
of motion, we need to include at least the following odd flux:
\eqn\oddflux{\hbox{Odd flux:}\quad\tilde F_{(3)mnp}.}
We will limit our investigation to those backgrounds in which the
other odd fluxes can be consistently set to zero, and in which all
fields have functional dependence on the base coordinates only.  (In
other words, we consider the low energy effective field theory at
energy scales below $1/R_\fib$; see Sec.~2.3 for further discussion.)
In addition, we assume vanishing field strengths of the D5 worldvolume
gauge fields.

The fluxes $\tilde F_{(p)}$ are the gauge-invariant fluxes that appear
in the kinetic terms of the bulk supergravity action.  In the
conventions in which the T-duality action on RR potentials is
simplest,\foot{There is different convention, $\tilde F_{(5)} =
F_{(5)}+\half B_{(2)}\w F_{(3)} -\half C_{(2)}\w H_{(3)}$, in which
the potentials $B_{(2)}$ and $C_{(2)}$ transform simply under
$SL(2,\IZ)$ duality.} the relation between the RR potentials $C_{(p)}$
and gauge-invariant fluxes $\tilde F_{(p)}$ is
\eqn\twoRRfluxes{\tilde F_{(p)} = 
\cases{F_{(p)} - C_{(p-3)}\w H_{(3)}&\quad $p\ge 3$,\cr
F_{(p)}&\quad $p=1,2$,}}
where
\eqn\RRfluxpot{F_{(p)} = dC_{(p-1)}.}
In Sec.~3, we will see that the odd flux \oddflux\ is completely
determined by the equations of motion.  On the other hand, the even
fluxes \evenfluxes\ contain both moduli-dependent and moduli-independent
components, with the latter constrained only by Dirac quantization.
Therefore, the even fluxes, like $\CF^\a$ and $M$, contain discrete
data that needs to be specified in order to fully define the model.

For trivial fibration and in the absence of even flux, this
orientifold is dual to type~I on $T^6$ via T-duality in the four
$\Tbase$ directions.  In this case, the orientifold preserves 4D
$\CN=4$ supersymmetry.  Also, since the D5 branes and O5 planes fill
4D spacetime, there is a Gauss's law constraint (RR tadpole
cancellation condition in worldsheet language) that requires $2M=32$
D5 branes.  This constraint and its generalization to the case of
nontrivial fibration and flux are discussed in Sec.~3.2 below.

\subsec{Low Energy Bosonic Field Content}

For the class of 10D supergravity backgrounds just described, there
exist deformations that continuously connect different consistent
supergravity solutions.  Let us restrict to deformations with no
functional dependence on $x^\m$.  For the case of trivial fibration
and in the absence of even flux, one class of such deformations
consists of the zero modes on $X_6$ of the even bosonic fields
\eqna\approxmoduli
$$\eqalign{\hbox{RR sector:}\quad &
C_{(2)\a\b},\quad C_{(2)\m\n}\leftrightarrow{\rm scalar},
\quad C_{(2)mn},\quad C_{(4)\a mnp},\cr
\hbox{NSNS sector:}\quad & g_{\a\b},\quad g_{mn},\quad B_{(2)\a m},
\quad \phi,}\eqno\approxmoduli a$$
together with the zero modes on the $T^2$ fiber of the D5 worldvolume
fields
$$\hbox{D5 worldvolume:}\quad  A_{I\,\a},\quad \Phi_I{}^m,\quad
I = 1,\ldots, M.\eqno\approxmoduli b$$
These are the deformations which, when promoted to 4D fields, become
the moduli of the 4D $\CN=4$ low energy effective field theory.

The quantity that plays the role of the axion-dilaton is
(cf.~Sec.~4.2)
\eqn\newaxdil{\tdil = {1\over(2\pi)^2\a'} \biggl(-C_{(2)\,45}
+ {i\over g_s}V_\fib\biggr),}
where the volume of the $T^2$ fiber is
\eqn\Vfib{V_\fib = \bigl(g^{(\Tfib)}\bigr)^{1/2}.}
This expression for $\tdil$ is T-dual to the more familiar expression
given App.~D that is common to O3 orientifolds, O7 orientifolds, and
pure IIB string theory.

A second class of such deformations consists of the zero modes on
$X_6$ of the even fields
\eqn\approxgauge{\eqalign{\hbox{RR sector:}\quad &
C_{(2)\a\m},\quad C_{(4)\a\b m\m},\cr
\hbox{NSNS sector:}\quad & V^\a{}_\m,\quad B_{(2)m\m},\cr
\hbox{D5 worldvolume:}\quad & A_{I\,\m},\quad
I = 1,\ldots, M.}}
These are the deformations which, when promoted to 4D fields, become
the massless gauge bosons of the 4D $\CN=4$ low energy effective field
theory.  Here, $V^\a{}_\m$ is the deformation that corresponds to the
Kaluza-Klein gauge boson for the translation isometry in the
$\a$-direction.  It arises by replacing $\CA^\a$ with $\CA^\a +
V^\a{}_\m dx^\m$ in the 10D metric~\dsstring{b}.

Finally, a third class of such deformations shifts the 4D metric away
from $\eta_{\m\n}$ and parametrizes the space of constant metrics on
$\IR^{3,1}$.  When promoted to 4D fields, these deformations become
the 4D graviton.

Note that there is an important distinction between the allowed
$x^\m$-independent deformations of the supergravity backround and the
fields of the 4D effective action.  While it is easy to identify the
former, the massless fields in the dimensional reduction to 4D are
more complicated than those obtained by simply endowing these
deformations with $x^\m$-dependence.  The correct low energy 4D fields
must also include warp factor dependence to prevent mixing with higher
Kaluza-Klein modes \refs{\FP,\FreyThesis,\GM}.

In the case of nontrivial fibration and flux, the supersymmetry and
massless field content is reduced.  The $U(1)^M$ $\CN=4$ vector
multiplets on the D5 branes remain massless, but many of the RR and NS
fields in \approxmoduli{}\ and \approxgauge\ are lifted.  The lifting
occurs through a supersymmetric Higgs mechanism.  For the bosons, the
equations of motion impose certain metric and axion-dilaton dependent
Hodge duality relations on the fluxes that lift a subset of the NS
moduli and the zero mode of the RR axion $C_{(2)\,45}$.  The vevs of
some of these moduli then break a subset of the gauge symmetries, and
their axionic partners (zero modes of some of the $C_{(2)mn}$ and
$C_{(2)\a mnp}$) are eaten by massive vectors.  The fermions also take
part in the superHiggs mechanism, and the various possibilities for
the resulting $\CN<4$ low energy spectrum are determined by the masses
of the ($4-\CN$) massive gravitini, as described in Ref.~\SUSYhiggs.
(See also Ref.~\gaugedsugraI).

\subsec{Energy Scales}

There are two points of view that we can adopt regarding the moduli
stabilization and symmetry breaking just described.  If we assume that
the fiber and base have roughly the same length scale $R$, then these
effects occur at the Kaluza-Klein scale $1/R$.  However, if we assume
a hierarchy $R_\fib\ll R_\base$, then these effects occur at the scale
$m=R_\fib/R_\base^2$ \Louis, which is much smaller than $m^{\rm
KK}_\base = 1/R_\base$ and $m^{\rm KK}_\fib = 1/R_\fib$:
\eqn\Rhierarchy{m : m^{\rm KK}_\base : m^{\rm KK}_\fib = 
1 : R_\base/R_\fib : \bigl(R_\base/R_\fib\bigr)^2.}
In the former case, it only makes sense to talk about a low energy
$\CN<4$ theory with all massive fields integrated out.  In the latter,
there is a consistent low energy 4D supergravity theory within which
$\CN=4$ is softly broken to $\CN<4$ by a supersymmetric Higgs
mechanism.  We will adopt the latter point of view below.  In either
case, we need $R_\base,R_\fib \gg\a'$ in order to be able to neglect
higher string modes.


\newsec{Equations of Motion and Integrability Constraints}

Before imposing the supersymmetry conditions, let us first determine
the constraints that follow from equations of motion alone.

\subsec{Equations of Motion}

The trace-reversed Einstein equations are
\eqn\trrev{R_{MN} = \tilde T_{MN},}
where $\tilde T_{MN} = T_{MN} - {1\over8}g_{MN} T$.

The components of the Ricci tensor in the metric \dsstring{}\ are
\eqn\Ricci{\eqalign{%
R_{\m\n} &= \nabla_\m\nabla_\n\log Z
+\quarter Z^{-1/2}\nabla_\CB^2\log Z\eta_{\m\n},\cr
R_{\a\b} &= \nabla_\a\nabla_\b\log Z 
+\quarter Z^{-1/2}\nabla_\CB^2\log Z\delta_{\a\b}
+ \quarter \CF^2_{\a\b},\cr
R_{mn} &= \nabla_m\nabla_n\log Z
-\quarter Z^{-1/2}\nabla_\CB^2\log Z\delta_{mn}
- \half \CF^2_{mn}\cr &\phantom{=}
- \half (\nabla_{\mkern-2mu \CB\ssk m}\log Z)
(\nabla_{\mkern-2mu \CB\ssk n}\log Z)
+ R_{\CB\,mn}.}}
Here $\CB$ denotes the 4D base of the torus fibration.  Most of the
results of Secs.~3 and~4 are independent of the choice $\CB=T^4$, so we
work with arbitrary $\CB$.  The operator $\nabla_\CB$ is the Levi-Civita
covariant derivative on the base, and the Laplacian $\nabla^2_\CB$ is
contracted using the base metric
\eqn\basem{ds^2_\CB = g_{\CB\ssk mn}dx^m dx^n}
that generalizes Eq.~\dsstring{c}.
 
In string frame, the trace-reversed stress tensor is $\hat T_{MN} =
\tilde T^{\rm NS}_{MN} + \tilde T^{\rm RR}_{MN} + \tilde T^{\rm
local}_{MN}$, where
\def\Tstrut{{\vphantom{\sum_{p=1,3}\Bigl(\Bigr)}}}
\eqn\stressT{\eqalign{\Tstrut\tilde T^{\rm NS}_{MN} 
&= -2\nabla_M\nabla_N\phi
+ \half\smallfrac{1}{2!}H^{\hphantom{(3)}2}_{(3)MN},\cr
\tilde T^{\rm RR}_{MN} &= \half e^{2\phi}
\sum_{p=1,3}\Bigl(\smallfrac{1}{(p-1)!}\tilde F^{\hphantom{(p)}2}_{(p)MN}
-\half g_{MN}\smallfrac{1}{p!}\tilde F_{(p)}{}^2\Bigr)
+\quarter e^{2\phi}\Bigl(\smallfrac{1}{4!}\tilde F^{\hphantom{(5)}2}_{(5)MN}
-\half g_{MN}\smallfrac{1}{5!}\tilde F_{(5)}{}^2\Bigr),\cr
\Tstrut\tilde T^{\rm local}_{MN} &= -(2\pi)^2\a'
\half e^{\phi}\sum_i (h_{MN}(x_i)-\half g_{MN})
Q_i{\d^4(x-x_i)\over Z\sqrt{g_\CB}}.}}
Here, $h_{MN}(x_i)$ is the restriction of the 10D metric \dsstring{}\
to the point $x=x_i$ in the base,
\eqn\hmetric{ds^2_{h(x_i)} = Z^{-1/2}\bigl(\eta_{\m\n}dx^\m dx^\n +
g^{(\Tfib)}_{\a\b}\eta^\a\eta^\b\bigr)\big|_{x^m=x_i^m}.}
The sum on $i$ runs over D5 brane and O5 plane sources of charge $Q_i$
in units of D5 brane charge, localized at points $x=x_i$ on the base.
For $\CB=T^4$, this includes $2M$ D5 branes of charge $Q_i=+1$ and 16
O5 planes of charge $Q_i=-2$ located at the $\IZ_2$ fixed points.  We
work on the covering space of the orientifold, so that there are $M$
independent D5 branes at $x_I$, $I=1,\ldots,M$, and $M$ image D5
branes located at $-x_I$.  The O5 plane charge on the covering space
is twice the O5 charge of the ``downstairs'' picture.

The dilaton equation of motion is
\eqn\dilEOM{\nabla^2\phi - 2(\nabla\phi)^2 = e^{2\phi}
\bigl(F_{(1)}{}^2 + \half\smallfrac{1}{3!}\tilde F_{(3)}{}^2\bigr)
-\half\smallfrac{1}{3!}H_{(3)}{}^2 + (2\pi)^2\a'\half e^{\phi}\sum_i
Q_i {\delta^{4}(x-x_i)\over Z\sqrt{g_\CB}}.}

Finally, the Bianchi identity for $\tilde F_{(3)}$ is
\eqn\Bianchi{d\tilde F_{(3)} = -F_{(1)}\w H_{(3)} -
(2\pi)^2\a'\Vol_\CB\sum_i Q_i{\d^4(x-x_i)\over\sqrt{g_\CB}},}
where $\Vol_\CB$ is the volume form on the base,
\eqn\VolB{\Vol_\CB = (g_\CB)^{1/2}dx^6\w dx^7\w dx^8\w dx^9.}
There are additional equations of motion and Bianchi identities for
the fluxes.  These are discussed in Sec.~3.8 below.

In order to make the equations of motion tractable, we will now impose
one further restriction on supergravity backgrounds that we consider.
We demand that $H_{(3)\a\b m}~=~0$.  This condition is equivalent to
restricting to the subset of vacua of this orientifold that are
related to D3/O3 orientifold vacua via T-duality in the fiber
directions.\foot{The result of T-dualizing $H_{(3)}$ in the directions
of two of its indices is a nongeometrical compactification
\refs{\nongeo,\KSTT}.}{\footlabel\nongeofootII}  We have not
succeeded in constructing supergravity solutions with $H_{(3)\a\b m}
\ne 0$.  One of the complications associated with this case is that
the pseudo-BPS constraints of Sec.~3.3 apparantly no longer hold.  The
flux $H_{(3)\a\b m}$ contributes negatively to an otherwise positive
semidefinite sum of squares.  (See App.~C for further discussion of
this point).  More significant qualitative differences implied by
$H_{(3)\a\b m}$ nonzero are (i) $B_{(2)\a\b}$ is nonzero, so that
there is noncommutativity on the D5 branes, and/or (ii) there does not
exist a consistent truncation of the supergravity action to the lowest
Kaluza-Klein modes.  See Ref.~\gaugedsugraII\ for a discussion of the
gauge algebra for certain supergravity backgrounds with $H_{(3)\a\b
m}\ne0$ and positive 4D vacuum energy.

\subsec{Gauss's Law Constraint}

Since the O5 planes and D5 branes fill all of the noncompact Minkowski
directions, we expect the Bianchi identity \Bianchi\ to imply an
integrability condition that roughly demands that the total D5 charge
vanish.

In order to derive this constraint, let us first introduce a notation
that will be useful throughout the paper.  For any $p$-form $\o_{(p)}$
on $X_6$, we write
\eqn\pformdecomp{\o_{(p)} = \o^{\bsk0}_{(p)}+\o^{\bsk1}_{(p)}
+\o^{\bsk2}_{(p)},}
where the component $\o^{\bsk i}_{(p)}$ is of rank $i$ on the $T^2$
fiber.  That is,
\eqna\omegapi
$$\eqalignno{&\o^{\bsk 0}_{(p)} = \smallfrac{1}{p!}
\o_{(p)m^1\ldots m^p}dx^{m^1}\w\ldots\w dx^{m^p}, &\omegapi a\cr
&\o^{\bsk 1}_{(p)} = \eta^\a\w\o_{(p)\a},\quad\o^{\bsk 2}_{(p)}
= \half\eta^\a\w\eta^\b\w\o_{(p)\a\b},&\omegapi b}$$
where
\eqn\omegapa{\eqalign{\o_{(p)\a} &= \smallfrac{1}{(p-1)!}
\o_{(p)\a m^2\ldots m^p}dx^{m^2}\w\ldots\w dx^{m^p}\cr
\o_{(p)\a\b} &=\smallfrac{1}{(p-2)!}
\o_{(p)\a\b m^3\ldots m^p}dx^{m^3}\w\ldots\w dx^{m^p}.}}
From our restrictions on the flux (including the additional assumption
$H_{(3)\a\b m}=0$ made at the end of the last section) we have
\eqn\Fdecomp{F_{(1)} = F^{\bsk0}_{(1)},\quad
H_{(3)} = H^{\bsk0}_{(3)},\quad{\rm and}\quad
\tilde F_{(3)} = \tilde F^{\bsk0}_{(3)} + \tilde
F^{\bsk1}_{(3)}.}  
Since the first two decompositions are trivial, we drop the
superscripts on $F_{(1)}$ and $H_{(3)}$ everywhere except in App.~C,
where the assumption $H_{(3)\a\b m} = 0$ is temporarily relaxed.  We
also drop the tilde on $\tilde F_{(3)\a}$ since $\tilde
F^{\bsk1}_{(3)} = F^{\bsk1}_{(3)}$ for $H_{(3)}$ even.  We will not
need to assume anything about $\tilde F_{(5)}$ in order to show in the
next section that it must vanish.  We note only that $\o^{\bsk0}_{(5)}
= 0$ identically for any five-form on $X_6$.  Therefore,
\eqn\Ffivedecomp{\tilde F_{(5)} = \tilde F^{\bsk1}_{(5)}+ \tilde
F^{\bsk2}_{(5)}.} 

The utility of these definitions is that forms \omegapi{a} and
\omegapa\ can be interpreted as (pullbacks of) forms on the base.
Noting that $d\tilde F_{(3)}^{\bsk1} = \CF^\a\w F_{(3)\a} - \eta^\a\w
dF_{(3)\a}$, the Bianchi identity \Bianchi\ becomes
\eqna\newBianchi
$$d\tilde F_{(3)}^{\bsk0} = -\CF^\a\w F_{(3)\a}
- F_{(1)}\w H_{(3)} - (2\pi)^2\a'\Vol_\CB\sum_iQ_i
{\d^4(x-x_i)\over\sqrt{g_\CB}},\eqno\newBianchi a$$
together with
$$dF_{(3)\a} = 0.\eqno\newBianchi b$$
All forms in \newBianchi{a}\ are now (pullbacks of) forms on the base.
So, the entire equation can be integrated over the base to give the
Gauss's law constraint
\eqn\Gauss{0 = N_\flux +\sum_i Q_i,} 
where
\eqn\Nflux{N_\flux = {1\over(2\pi)^2\a'}\int_\CB
\bigl(\CF^\a\w F_{(3)\a} + F_{(1)}\w H_{(3)}\bigr).}
For the case that the base $\CB$ is $T^4$,
\eqn\sumQ{\sum_i Q_i = 2M - 32.}
Note that this implies that $N_\flux$ must be an even integer.  This
guaranteed by the Dirac quantization conditions, as will be shown in
Sec.~3.9.

The results of this subsection resolve a naive paradox concerning the
Bianchi identity.  In all but one of the examples that we consider in
Secs.~6--8, the only nonzero flux is RR three-form flux and $2M < 32$.
Naively, we can integrate both sides of Eq.~\Bianchi\ over a
transverse four-cycle linking all of the sources of D5 charge, to give
zero on the LHS and $(2M-32)(2\pi)^2\a' < 0$ on the RHS.  The
resolution is that no such linking {\it cycle} exists.  The desired
linking cycle would be a global section embedding the base in the
fibration.  However, unless the fibration is trivial ($\CF^\a=0$),
there does not exist such global section.\foot{This statement relies
on the particular form of the $T^2$ fibration \dsstring{}, in which
there are no fiber degenerations associated with the shrinking of a
$(p,q)$ $S_1$ in the fiber.  It is true that there exist, for example,
elliptically fibered Calabi-Yau manifolds with global section, but
this relies on the existence of such degenerations.}  There only exist
four-{\it chains} $C$ that locally approximate the base in the sense
that they project to the base, $\pi C = \CB$, but globally necessarily
have boundary.  Consequently, $\int_C d\tilde F_{(3)} = \int_{\partial
C} \tilde F_{(3)} \ne 0$.  On the other hand, it is true that $\int_C
d\tilde F_{(3)}^{\bsk0} = 0$, since this integral is the pullback of
$\int_\CB d\omega_{(3)}$, where $\tilde F_{(3)}^{\bsk0} =
\pi^*\omega_{(3)}$.  Note that the nonexistence argument just given
only applies to cycles linking all sources of D5 charge.  There do
exist (homologically trivial) four-cycles linking only individual D5
branes or O5 planes.  Therefore, the charge of these objects is still
well-defined.

\subsec{Pseudo-BPS Constraints}

As a consequence of the dilaton equation \dilEOM, the Bianchi identity
\newBianchi{a}, and the Einstein equation for $g^{\m\n}R_{\m\n} +
g^{\a\b}R_{\a\b}$, it is shown in App.~C that
\def\strutBPS{{\vphantom{\biggl(\biggr)}}}
\eqn\BPSlike{\eqalign{%
d\biggl(Z^{-2}g_s^{-2} \Bigl(\tilde F^{\bsk0}_{(3)} 
& + Z^2 *_\CB d\bigl(g_s^{-1}Z^{-1})\Bigr)\biggr) =\cr
& \ha Z^{-2}g_s^{-1} \biggl( \bigl(F_{(1)} + *_\CB g_s^{-1}
H_{(3)}\bigr)\w *_\CB \bigl(F_{(1)} + *_\CB g_s^{-1} H_{(3)}\bigr)\cr
& \strutBPS + 2Z^{-1}\Bigl(\tilde F_{(3)}^{\bsk0} + Z^2*_\CB
d\bigl(g_s^{-1}Z^{-1}\bigr)\Bigr)\w *_\CB \Bigl(\tilde 
F_{(3)}^{\bsk0} + Z^2 *_\CB d\bigl(g_s^{-1}Z^{-1}\bigr)\Bigr)\cr
& \strutBPS + g^{(\Tfib)}_{\a\b}\Bigl(F_{(3)\a} 
- g^{(\Tfib)}_{\a\g} *_\CB g_s^{-1} \CF^{\g}\Bigr) \w *_\CB 
\Bigl(F_{(3)\b}
-  g^{(\Tfib)}_{\b\d} *_\CB g_s^{-1}\CF^{\d}\Bigr)\cr
& + g^{(\Tfib)}_{\a\b}\Bigl(
2Z^{-1}\tilde F_{(5)\a}^{\bsk1}\w *_\CB \tilde F_{(5)\b}^{\bsk1}
 + g^{(\Tfib)}_{\g\d}
\tilde F_{(5)\a\g}^{\bsk2}\w *_\CB \tilde F_{(5)\b\d}^{\bsk2}\Bigr)
\biggr).}}
Here, $g_s$ is defined by the equation
\eqn\gs{e^\phi = g_s Z^{-1/2}}
(cf.~Eq.~\Dpsoln{b}) and is not necessarily constant at this point,
although that will be shown soon.  Since Eq.~\BPSlike\ involves only
(pullbacks of) forms on the base $\CB$, we can integrate both sides over
$\CB$.  The integral of the LHS vanishes, while the RHS is positive
semidefinite and vanishes only if all terms vanish individually.
Therefore, we obtain the Hodge duality relations
\eqna\BPScond
$$\eqalignno{F_{(1)} &= -*_\CB g_s^{-1} H_{(3)}, & \BPScond a\cr
\tilde F_{(3)}^{\bsk0} &= - Z^2 *_\CB d\bigl(g_s^{-1}Z^{-1}\bigr),
& \BPScond b\cr
F_{(3)\alpha} &= g^{(\Tfib)}_{\a\b} *_\CB g_s^{-1} \CF^\b,
& \BPScond c\cr
\tilde F_{(5)} &= 0. & \BPScond d}$$

We can alternatively write the first three relations as
\eqna\Fcalib
$$\eqalignno{* \tilde F_{(3)}
&= d\bigl(e^{-\phi} \Vol_{h(x)}\bigr), &\Fcalib a\cr
* F_{(1)} &= e^{-\phi} \Vol_{h(x)}\w H_{(3)}, &\Fcalib b}$$
where $*$ is 10D Hodge star operator in the metric \dsstring{}, and
$\Vol_{h(x)}$ is the volume form in the metric \hmetric\ with the
point $x$ on the base allowed to vary:
\eqn\Volh{\Vol_{h(x)} = Z^{-3/2} dx^{0123}\w\Vol_\fib.}
Here,
\eqn\fourDwedge{dx^{0123} = dx^0\w dx^1 \w dx^2\w dx^3,}
and
\eqn\Volfib{\Vol_\fib = V_\fib\eta^4(x)\w\eta^5(x),}
with $V_\fib$ given by Eq.~\Vfib.  We will refer to the constraints
\BPScond{}\ and \Fcalib{}\ as pseudo-BPS conditions, since (i) they
arise from the equations of motion alone, with no supersymmetry
conditions imposed, and (ii) they give a proper subset of the
supersymmetry conditions.  The existence of such pseudo-BPS conditions
is a consequence of the metric ansatz with 4D Minkowski space in the
noncompact directions, together with a property of the low energy
supergravity theory known as no-scale structure: consistency of the
ansatz requires vanishing 4D vacuum energy, and the no-scale structure
implies that the scalar potential is positive semidefinite.\foot{The
canonical example of no-scale structure is the $\CN=1$ no-scale
structure of a theory whose superpotential is independent of K\"ahler
moduli.  Then, modulo $D$-terms, the scalar potential is $V = \sum
|\partial_i W|^2 - 3|W|^2 = \sum'|\partial_iW|^2$, where $\sum$ runs
over all moduli and $\sum'$ runs over all moduli other than the
K\"ahler moduli.  The models of Ref.~\GKP\ possess exactly this type
of no-scale structure.  In the O3 orientifolds of
Refs.~\refs{\FP,\KST}, the no-scale structure is an $\CN>1$ analog of
this.  For the O5 theory discussed here, the only difference is that
the O3 K\"ahler moduli are replaced by a more complicated subset of
the O5 moduli. See Refs.~\refs{\oldnoscale,\noscale,\GKP}.}

Though the pseudo-BPS conditions were derived from the closed string
sector equations of motion, they have a clear interpretation in the
open string sector.  The interpretation of Eq.~\Fcalib{a}\ is that
$\Vol_\fib$ is a generalized calibration for fiber-class cycles
wrapped by the D5 branes.  The mass of a D5 brane that wraps a
two-cycle in the homology class of the $T^2$ fiber is minimized when
the two-cycle is the vertical fiber over a point in the base, as
opposed to another cycle in the same homology class with horizontal
components.  For certain choices of flux, such as those in the
examples that we present in Secs.~6.1 and 8.1, the S-dual of this
calibration condition has already appeared elsewhere
\refs{\GauntSLAG,\GauntIntr}.  The S-duality is discussed briefly in
Sec.~4.3.  In the earlier treatment, the generalized calibration
$\Vol_\fib$ was given a further interpretation as deriving from an
SU(2) structure on the 6D internal manifold.\foot{The SU(2) structure
on $X_6$ is in this case defined by a triple of two-forms $J$,
$\Re\O_{(2)}$ and $\Im\O_{(2)}$ on the 4D base, together with a pair
of Killing one-forms $K_1$ and $K_2$, such that $K_1\w K_2
=\Vol_\fib$.}

Eq.~\Fcalib{b} is the condition that the Myers D5 worldvolume coupling
\Myers
\eqn\Myerscoupling{S_{\Phi^3} = 
-T_5\int_{\rm wv}\bigl(e^{-\phi}\Vol_{\rm wv}H_{(3)mnp}
-F_{(9)012345mnp}dx^{0123}\w\eta^4\w\eta^5\bigr)
 \Tr(\Phi^m\Phi^n\Phi^p)} 
vanish.  Here $\Vol_{\rm wv}$ is the volume form on the D5
worldvolume.  When this condition is not satisfied, the D5 branes are
polarized by the external $F_{(1)}$ and $H_{(3)}$ fluxes and blow up
into D7 branes.

\subsec{Dilaton}

The dilaton equation of motion \dilEOM\ can be combined with the
$g^{\m\n}R_{\m\n}$ Einstein equation to give
\eqn\gsconst{\eqalign{-\nabla_\CB^2\bigl(Z^{-1}e^{-2\phi}\bigr)
&= Z^{-1/2}e^{-2\phi}\Bigl(g^{\m\n}
\bigl(R_{\m\n}+2\nabla_\m\nabla_\n\phi\bigr)
+2\bigl(\nabla^2\phi-2(\nabla\phi)^2\bigr)\Bigr)\cr
&=Z^{-1/2}\Bigl(F_{(1)}{}^2+\half\smallfrac{1}{5!}\tilde F_{(5)}{}^2
-e^{-2\phi}\smallfrac{1}{3!}H_{(3)}{}^2\Bigr).}}
(See App.~C, Eqs.~\pretris\ and \pretriF).  The RHS vanishes by
Eqs.~\BPScond{}, so $g_s$ as defined in Eq.~\gs\ is indeed constant.

\subsec{Warp Factor}

The warp factor $Z$ is a solution to Poisson's equation with localized
sources at the D5 branes and O5 planes, and constant D5 charge density
from the flux and fibration.  From the result that $g_s$ is constant,
\eqn\dFLaplace{d\tilde F_{(3)}^{\bsk0} = {1\over g_s}d*_\CB dZ
= {1\over g_s}\nabla_\CB^2 Z\Vol_\CB.}
It will be shown in Sec~3.7 and~3.8 that $\CF^\a$, $F_{(3)\a}$,
$H_{(3)}$, and $F_{(1)}$ are (pullbacks of) harmonic forms on $\CB$ as
a consequence of the geometrical conditions on $\CF^\a$ and the
Bianchi identities/equations of motion for the fluxes.  Therefore,
the integrand in Eq.~\Nflux\ is a harmonic representatives of
$H^4(\CB,\IR)$, that is, a constant multiple of $\Vol_\CB$.  This
allows us to replace Eq.~\Nflux\ with the stronger relation
\eqn\FwedgeF{\CF^\a\w F_{(3)\a} 
+ F_{(1)}\w H_{(3)} = (2\pi)^2\a'N_\flux V_\CB{}^{-1}\Vol_\CB.}
Here, $V_\CB=\int_\CB\Vol_\CB$ is the volume of the base.  Note that
from Eqs.~\BPScond{a,c}, we then have $N_\flux\ge0$.  Therefore, the
number of D5 branes is bounded above by (the absolute value of) the
total charge from O5 planes, due to the constraint
\Gauss.

Combining Eqs.~\dFLaplace\ and \FwedgeF, the Bianchi identity
\newBianchi{a}\ becomes
\eqn\Zpoisson{\eqalign{-{1\over g_s}\nabla_\CB^2 Z
&= (2\pi)^2\a'\biggl({N_\flux\over V_\CB} 
+ \sum_i Q_i{\d^4(x-x_i)\over\sqrt{g_\CB}}\biggr)\cr
&= (2\pi)^2\a'\sum_iQ_i
\biggl({\d^4(x-x_i)\over\sqrt{g_\CB}}-{1\over V_\CB}\biggr).}}
The solution is
\eqn\Zsoln{Z = 1+(2\pi)^2\a'g_s\sum_iQ_iG_\CB(x,x_i),}
where $G_\CB$ is the Green's function for Poisson's equation on the base
$\CB$,
\eqn\Green{-\nabla_\CB^2 G_\CB(x,x') = {\d^4(x-x')\over\sqrt{g_\CB}}
-{1\over V_\CB}.}
As part of our definition of the Green's function, we fix the constant
component of $G_\CB$ by requiring that $\int_\CB d^4x \sqrt{g_\CB}
G_\CB(x,x') = 0$.  The leading integration constant of unity in
Eq.~\Zsoln\ is conventional and ensures that the warp factor
completely drops out of the metric \dsstring{}\ in the limit of zero
gravitational coupling $g_s\to0$, at fixed $V_\CB$.  This fixes the
rescaling ambiguity
\eqn\ambiguity{g_s\to\l g_s,\quad Z\to\l^2 Z\quad
\hbox{at fixed }e^\phi}
in Eq.~\gs.

\subsec{Base Geometry}

The only other constraint that follows from the Einstein, dilaton, and
$\tilde F_{(3)}$ equations, and is not automatically satisfied as a
consequence of Eqs.~\BPScond{}, \gsconst, \newBianchi{b}, and \Zsoln\
is
\eqn\Ricciflat{R_{\CB\ssk mn} = 0,}
from the $mn$ Einstein equations.  For compact base $\CB$, this comes
close to implying that $\CB$ is $T^4$~or~$K3$.  Where it falls short
is that Ricci-flatness implies only that $c_1(\CB)$ vanishes in
$H^2(\CB,\IR)$.  There can still be a torsion component in
$H^2(\CB,\IZ)$, as is the case for an Enriques surface.  Such
manifolds are ruled out only after we impose supersymmetry.  The
supersymmetry conditions $\partial_\a\hat\chi =
\nabla_{\mkern-3mu \CB\ssk m}\hat\chi = 0$ of Sec.~4.4 give
$c_1(\CB)=0$ as an integrability condition.

\subsec{Geometrical Bianchi Identity and Quantization of the Fibration
Curvature} 

For the fibration \dsstring{}\ to be globally well-defined, the
fibration curvature must satisfy
\eqn\Faclosed{d\CF^\a = 0.}
This is equivalent to the triple-overlap condition on the transition
functions that relate coordinate patches.  Under a fiber coordinate
redefinition, $x^\a\to x^\a+\L^\a\bigl(\{x^m\}\bigr)$, we have
$\CF^\a\to\CF^\a-d\L^\a$ (cf.~App.~B).  Therefore, the exact part of
$\CF^\a$ contains coordinate-gauge information, and the topology of
the fibration is characterized by the cohomology class of $\CF^\a$.

We require that $[\CF^\a]$ be $2\IZ$ valued (cf.~Eq.~\Chern) for the
following reason.  First, ignore the orientifold operation.  The
subgroup of $H^2(X_6,\IZ)$ that comes directly from $H^2(\CB,\IZ)$ is
the quotient group $H^2(\CB,\IZ)/\{[\CF^\a]\}$.  This quotient is
well-defined only if $[\CF^\a]\in H^2(\CB,\IZ)$.  Then, in order to
define the orientifold, we further require that $\CF^\a$ describe a
fibration over $\CB/\IZ_2$.  This is equivalent to the condition
\eqn\ChernB{[\CF^\a]\in H^2(\CB,2\IZ),}
which guarantees integer periods of $[\CF^\a]$ over the cycles in
$\CB/\IZ_2$ that descend from half-cycles in $\CB$.  It is possible
that the $2\IZ$ quantization condition can be replaced by a $2\IZ+1$
quantization condition if one includes localized fibration curvature
at some or all of the $\IZ_2$ fixed points.  However, we do not
consider such localized curvature here.

\subsec{Flux Bianchi Identities/Equations of Motion}

In addition to Eqs.~\newBianchi{a,b} on $\tilde F_{(3)}$, the Bianchi
identities/equations of motion that the fluxes must satisfy are
\eqn\fluxBianchi{dF_{(1)}=0,\quad d\tilde F_{(5)} = 0,\quad{\rm and}
\quad dH_{(3)}=0,}
together the conditions
\eqna\fluxEOM
$$\eqalignno{d\tilde F_{(9)} &= H_{(3)}\w\tilde F_{(7)}, &\fluxEOM a\cr
d\tilde F_{(7)} &= H_{(3)}\w\tilde F_{(5)}, &\fluxEOM b\cr
d\bigl(e^{-2\phi}H_{(7)}\bigr) &= F_{(1)}\w\tilde F_{(7)},
&\fluxEOM c}$$
on the dual fluxes
\eqn\dualflux{\tilde F_{(9)} = * F_{(1)},\quad
\tilde F_{(7)} = * \tilde F_{(3)},\quad{\rm and}\quad
H_{(7)} = * H_{(3)}.}
The middle dimensional flux $\tilde F_{(5)}$ is selfdual, $\tilde
F_{(5)} = *\tilde F_{(5)}$.  Here, $*$ is the 10D Hodge star operator
in the metric \dsstring{}.  In writing these equations, we have used
the fact that the only local sources in our model are O5 planes and D5
branes.

The conditions on the dual fluxes are automatically satisfied as a
consequence of Eq.~\fluxBianchi\ and the Hodge duality constraints
\BPScond{}.  For example,
\eqn\Fnine{\tilde F_{(9)} = *F_{(1)} = -Z^{-1}
dx^{0123}\w\eta^4\w\eta^5\w\bigl(*_\CB F_{(1)}\bigr).}
Using Eqs.~\gs, \gsconst, \BPScond{a}, and the Bianchi identity
$dH_{(3)} = 0$, this becomes
\eqn\dFnine{d\tilde F_{(9)} = d\bigl(e^{-\phi}\Vol_{h(x)}\bigr)
\w H_{(3)},}
which, via Eq.~\Fcalib{a}, is the desired Bianchi identity \fluxEOM{a}
for $\tilde F_{(9)}$.  Eqs.~\fluxEOM{b,c} follow analogously.

Note that Eqs.~\fluxBianchi\ and \BPScond{a}\ imply that after
imposing the equations of motion, $F_{(1)}$ and $H_{(3)}$ are harmonic
on $\CB$, and $F_{(5)\a\b}$ is closed.  Similarly,
Eqs.~\newBianchi{b}, \Faclosed, and \BPScond{c}\ imply that $\CF^\a$
and $F_{(3)\a}$ are harmonic on $\CB$.

\subsec{Discrete Data versus Moduli, and Additional Moduli Constraints}

Beyond the choice of base manifold $\CB$ and orientifold operation,
the complete set of discrete data needed to define the model is the
number $2M$ of D5 branes and their images, together with the quantized
parts of the even fluxes and the fibration curvature \ChernB.

For later discussion of moduli stabilization, it is necessary to
decompose the fluxes into a part that is moduli-dependent and a
quantized part that only involves the discrete choice.  For simplicity
of exposition, we restrict to the case that $\CB=T^4$.  (The results
of this section are analogous in the case that $\CB=K3$, except that
the absence of a first cohomology class for $K3$ implies that
$F_{(1)}=0$, and then from Eq.~\BPScond{a}, $H_{(3)}=0$.)  In this
subsection, we also drop all functional dependence of the moduli on
$x^\m$.  That is, we analyze deformations of the supergravity
background, and do not yet promote these deformations to 4D fields.
We restrict to deformations that correspond to moduli and not to gauge
bosons.  In the next subsection, when we discuss the kinetic terms and
gauge couplings, the $x^\m$-dependence and gauge bosons will be
reintroduced.

Let us write all of the internal fluxes and potentials as the sum of a
background value and a deformation.  In the NS sector we write
\eqn\BHdefs{\eqalign{B_{(2)} &= B^\bg_{(2)}+ b_{(2)},\cr
H_{(3)} &= H^\bg_{(3)}+ h_{(3)}.}}
Since we have assumed that $H_{(3)}=H^{\bsk0}_{(3)}$ (in the notation
of Sec.~3.2), we take $B^\bg_{(2)}$ to have purely base components,
$B^\bg_{(2)} = B^{\bg\,0}_{(2)}$.  The deformation permitted by the
orientifold projection is a shift $b_{(2)\a m}$ in the zero-mode, or
constant component,\foot{To justify this usage of the term zero-mode,
note that the $p$-forms with constant coefficients are annihilated by
a Laplacian operator on $X_6$ formed from a torsionful connection that
forgets about the fibration and warping (cf.~Sec.~4.4).} of the even
potential $B_{(2)\a m}$.  Therefore,
\eqn\Bbdef{\eqalign{b_{(2)} &= \eta^\a\w b_{(2)\a},\cr
h_{(3)} &= db_{(2)} = \CF^\a\w b_{(2)\a}.}}

In the RR sector, we similarly write
\eqn\CFdefs{\eqalign{C_{(p)} &= C^\bg_{(p)}+ c_{(p)},\cr
F_{(p+1)} &= F^\bg_{(p+1)} + f_{(p+1)},\quad
\tilde F_{(p+1)} = \tilde F^\bg_{(p+1)} + \tilde f_{(p+1)}.}}
In this case, the moduli are not quite the $c_{(p)}$.  The latter are
in general multivalued quantities on $X_6$ due to the action of the
$C_{(p-2)}$ transition functions on $C_{(p)}$ when $H_{(3)}\ne0$.
This subtlety is discussed in Ref.~\FP.  Instead, the moduli are the
combinations
\eqna\ctilde
$$\eqalignno{\tilde c_{(2)}
&= c_{(2)}-b_{(2)}\msk C^\bg_{(0)}, &\ctilde a\cr
\tilde c_{(4)} &= c_{(4)} - b_{(2)}\w\bigl(C^{\bg\,1}_{(2)} + \tilde
c_{(2)}\bigr) -\half b_{(2)}\w b_{(2)}\w C^\bg_{(0)}, &\ctilde b\cr
\tilde c_{(6)} &= c_{(6)} - b_{(2)}\w \tilde c_{(4)}
-\half b_{(2)}\w b_{(2)}\w\tilde c_{(2)}, &\ctilde c}$$
as can be verified by arguments analogous to those in Ref.~\FP.  Here,
Eqs.~\ctilde{a,b,c}\ are relations between quantities with internal 6D
indices only.

The nonvanishing components of the background are
\eqn\nvCbg{C^\bg_{(0)},\quad C^\bg_{(2)mn},\quad
C^\bg_{(2)\a m},\quad C^\bg_{(4)mnpq},\quad C^\bg_{(4)\a npq},}
or a subset thereof, and the moduli are the zero-modes
\eqn\nvctilde{\tilde c_{(2)mn},\quad \tilde c_{(2)\a\b},\quad \tilde
c_{(4)\a mnp},\quad \tilde c_{(6)\a\b mnpq}.}
The last modulus is $\tilde c_{(6)\,456789}$.  This is the scalar that
is dual to the corresponding deformation $\tilde c_{(2)\m\n}$ of
$C_{(2)\m\n}$, and will be more convenient to work with than $\tilde
c_{\m\n}$ when we discuss gauge couplings in Sec.~3.10 and the
T-duality map in Sec.~5.2.

The flux deformations that follow from Eqs.~\ctilde{a,b,c}\ are
\eqna\fluxdeform
$$\eqalignno{f_{(3)} &= \tilde f_{(3)} = d\tilde c_{(2)} + b_{(2)}\w
F_{(1)}, &\fluxdeform a\cr
\tilde f_{(5)} &= -\tilde c_{(2)}\w H^\bg_{(3)}
+ b_{(2)}\w\bigl(F^{\bg\,1}_{(3)}+d\tilde c_{(2)}\bigr)
+ \half b_{(2)}\w b_{(2)}\w F_{(1)}, &\fluxdeform b}$$
expressable in terms of the moduli and the gauge invariant fluxes.
The second equality in Eq.~\fluxdeform{a}\ implies that the
only nonvanishing component of $\tilde f_{(3)}$ is $\tilde f_{(3)\a}$,
with one fiber index; this, combined with the earlier observation
that $\tilde F_{(3)\a} = F_{(3)\a}$, gives the first equality.  In
Eq.~\fluxdeform{b}, $F_{(1)} = F^\bg_{(1)}$, so we have dropped the
superscript. 

One conseqence of Eqs.~\Bbdef\ and \fluxdeform{a}\ is that
\eqn\diffNflux{\CF^\a\w f_{(3)\a} + F_{(1)}\w h_{(3)} = 0,}
from which Eq.~\Nflux\ becomes
\eqn\Nfluxbg{N_\flux = {1\over(2\pi)^2\a'}\int_\CB
\bigl(\CF^\a\w F^\bg_{(3)\a} + F_{(1)}\w H^\bg_{(3)}\bigr),}
independent of the moduli.

Dirac quantization constrains the closed moduli-independent fluxes
$F_{(1)}$, $F^\bg_{(3)\a}$, $F^\bg_{(5)\a\b}$, and $H^\bg_{(3)}$ to be
representatives of $2\IZ$-valued cohomology:
\eqna\DiracCohomB
$$\eqalignno{&[F_{(1)}]\in  H^1(\CB,2\IZ),\quad
[F^\bg_{(3)\a}] \in (2\pi)^2\a' H^2(\CB,2\IZ), &\DiracCohomB a\cr
&[H^\bg_{(3)}] \in (2\pi)^2\a' H^2(\CB,2\IZ),\quad
[F^\bg_{(5)\a\b}] \in (2\pi)^4\a'^2 H^3(\CB,2\IZ). &\DiracCohomB b}$$
As in Eq.~\ChernB, the $2\IZ$ quantization of periods on the covering
space $\CB$ ensures $\IZ$ quantization of periods over half cycles in
$\CB$ that descend to proper boundaryless cycles in $\CB/\IZ_2$.  We
can now return to the issue of the even integer quantization of
$N_\flux$ mentioned in Sec.~3.2. From Eq.~\Nfluxbg, we see that
$N_\flux$ depends only on the discrete choice \ChernB\ and
\DiracCohomB.  As a result, $N_\flux$ satisfies $N_\flux\in4\IZ$,
which is indeed an even integer, and is moduli independent, so no
moduli constraints arise from the Gauss's law constraint.

While Dirac quantization alone permits a quantized five-form flux
$F^\bg_{(5)\a\b}$, it can be shown that $F^\bg_{(5)\a\b}=0$ in order
to satisfy the equations of motion.  We omit the direct proof, but
Sec.~5.2 contains a proof via T-duality.  From the constraint
\BPScond{d}, we then have
\eqn\noffivetilde{\tilde f_{(5)} = 0.}

Without imposing any supersymmetry conditions, the complete set of
moduli constraints that follow from the equations of motion consists
of Eqs.~\BPScond{a,c}\ and \noffivetilde.  When $F_{(1)} = H^\bg_{(3)}
= 0$, the constraints simplify.  Eq.~\BPScond{a} then implies that
\eqna\FthreeBconstr
$$0 = h_{(3)} = \CF^\a\w b_{(2)\a},\eqno\FthreeBconstr a$$
and the constraint \noffivetilde\ becomes
$$b^{\vphantom{\bg}}_{(2)[\a}\w F^\bg_{(3)\b]} = 0.
\eqno\FthreeBconstr b$$ 
In this case, the complete set of moduli constraints is \BPScond{c}\
and Eqs.~\FthreeBconstr{a,b}.

\subsec{Couplings to Gauge Bosons}

In addition to the deformations of the supergravity background just
discussed, which when promoted to 4D fields become moduli, there are
other deformations of the supergravity background which when promoted
to 4D fields become gauge bosons.  In the closed string sector, these
deformations are the zero modes on $X_6$ of $V^\a{}_\m$, $B_{(2)m\m}$,
$C_{(2)\a\m}$, and $C_{(4)\a\b m\m}+6b_{(2)[\a m|}C_{(2)|\b\m]}$, and
will be denoted by the lowercase symbols
\eqn\zmgauge{v^\a{}_\m,\quad b_{(2)m\m},\quad c_{(2)\a\m},\quad
{\rm and}\quad \tilde c_{(4)\a\b m\m},}
respectively.  When the couplings to these gauge bosons are included
and all deformations are promoted to 4D fields, the kinetic term for
$\tilde c_{(2)mn}$ is the square of
\eqna\gaugecouplings
$$\partial_\m\tilde c_{(2)mn} - F^\bg_{(3)\a mn}v^\a{}_\m
+ 2F_{(1)[m} b_{(2)n]\m} + \CF^\a{}_{mn}c_{(2)\a\m},
\eqno\gaugecouplings a$$
that for $\tilde c_{(4)\a mnp}$ is the square of
$$\partial_\m\tilde c_{(4)\a mnp} + 3\CF^\b_{[mn|}
\tilde c^{\vphantom{\bg}}_{(4)\b\a|p]\m} + H^\bg_{(3)mnp}
c^{\vphantom{\bg}}_{(2)\a\m} + 3F^\bg{}_{(3)\a[mn}b_{(2)p]\m},
\eqno\gaugecouplings b$$
and that for $\tilde c_{(6)\a\b mnpq}$ (the scalar that is dual to
$\tilde c_{(2)\m\n}$) is the square of
$$\partial_\m\tilde c_{(6)\a\b mnpq} - 3H^\bg_{(3)[mnp|}
\tilde c^{\vphantom{\bg}}_{(4)\a\b|q]\m}. \eqno\gaugecouplings c$$
The axion $\tilde c_{(2)\a\b}$ of $\tdil$ does not couple to gauge
bosons.

Beyond the closed string sector gauge bosons, there are also gauge
bosons that arise from the lowest Kaluza-Klein mode of the D5
worldvolume gauge fields $A_{I\,\m}$ on the wrapped $T^2$ fiber.


\newsec{Supersymmetry Conditions}

The 10D type IIB dilatino and gravitino variations are given in
App.~D.  From these fermion variations, we will now determine the
conditions on the 6D geometry and internal flux for unbroken 4D
$\CN\ge1$ supersymmetry.

\subsec{Decomposition of 10D Supersymmetry Parameters}

In 10D, the IIB supersymmety transformations are parametrized by two
Majorana-Weyl spinors $\e_{L,R}$ (real and negative chirality in our
conventions), which combine to form a single Weyl spinor $\e=\e_L +
i\e_R$.  For compactification to 4D, it is desirable to decompose $\e$
into 4D and 6D spinors.\foot{In this discussion, we follow Gra\~na and
Polchinski \GP.}  An arbitrary 10D negative chirality Weyl spinor $\e$
can be written
\eqn\tenDWeyl{\e = u\otimes\chi_1 + u^*\otimes\chi^*_2,}
where $\chi_1$ and $\chi_2$ are negative chirality 6D Weyl spinors,
and $u$ is a positive chirality 4D Weyl spinor.\foot{Note also that
for $\e$ and $u$ to be standard anticommuting spinors, $\chi_1$ and
$\chi_2$ must be commuting spinors \BeckerM.}  For each pair
$(\chi_1,\chi_2)$ such that \tenDWeyl\ gives vanishing dilatino and
gravitino variation for all $u$, we obtain one 4D $\CN=1$
supersymmetry generated by $u$.

Orientifold planes impose additional constraints on the spinors.  Two
broad classes of constraints are
\eqn\BeckerAndy{\eqalign{\chi_1=0\quad\hbox{or}\quad\chi_2 = 0
&\qquad\hbox{Becker-type or chiral constraints,}\cr
\chi_2 = e^{i\a}\chi_1,\quad \a=\hbox{ const}
&\qquad\hbox{Andy-type or Majorana-Weyl constraints.}}}
Becker-type spinors are of definite 6D and 4D chirality, while
Andy-type spinors are real up to an overall constant phase.  Roughly
speaking, Becker-type constraints are associated with O3 or O7 planes,
and Andy-type constraints are associated with O5 or O9 planes or the
heterotic theories. More precisely, O3 and O9 constraints require that
the spinors be of pure Becker- or Andy-type; O7 and O5 constraints
require that the spinors be a particular linear combination of two
Becker- or two Andy-type spinors, such that for certain choices of
flux we can decompose the supersymmetry algebra into $\CN=1$
subalgebras, each generated by a spinor for which one of the two terms
in the linear combination vanishes.  We will demonstrate this for the
O5 case which is the focus our investigation, and for the O3 case
which is needed in Sec.~5.

Independent of the details of the orientifold projection, $\e_R$ is a
real Majorana-Weyl spinor, and can therefore be written as
\eqn\Repsilon{\e_R={1\over\sqrt2}\bigl(u\otimes\chi
+ u^*\otimes\chi^*\bigr),}
for some $u$ and $\chi$, where the leading factor of $1/\sqrt2$ is for
compatibility with the normalization conventions
\eqn\spinnorm{\e_{L,R}^\dagger\e_{L,R}^{\vphantom{\dagger}}
= u^\dagger u = \chi^\dagger\chi = 1.}

For the O5 planes of interest, the 10D supersymmetries preserved by
the orientifold projection are generated by $\e_{L,R}$ such that
\eqn\eOfive{\e_L = \G_\CB\e_R.}
Here, $\G_\CB$ is the chirality operator formed from product of Dirac
matrices in the directions transverse to the orientifold planes,
normalized so that $\G_\CB{}^2=1$.  Similarly, we define $\G_\fib$
to be the chirality operator in the two compact directions wrapped by
the O5 plane.  That is, $\G_\fib$ is the chirality operator on the
$T^2$ fiber, and $\G_\CB$ is the chirality operator on the base $\CB$.
Using these chirality operators, we can decompose the 6D spinor $\chi$
into components of definite fiber and base chirality,
\eqn\FibBaseChi{\chi = \chi_+ + \chi_-,\quad\hbox{where}\quad
\g_\fib\chi_\pm = \pm\chi_\pm,\quad 
\g_\CB\chi_\pm = \mp\chi_\pm.}
Here, we have gone from uppercase 10D to lowercase 6D internal Dirac
matrices using the relations $\G_\fib = 1\otimes\g_\fib$ and $\G_\CB=
1\otimes\g_\CB$.  (See App.~A for a precise statement of our
conventions for Dirac matrices, chirality operators, and their
decompositions under 10D $\to$ 4D$\times$6D $\to$
4D$\times$2D$\times$4D).

Applying the decomposition \FibBaseChi\ to Eq.~\eOfive, we obtain
\eqn\Lepsilon{\e_L={1\over\sqrt2}\bigl(u\otimes\chi_L
+ u^*\otimes\chi_L^*\bigr),}
where
\eqn\OfivechiL{\chi_L = \g_\CB\chi = -\chi_+ + \chi_-.}
The 10D Weyl spinor $\e= \e_L + i\e_R$ becomes
\eqn\OfiveAndy{\e = 
-e^{-i\pi/4}\bigl(\chi_+\otimes u + \chi_+^*\otimes u^*\bigr)
+e^{i\pi/4}\bigl(\chi_-\otimes u + \chi_-^*\otimes u^*\bigr),}
which is the desired linear combination of two Andy-type spinors.  In
the case that $F_{(1)} = H_{(3)}=0$, we will show in Sec.~4.2 that the
supersymmetry conditions on $\chi_+$ and $\chi_-$ decouple from one
another, so that the space of 6D Killing spinors $\chi$ decomposes
into a subspace on which $\chi_-=0$ and a subspace on which
$\chi_+=0$.  On either subspace, $\e$ is of pure Andy-type.

For spacetime filling O3 planes, the 10D supersymmetries are generated by
$\e_{L,R}$ such that
\eqn\eOthree{\e_L = -i\G^{(6)}\e_R.}
Here, $\G^{(6)}$ is the chirality operator formed from the product of
Dirac matrices in the six internal directions transverse to the
orientifold planes.  Using Eq.~\Repsilon, this becomes Eq.~\Lepsilon\
with
\eqn\OthreechiL{\chi_L = -i\g^{(6)}\chi = i\chi.}
So, the 10D Weyl spinor is
\eqn\OthreeBecker{\e = \sqrt2 i\chi\otimes u,}
of Becker-type, as claimed.  

Although it is not directly relevant to this investigation, it is
interesting to note that there exist supergravity backgrounds that are
neither Andy-type or Becker type, but rather interpolate between the
two \refs{\FG,\FreySU}.  These solutions are D3-like in some regions
and D5/NS5-like in others.  Similarly, there exist dielectric flow
solutions, involving D3 branes that become polarized into D5/NS5
branes under radial flow in IIB, or M2 branes that become polarized
into M5 branes under radial flow in M~theory \refs{\PS,\Warneretal}.
In Ref.~\Warneretal, the technique of ``Algebraic Killing spinors''
was developed and repeatedly applied as an efficient method for
obtaining solutions to the supergravity equations of motion.  In the
M~theory case, these supergravity solutions fit nicely into
classification of Ref.~\MartelliSparks, which provides a general
framework in the language of $G$-structures for describing arbitrary
compactifications of M~theory to $\IR^{2,1}$ and $AdS_3$
\MartelliSparks.  This framework also accomodates the M theory dual of
Refs.~\refs{\FG,\FreySU}.  A more restrictive treatment with
applications, in particular, to the M~theory lift of IIB pp-wave
backrounds has also been given \BeckerNew.  Finally, recent work has
shown that the correct description of the most general $\IR^{3,1}$ or
$AdS_5$ compactifications of type IIB or M~theory, should be given in
terms of $SU(2)$ structures
\refs{\DallAgataIIA,\GauntAdS,\DallAgataIIB}.  This work has brought
us closer to understanding, for example, what the complete
supergravity solution for the Polchinski-Strassler background might
be, a subject that is currently under investigation \HumboldtPSflow.

\subsec{Supersymmetry Conditions for RR Three-Form Flux Only}

In the case that there is only $F_{(3)}$ RR flux and nontrivial
fibration, the supersymmetry conditions are very similar to those for
the heterotic string with NS flux.  By substituting the expression
\OfiveAndy\ for $\e$ into the Eqs.~\preIIBfermvar\ for the IIB fermion
variations, and then demanding that the result vanish, we obtain
\eqn\Fdilvars{\eqalign{\bigl(\half\dsl\phi 
\mp \quarter e^\phi\Fsl_{(3)}\bigr)u\otimes\chi_\pm & = 0,\cr
\bigl(\nabla_M -\quarter \G_M\dsl\phi 
\pm \quarter e^\phi\Fsl_{(3)M}\bigr)u\otimes\chi_\pm &= 0.}}
After performing a Weyl rescaling
\eqn\WeylFthree{ds_{\rm string}^2 = e^{(\phi-\phi_0)}\widehat{ds}^2,
\quad \chi_\pm = e^{(\phi-\phi_0)/4}\hat\chi_\pm,
\quad \G^M = e^{-(\phi-\phi_0)/2}\hat\G^M,}
where $e^{\phi_0} = g_s$, these equations become
\eqna\hateqs
$$\eqalignno{\bigl(\half\hat{\dsl}\phi \mp \quarter 
g_s \Fslhat_{(3)}\bigr)u\otimes\hat\chi_\pm & = 0, 
& \hateqs a\cr
\bigl(\hat\nabla_M  \pm \quarter g_s \Fslhat_{(3)M}\bigr)
u\otimes\hat\chi_\pm &= 0. & \hateqs b}$$
Here, the hatted metric is
$${\widehat{ds}}^2 = Z^{-1/2}e^{-(\phi-\phi_0)}
\bigl(\eta_{\m\n}dx^\m dx^\n + ds_6^2\bigr). \eqno\hateqs c$$

\eqna\dblresults\eqna\dblACS
Eqs.~\hateqs{}\ are a doubled version of the equations that formed the
starting point for Strominger's ``Superstrings with Torsion'' analysis
\Strom.  Since the constraints on $\chi_+$ and $\chi_-$ are decoupled
from one another, we are free to take $\chi = \chi_+$ or $\chi_-$, and
set the other spinor to zero.  Then the starting points are identical.
The results that follow from the same analysis are just Strominger's
results, with $H_{(3)}\to \mp g_s F_{(3)}$, $\chi\to\hat\chi_\pm$, and
$\phi\to-\phi$ compared to Sec.~1.1:\foot{The heterotic analogs of
Eqs.~\dblresults{a,d} and \dblACS{b} were suppressed from our short
summary in Sec.~1.1, but follow from Strominger's analysis
\refs{\Strom,\GauntSLAG,\GauntIntr}.}
$$\eqalignno{Z^{-1/2} e^{-(\phi-\phi_0)} &= 1, & \dblresults a\cr
e^{-2\phi}d\bigl(e^{2\phi}J_\pm\bigr) & = \pm *_6 g_s F_{(3)}, 
&\dblresults b\cr
d\bigl(e^{2\phi} *_6 J_\pm\bigr) & = 0, &\dblresults c\cr
d\bigl(e^{2\phi} \O_\pm\bigr) & = 0, &\dblresults d\cr
}$$
where the $SU(3)_\pm$ structures are\foot{The fact that $\O_\pm$
depends antiholomorphically on $\hat\chi_\pm$ is a consequence of the
negative $\g^{(6)}$ chirality of $\hat\chi_\pm$.}
$$\eqalignno{J_{\pm a}{}^b 
&= i\hat\chi_\pm^\dagger\g_a{}^b\hat\chi_\pm, & \dblACS a\cr
\O_{\pm abc} 
&= \hat\chi_\pm^\dagger\g_{abc}\hat\chi^*_\pm. & \dblACS b}$$
Just as for the heterotic string, $e^{2\phi}\O_\pm$ is a holomorphic
(3,0) form and the Nijenhuis tensor vanishes, so that the complex
structure \dblACS{a}\ is integrable.  Eq.~\dblresults{a} reproduces
the relation \gs\ between the dilaton and warp factor.
Eqs.~\dblACS{a,b} imply the relations~\refs{\Strom,\Louis}
\eqna\Jrels
$$\eqalignno{J_{\pm a}{}^b J_{\pm b}{}^c &= -\d_a{}^c, & \Jrels a\cr
{1\over3!} J_\pm\w J_\pm\w J_\pm &= {i\over8} \O_\pm\w\bar\O_\pm 
= \Vol_6, & \Jrels b}$$
where $\Vol_6$ is the volume form associated with the metric \sixDg,
\eqn\volsix{\Vol_6 = Z^2 \bigl(g^{(\Tfib)}g_\CB\bigr)^{1/2}
\eta^4\w \eta^5\w dx^6\w dx^7\w dx^8\w dx^9.}

In the next two paragraphs, we derive a number of results concerning
the fiber and base decomposition of $J_\pm$ and $\O_\pm$ that will be
useful for the examples in Sec.~6-8.  These paragraphs can be skipped
over, if desired, without much loss to the understanding of this
section.
\bigskip

Since $\hat\chi_+$ and $\hat\chi_-$ have definite fiber and base
chirality, these spinors factorize into the product of a 2D Weyl
spinor on the fiber and a 4D Weyl spinor on the base,
\eqn\chizeta{\hat\chi_\pm = \z^\fib_\pm\otimes\z^\CB_\pm,\quad
\hbox{where}\quad\tilde\g_\fib\z^\fib_\pm = \pm\z^\fib_\pm,
\quad \tilde\g_\CB\z^\CB_\pm = \mp\z^\CB_\pm.}
(The tilded Dirac matrices are defined in App.~A).  Consequently, the
K\"ahler form and (3,0) form decompose as
\eqn\OJpm{J_\pm = J^{(\Tfib)}_\pm + Z J^\CB_\pm,\quad
\O_\pm = \bigl(2g^{(\Tfib)}_{1\bar1}\bigr)^{1/2}
\eta^{z^1}\w Z\O^\CB_\pm.}
Here, the forms $J^\CB_\pm$ and $\O^\CB_\pm$ are the K\"ahler form and
holomorphic (2,0) form on the base, defined via
\eqn\OJbase{J^\CB_{\pm m}{}^n = i{\z^\CB}^\dagger_\pm\tilde
\g_m{}^n\z^\CB_\pm,\quad
\O^\CB_{\pm mn} = {\z^\CB_\pm}^\dagger\tilde\g_{mn}\z^{\CB*}_\pm.}
Also, in complex coordinates, the fiber metric \dsstring{b} is
\eqn\cpxfibm{ds^2_{\Tfib} = g^{(\Tfib)}_{1\bar1}\eta^{z^1}
\eta^{\bar z^1} + g^{(\Tfib)}_{\bar1 1}\eta^{\bar z^1}
\eta^{z^1},\quad g^{(\Tfib)}_{1\bar1} = g^{(\Tfib)}_{\bar1 1}
= {V_\fib\over2\bigl|\Im\t_1\bigr|},}
where $V_\fib$ is given by Eq.~\Vfib, and the fiber (1,0) form is
\eqn\etaz{\eta^{z^1} = dz^1+\CA^{z^1},\quad\hbox{where}\quad
z^1 = x^4 + \t_1 x^5,\quad \CA^{z^1} = \CA^4+\t_1 \CA^5,}
in terms of the fiber complex structure modulus $\t_1$.

The fiber complex structure is related to the spinors by
\eqn\Jfib{J^{(\Tfib)}_{\pm\a}{}^\b = i{\z^\fib_\pm}^\dagger
\tilde\g_\a{}^\b\z^\fib_\pm,}
which implies that
\eqna\fibpm
$$J^{(\Tfib)}_\pm = \pm\Vol_\fib,\eqno\fibpm a$$
where $\Vol_\fib$ was defined in Eq.~\Volfib, and
$$\pm = \sign(\Im\t_1),\eqno\fibpm b$$
Then, from Eqs.~\Jrels{b} and \OJpm,
$$\ha J^\CB_\pm\w J^\CB_\pm = \pm \Vol_\CB.\eqno\fibpm c$$
So, the geometrical origin of the $\pm$ doubling in
Eqs.~\dblresults{}\ and \dblACS{}\ is that positive orientation of
$X_6$ \Jrels{b}\ can correspond to either positive-positive or
negative-negative orientation of the fiber and base \fibpm{a,c}.
\bigskip

As a consequence of Eqs.~\dblresults{a} and \dblACS{b},
Eq.~\dblresults{d} is automatically satisfied.  The remaining
conditions \dblresults{b,c} are equivalent to \refs{\CardTor,\Keshav}
\eqn\cpxfluxI{\tilde F_{(3)}\mp {i\over g_s} e^{-2\phi}
d\bigl(e^{2\phi}J_\pm\bigr)\quad\hbox{(2,1) and primitive.}}
Using Eqs.~\OJpm\ and \fibpm{a}, and the decomposition \Fdecomp, this
becomes
\eqna\cpxfluxII
$$\eqalignno{\tilde F^{\bsk0}_{(3)} &= *_\CB g_s^{-1} dZ, &\cpxfluxII a\cr
G_{(3)}&\quad\hbox{(2,1) and primitive,}& \cpxfluxII b}$$
where
$$G_{(3)} = F^{\bsk1}_{(3)} - {i\over g_s} d\Vol_\fib.
\eqno\cpxfluxII c.$$
From the results of Sec.~3.9 applied to the case that $H_{(3)} = 0$, we
can also write $G_{(3)}$ as
$$G_{(3)} = F^{\bg\,1}_{(3)} - (2\pi)^2\a'\tdil\,
d\bigl(\eta^4\wedge\eta^5\bigr), \eqno\cpxfluxII d$$
where 
\eqn\newtdil{\tdil = {1\over(2\pi)^2\a'} \biggl(-\tilde c_{(2)\,45}
+ {i\over g_s}V_\fib\biggr),}
which is the suitable generalization of Eq.~\newaxdil\ to the case of
nonvanishing background flux.

Eqs.~\cpxfluxII{a,b,d}\ express the supersymmetry conditions in the
form that we will find most convenient to apply in Secs.~6--8.  Given
a discrete choice of $F^\bg_{(3)\a}$ and $\bigl[\CF^\a\bigr]$, we can
easily deduce the constraints on moduli via these equations.  The
number $\CN$ of supersymmetries preserved is
\eqn\Nsusy{\CN = \CN_+ + \CN_-,}
where $\CN_\pm$ is the number of independent $\pm$ complex structures
(i.e., the number of $\hat\chi_\pm$) such that these conditions are
satisfied.  Eqs.~\cpxfluxII{}\ also imply the equations of motion
\BPScond{}.\foot{The equations of motion \BPScond{b,c} require that
$G_{3}$ be imaginary-selfdual (ISD): $*_6 G_{(3)} = i G_{(3)}$.  This
is weaker than \cpxfluxII{b}.  The space of ISD three-forms includes
not only primitive (2,1) forms, but also (0,3) forms and non-primitive
(1,2) forms $J\w\o$, where $\o$ is a (0,1) form.}\footlabel\ISDfoot\
The maximum amount of supersymmetry is $\CN=4$, $\CN_+=\CN_-=2$, since
there are two 4D spinors of each chirality and one 2D spinor of each
chirality, from which we must construct the negative chirality 6D
spinors $\hat\chi_\pm$.  For nonvanishing flux, the supersymmetry is
strictly less than this.  That is, either $\CN_+<2$ and $\CN_-\le2$,
or $\CN_+\le2$ or $\CN_-<2$.

\subsec{S-dual Orientifolds with NS Flux Only}

The class of supersymmetric vacua just discussed is S-dual to a class
of vacua with NS flux $H_{(3)}$ only.  Such vacua have received
considerable attention, so it is desirable to state the precise
connection between Sec.~4.2 and known results.

The result of applying S-duality to the class of orientifolds analyzed
here is a class of dual orientifolds in which O5 planes are replaced
by ON5 planes and D5 branes are replaced by NS5 planes \HananyKol.
The orientifold operation $\O\CI_4$ in the original theory becomes
$(-1)^{F_L}\CI_4$ in the dual theory.  In our S-duality conventions,
the dual NS flux is related to the original RR three-form flux by
\eqn\SdualH{H'_{(3)} = -F_{(3)}.}
The S-dual metric is
\eqn\Sdualm{\eqalign{ds'^2_{\rm string}
&= e^{-(\phi-\phi_0)}ds_{\rm string}^2\cr
&= \eta_{\m\n}dx^\m dx^\n + ds_6{}^2,}}
where we have used Eq.~\gs\ (or equivalently, Eq.~\dblresults{a}) in
the second equality.  We recognize this as the hatted metric
\hateqs{c} that appeared in the previous section.

The S-dual dilaton is given by
\eqn\Sdualdil{e^{(\phi'-\phi'_0)} = e^{-(\phi-\phi_0)}
= Z^{1/2},\quad g'_s = g_s^{-1}.}
So, the 6D metric can be written
\eqn\Gauntm{ds_6{}^2 = g^{(\Tfib)}_{\a\b}(dx^\a+\CA^\a)(dx^\b+\CA^\b)
+e^{2(\phi'-\phi'_0)}g_{\CB\ssk mn}dx^m dx^n,}
which is a form that has appeared previously in the literature.
Nearly identical vacua were described by Gauntlett, Martelli, and
Waldram, in their classification of static supersymmetric backgrounds
with NS flux only \GauntIntr.  The one minor difference is that the
local ON5 and NS5 sources were excluded from their discussion, and as
a result, the base geometry was required to be noncompact to avoid
contradicting the Gauss's law constraint \Gauss.

In the conventions of Eq.~\SdualH, the 10D IIB supersymmetry
parameters after the S-duality are
\eqn\SdualeLR{\eqalign{\e'_L &= {1\over\sqrt{2}}\bigl(
u\otimes\chi_- + u^*\otimes\chi_-^*\bigr),\cr
\e'_R &= {1\over\sqrt{2}}\bigl(
u\otimes\chi_+ + u^*\otimes\chi_+^*\bigr),}}
up to a possible overall sign in $\e'_L$ and/or $\e'_R$ that will not
concern us here.  Thus, $\CN_+$ and $\CN_-$, which before the
S-duality counted the number of 6D Killing spinors of $+$ and $-$
fiber chirality, map to the number of right and left Killing spinors,
respectively, in the S-dual theory:
\eqn\SdualN{\CN'_R = \CN_+,\quad \CN'_L = \CN_-.}

In Ref.~\GauntIntr, the metric \Gauntm\ was shown to describe
supergravity backgrounds (with $SU(2)$ structure in 6D) such that
$\CN'_R >0$ and $\CN'_L=0$.  On the other hand, for $\CN'_R=\CN'_L
=1$, the metric was shown to possess an almost product structure and
take the form
\eqn\Gauntmii{ds'^2_6 = g^{(4)}_{ab}dx^a dx^b +
e^{2(\phi-\phi_0)}g^{(2)}_{cd}dx^c dx^d,\quad
a,b=4,5,6,7,\quad c,d=8,9.}
Here, $g^{(4)}_{ab}$ and $\phi$ can depend on all six coordinates, but
$g^{(2)}_{cd}$ can only depend on $x^8$ and $x^9$.

In contrast, for the orientifold backgrounds studied here, we have
argued that the S-dual metric is of the form \Gauntm, without
reference to $\CN_\pm$.  For the two results to agree, it must be true
that for $\CN_+,\CN_- \ge1$, the 6D metric of the O5 theory takes a
form compatible with both \Gauntm\ and \Gauntmii.  In Sec.~6.2, we
present an example with $\CN_+=\CN_-=1$, in which this is indeed the
case.  In this example, the base metric $g_{\CB\ssk mn}$ reduces to
the product metric on flat $T^2_{\{67\}}$ times flat $T^2_{\{89\}}$,
and the fibration is such that we can take
\eqn\SAchoice{\CA^4 = -2nx^8dx^6,\quad \CA^5 = -2nx^8dx^7.}
Also, as is true throughout our investigation, $g^{(\Tfib)}_{\a\b}$ is
constant and the dilaton depends only on the base coordinates.
Therefore, the metric \Gauntm\ can be cast in the form \Gauntmii\ by
writing
\eqn\GauntmMap{\eqalign{g^{(4)}_{ab} dx^a dx^b
&= g^{(\Tfib)}_{\a\b}(dx^\a+\CA^\a)(dx^\a+\CA^\a)
+ e^{2(\phi'-\phi'_0)}ds^2_{T^2{\{67\}}},\cr
g^{(2)}_{cd} dx^c dx^d &= ds^2_{T^2{\{89\}}}.}}
Similar remarks apply to the example in Sec.~7.1, which preserves
$\CN_+=1$, $\CN_-=2$ supersymmetry.

\subsec{Supersymmetry Conditions for More General Flux}

The generalization of Eqs.~\hateqs{a,b}\ to the case of arbitrary flux
\evenfluxes\ and \oddflux\ is
\eqna\gensusy
$$\eqalignno{&\bigl(\half\hat{\dsl}\phi \mp \quarter 
g_s \hat{\tilde\Fsl}_{(3)}\bigr)u\otimes\hat\chi_\pm
+\Bigl(\quarter e^{-(\phi-\phi_0)}\hat\Hsl_{(3)}
\mp\half g_s e^{(\phi-\phi_0)}\Fslhat_{(1)}\Bigr)
u\otimes\hat\chi_\mp = 0, & \gensusy a\cr
\noalign{\vskip5pt}&\bigl(\hat\nabla_M  \pm \quarter g_s \Fslhat_{(3)M}\bigr)
u\otimes\hat\chi_\pm - \Bigl(\eighth e^{-(\phi-\phi_0)}
\bigl(\hat\Hsl_{(3)}\hat\G_M + 2\hat\G_M\hat\Hsl_{(3)}\bigr) &\cr
&\quad \pm \eighth g_s e^{(\phi-\phi_0)}
\bigl(\Fslhat_{(1)}\hat\G_M + 2\hat\G_M\Fslhat_{(1)}\bigr)
\mp\smallfrac{1}{16} g_s e^{-(\phi-\phi_0)}
\hat{\tilde\Fsl}_{(5)}\hat\G_M\Bigr)
u\otimes\hat\chi_\mp 
= 0. & \gensusy b}$$
In analyzing the equations of motion in Sec.~3, we made the
simplifying assumption that $H_{(3)\a\b m} = 0$ and found that $\tilde
F_{(5)} = 0$ as a result.  It can be shown that the same conclusion
also follows if instead of using the equations of motion we impose the
supersymmetry conditions plus Bianchi identities.\foot{When there is
maximal unbroken supersymmetry, the supersymmetry conditions together
with the Bianchi identities imply the equations of motion.  For
reduced supersymmetry, this is not necessarily the case, nevertheless,
it does seem to be true for the backgrounds presented in
\refs{\GKP,\GauntIntr,\KST,\KSTT}, as well as those presented here.
See Ref.~\GauntPakis\ for a recent discussion.}  However, in this
section, we will simply take as a starting point that both $H_{(3)\a\b
m}$ and $\tilde F_{(5)}$ vanish.

Since the analysis below is somewhat involved, let us first summarize
the results.  We will find that the supersymmetry conditions become a
refinement of the Hodge duality conditions \BPScond{a,b,c}, together
with the condition that $\hat\chi = \hat\chi_+ + \hat\chi_-$ is
constant on the fiber and base,
\eqn\constchi{\partial_\a\hat\chi = 0,\quad
\nabla_{\mskip-2mu \CB\mskip1mu m}\hat\chi = 0.}
From the internal 6D point of view, Eq.~\constchi\ says that
$\hat\chi$ is covariantly constant with respect to a torsionful
connection that simply forgets about the warping ($Z$) and fibration
($\CA$) in the metric \sixDg.  The refinement of Eqs.~\BPScond{a,b,c}
is
\eqna\Gtildeconds
$$\eqalignno{& G_{(1)}\quad\hbox{of type (0,1),} &\Gtildeconds a\cr
&\tilde G_{(3)}\quad\hbox{(2,1) and primitive,} &\Gtildeconds b}$$
where
$$\eqalignno{G_{(1)} &= F_{(1)} - {i\over2}(Z J\w J)\lrcorner
\,\biggl(\ssk{1\over g_s}\mkern-3mu\Vol_\fib\w H_{(3)}\biggr),
& \Gtildeconds c\cr
\tilde G_{(3)} &= G_{(3)} - {i\over2}Z^{-1/2}J\w G_{(1)}, 
& \Gtildeconds d}$$
and $G_{(3)}$ is given by Eqs.~\cpxfluxII{c}.\foot{In case the reader
is bothered by the explicit appearance of the warp factor $Z$ in
Eqs.~\Gtildeconds{c,d}, we note that $J$ and $\lrcorner\,$ also
contain implicit $Z$ dependence, to that $Z$ drops out of the moduli
constraints implied by conditions \Gtildeconds{a,b}\ (cf.~Sec.~6.3).
We can also remove explicit reference to $Z$ from
Eqs.~\Gtildeconds{c,d}, if desired, by redefining $J$ and
$\lrcorner\,$ in terms of the rescaled metric $Z^{-1/2}ds_6{}^2$,
which is the restriction of the 10D string frame metric \dsstring{}\
to the internal space.}  Here, the symbol $\lrcorner\,$ denotes
contraction,
\eqn\contr{\bigl(A_{(p)}\lrcorner\, B_{(q)}\bigr)_{a_1\ldots a_{q-p}}
={1\over p!}A_{(p)}{}^{b_1\ldots b_p}B_{(q)b_1\ldots b_p a_1\ldots
a_{q-p}}.} 

To obtain these results, first note that Eqs.~\gensusy{}\ can be
further decomposed based on fiber and base chirality.  Using the
assumption that $H_{(3)\a\b m}$ and $\tilde F_{(5)}$ vanish, the
dilatino equation \gensusy{a}\ becomes
\eqna\gendileqs
$$\eqalignno{&\bigl(\half\g^m\partial_m\phi\mp\smallfrac{1}{24}g_s
\tilde F_{(3)mnp}\g^{mnp}\bigr)\hat\chi_\pm = 0,&\gendileqs a\cr
\noalign{\vskip5pt} &
\bigl(\pm\eighth g_s F_{(3)\a mn}\g^{\a mn}\big)\hat\chi_\pm
+ \bigl(\mp\half g_s e^{(\phi-\phi_0)}F_{(1)m}\g^m + \smallfrac{1}{24}
e^{-(\phi-\phi_0)} H_{(3)mnp}\g^{mnp}\bigr)\hat\chi_\mp = 0. &\cr
&&\gendileqs b}$$

Eq.~\gensusy{b} can be similarly decomposed.  In the $M=\m$ equation,
we can take the 4D spacetime spinor $u$ to be constant on $\IR^{3,1}$;
that is, $\partial_\m u = 0$.  (If the theory has 4D local
supersymmetry, then it also has 4D global Poincar\'e supersymmetry).
Then, this equation becomes
\eqna\gengravmu
$$\eqalignno{&\hphantom{-\eighth}\g_\m\bigl(\nabla_{\mkern-2mu
\CB\ssk m} (\phi+\half\log Z)\bigr)\g^m\hat\chi_\pm = 0,
&\gengravmu a\cr
&{-\eighth}\g_\m\Bigl(\smallfrac{1}{3!}e^{-(\phi-\phi_0)}H_{(3)mnp}
\g^{mnp} \mp g_s e^{(\phi-\phi_0)}F_{(1)m}\g^m\Bigr)\hat\chi_\mp = 0.
& \gengravmu b}$$
Here, the leading $\g_\m$ factors can be eliminated by contracting
with $\quarter\g^\m$.  Eq.~\gengravmu{a}\ is due to the spin
connection $(\hat w_\m)_{\n m}$ in the metric \hateqs{c}.  It
reproduces the condition that $g_s$ as defined in Eq.~\gs\ is
constant, and it is equivalent to Eq.~\dblresults{a}, which states
that the prefactor $Z^{-1/2} e^{-(\phi-\phi_0)}$ in the hatted metric
\hateqs{c}\ is unity.

For $M=\a$, Eq.~\gensusy{b} decomposes into $\partial_\a\hat\chi_\pm =
0$ and
\eqn\gengravalpha{-\eighth\Bigl(g^{(\Tfib)}_{\a\b}\CF^\b_{mn}
\mp g_s F_{(3)\a mn}\Bigr)\g^{mn}\hat\chi_\pm = 0} 
as a consequence of Eqs.~\dblresults{a}\ and \gengravmu{b}.

Finally, using Eqs.~\dblresults{a}, \gendileqs{a}, and \gengravmu{b},
the $M=n$ equation \gensusy{b} decomposes into $\nabla_{\mkern-2mu
\CB\ssk n}\hat\chi_\pm = 0$ and
\eqn\gengravn{{-\quarter}\Bigl(g^{(\Tfib)}_{\a\b}\CF^\b_{mn}
\mp g_s F_{(3)nm\a}\Bigl)\g^{m\a}\hat\chi_\pm
-\Bigl(\eighth e^{-(\phi-\phi_0)}H_{(3)nmp}\g^{mp}
\mp\quarter g_s e^{(\phi-\phi_0)}F_{(1)n}\Bigr)\hat\chi_\mp = 0.}

In Sec.~3, we found that the equations of motion and Bianchi
identities implied a set of Hodge duality constraints \BPScond{}.  We
can now reproduce two of these constraints from the supersymmetry
conditions as follows.  For any Hodge dual pair of odd degree forms
on~$\CB$,
\eqn\dualonethree{\o_{(3)} = *_\CB \o_{(1)},
\quad \o_{(1)} = -*_\CB \o_{(3)},}
we have
\eqna\threevsone
$$\eqalignno{\smallfrac{1}{6}\o_{(3)mnp}\g^{mnp}\hat\chi_\pm
&= \mp Z^{-1}\o_{(1)r}\g^r\hat\chi_\pm, &\threevsone a\cr
\half\o_{(3)mnp}\g^{np}\hat\chi_\pm
&= \pm Z^{-1}\o_{(1)r}\g^r{}_n\hat\chi_\pm. &\threevsone b}$$
From Eq.~\dblresults{a} and the first of these identities,
Eqs.~\gendileqs{a} and \gengravmu{b}\ become
\eqna\identone
$$\eqalignno{\bigl(dZ + g_s*_\CB\tilde F^{\bsk0}_{(3)}\bigr)_m
\g^m\hat\chi_\pm &= 0, &\identone a\cr
\bigl(g_s F_{(1)} + *_\CB H_{(3)}\bigr)_m
\g^m\hat\chi_\pm &= 0. &\identone b}$$
Therefore,
\eqna\identonesimp
$$\eqalignno{dZ &= -*_\CB g_s \tilde F^{\bsk0}_{(3)}, &\identonesimp a\cr
g_s F_{(1)} &= -*_\CB H_{(3)}, &\identonesimp b}$$
which are the first two constraints~\BPScond{a,b}.

To reproduce the third constraint and obtain the refining primitivity
condition, we will need to introduce $SU(3)_{L,R}$ structures.  The
three spinor conditions that remain are Eqs.~\gendileqs{b},
\gengravalpha, and \gengravn.  Using Eqs.~\dblresults{a},
\threevsone{a}, and \identonesimp{b}, the first of these spinor
conditions is implied by the other two.  So, Eqs.~\gengravalpha\ and
\gengravn\ are the complete set of equations that remain.  Since the
latter couples $\chi_+$ and $\chi_-$, we cannot in general expect to
find solutions with $\chi_+$ or $\chi_-$ set to zero.  Therefore, we
will write these constraints in terms of the spinor
\eqn\chihat{\hat\chi = e^{(\phi-\phi_0)/4}\chi = \hat\chi_+ +
\hat\chi_-}
(related to $\e_R$ by Eq.~\Repsilon), and the $SU(3)_R$ structures
constructed from this spinor,
\eqna\SUR
$$\eqalignno{J_a{}^b &= i\hat\chi^\dagger\g_a{}^b\hat\chi, &\SUR a\cr
\O_{abc} &= \hat\chi^\dagger\g_{abc}\hat\chi^*. &\SUR b}$$
Alternatively, we could express the constraints in terms of
\eqn\chiLhat{\hat\chi_L = e^{(\phi-\phi_0)/4}\chi_L = -\hat\chi_+ +
\hat\chi_-}
(cf.~Eqs~\Lepsilon\ and \OfivechiL), and the corresponding $SU(3)_L$
structures
\eqna\SUL
$$\eqalignno{J_{L\ssk a}{}^b &= i\hat\chi_L^\dagger\g_a{}^b
\hat\chi_L, &\SUL a\cr
\O_{L\ssk abc} &= \hat\chi_L^\dagger\g_{abc}\hat\chi_L^*.
&\SUL b}$$
However, since $\chi_L$ is related to $\chi$ by Eq.~\OfivechiL, the
two SU(3) structures are also related, so that there is no new
information gained in doing this.\foot{One can easily show that, for
example, $J_{L\ssk mn} = J_{mn}$, $J_{L\ssk \a\b} =
J_{\a\b}$, and $J_{L\ssk m\a} = - J_{m\a}$.}  Note that in the
case that $\chi_+$ or $\chi_-$ vanishes, we have $\chi$ and $\chi_L$
equal up to a sign, and the $SU(3)_R$ and $SU(3)_L$ structures are
identical.  This was the case in Sec.~4.2.
\eqna\remeqs

Using Eqs.~\dblresults{a}, and the identities \threevsone{}, the two
remaining equations become\foot{For $\chi\to\chi_L$, the sign of the
second term in Eq.~\remeqs{b} is reversed.}
$$\eqalignno{&\bigl(*_6 F_{(3)} 
- {1\over g_s}d\Vol_\fib\bigr)_{\a mn}\g^{mn}\hat\chi = 0,
&\remeqs a\cr
&\biggl(\bigl(F_{(3)} + {i\over g_s}d\Vol_\fib\bigr)_{nm\a}\g^{m\a}
+\Bigl(Z^{-1/2}F_{(1)n} - {1\over g_s}Z^{1/2}
\bigl(*_6(\Vol_\fib\w H_3)\bigr)_m\g^m{}_n\Bigr)\biggr)
\hat\chi = 0. &\cr
&&\remeqs b}$$

With the sign conventions
\eqn\Jjk{J_{j\bar k} = (ig_6)_{j\bar k},}
(equivalent to Eq.~\presignconv{b}), the nonzero components of the ACS
are
\eqn\ACSjk{J_j{}^k = i\d_j{}^k,
\quad J_\jbar{}^{\bar k} = -i\d_\jbar{}^{\bar k}.}
Together with Eq.~\SUR{a}, this implies that $\hat\chi$ is annihilated
by $\g_j$ and $\g^{\bar k}$, while $\g_\jbar$ and $\g^k$ act as
creation operators.  The spinors $\hat\chi$, $\g^\ibar\hat\chi^*$ form
a basis for the space of 6D spinors of negative chirality, and
$\hat\chi^*$, $\g^i\hat\chi$ form a basis for positive chirality 6D
spinors.  Using Eq.~\SUR{b} as well, it is then possible to prove the
identities\foot{Similar identities have appeared in numerous places.
See, for example ref.~\Marino.}
\eqna\gchiid
$$\eqalignno{\g^{bc}\hat\chi &= -\half\bar\O^{bc}{}_d\g^d\hat\chi^*
-iJ^{bc}\hat\chi, & \gchiid a\cr
\g^a\g^{bc} &= -\bar\O^{abc}\hat\chi^*
+\bigl(\quarter\bar\O_d{}^{bc}\O^{da}{}_f - iJ^{bc}\d^a_f\bigr)
\g^f\hat\chi. & \gchiid b}$$
The latter implies that for an arbitrary three-form $\o_{(3)}$,
\eqn\ogchi{{1\over3!}\o_{(3)abc}\g^{abc}\hat\chi =
- \bigl(\bar\O\lrcorner\,\o_{(3)}\bigr)\hat\chi^*
-i\bigl(J\lrcorner\,\o_{(3)}\bigr)_i\g^i\hat\chi,}
where the contraction operater $\lrcorner\,$ was defined in
Eq.~\contr.

Now, define $G_{(3)}$ as in Eq.~\cpxfluxII{d}, and define the complex
one-form flux
\eqn\altGone{G_{(1)a} = F_{(1)a} - {i\over
g_s}Z\bigl(*_6(\Vol_\fib\w H_{(3)})\bigr)_b J^b{}_a,}
which is equivalent to Eq.~\Gtildeconds{c}.  Using the Hodge duality
constraint \BPScond{b}, the remaining spinor conditions \remeqs{a,b}
contracted with $\g^\a$ and $\g^n$, respectively, become
\eqna\Gcontreqs
$$\eqalignno{\bar\O\lrcorner\, G_{(3)} = 0,\quad
& \bigl(J\lrcorner\, G_{(3)})^{(1,0)} = 0,\quad
\bigl(G_{(1)}\bigr)^{(1,0)} = 0,& \Gcontreqs a\cr
\bar\O\lrcorner\,\bar G_{(3)} = 0,\quad
& \bigl(-iJ\lrcorner\,\bar G_{(3)}
+ Z^{-1/2}\bar G_{(1)}\bigr)^{(1,0)} = 0,
& \Gcontreqs b\cr}$$
where we have made use of the identity \gchiid{b}. Eq.~\Gtildeconds{a}
implies the Hodge duality condition \BPScond{b}.  The constraints on
$G_{(3)}$ are equivalent to demanding that $\tilde G_{(3)}$ be
primitive and of type $(2,1)+(1,2)$, where $\tilde G_{(3)}$ is given
by Eq.~\Gtildeconds{d}.

Using this result, the original noncontracted constraints \remeqs{}
eliminate the (1,2) primitive piece via identity \gchiid{a}.  Thus,
our final condition on $G_{(3)}$ for supersymmetry becomes
Eq.~\Gtildeconds{b}.  This implies, among other things, that $G_{(3)}$
is imaginary-selfdual (ISD), which in turn reproduces the remaining
Hodge duality condition \BPScond{c}.

To summarize, the complete set of conditions for solutions to the
equations of motion with $\CN\ge1$ supersymmetry consists of the
constancy conditions \constchi, the constraints~\BPScond{a}\ and
\dblresults{a}\ relating the dilaton and $F^{\bsk0}_{(3)}$ to the warp
factor, vanishing five-form flux \noffivetilde, and finally the flux
constraints \Gtildeconds{a,b}.


\newsec{Relation to T-Dual O3 Orientifolds}

The O5 orientifolds just discussed are T-dual to O3 orientifolds on
$T^2\times K3$ for $\CB=K3$ and on $T^6$ for $\CB=T^4$.  For the rest
of the paper, we will restrict to the case that $\CB=T^4$.

\subsec{Review of O3 Orientifolds with Internal $T^6$}

In this subsection and the next, symbols with (without) a prime denote
quantities associated with the O3 (O5) orientifold.  The orientifold
projection for O3 planes has already been discussed in connection with
the decomposition of the 10D supersymmetry parameters $\e'_{L,R}$ in
Sec.~4.1.  The even internal fluxes preserved by the orientifold
projection are $F'_{(3)}$ and $H'_{(3)}$.  Except in the case that O3
and D3 charges cancel locally, the equations of motion also require
odd $F'_{(5)}$ flux.  For compact internal manifold $X'_6$, the most
general $\CN'\ge1$ O3 orientifold background compatible with 4D
Poincar\'e invariance is \GP
\eqna\Othreebg
$$\eqalignno{ds_{\rm string}^2 &= Z'^{-1/2}\eta_{\m\n}dx^\m dx^\n
+ Z'^{1/2} ds^2_6, &\Othreebg a\cr
e^{\phi'} &= g'_s = {\rm const}, &\Othreebg b\cr
-{1\over g'_s}\nabla^2_{X'_6} Z' &= (2\pi)^4\a'^2\sum_i Q'_i
\biggl({\d^6(x-x_i)\over\sqrt{g'_6}}
 - {1\over V'_6}\biggr),
&\Othreebg c\cr
\tilde F'_{(5)} &= (1+*_6) d\bigl(g'^{-1}_s Z'^{-1}
dx^0\w dx^1\w dx^2\w dx^3\bigr), &\Othreebg d\cr}$$
where $X'_6$ is a Calabi-Yau threefold, $K3\times T^2$, or $T^6$, and
$V'_6$ is the volume of $X'_6$.

The Gauss's law constraint is
$$N'_\flux+\sum_i Q'_i = 0,\quad{\rm where}\quad
N'_\flux = {1\over(2\pi)^4\a'^2}\int_{X'_6}F'_{(3)}\w H'_{(3)}.
\eqno\Othreebg e$$
Here, $\{(Q'_i,x_i)\}$ is the set of charges and positions of local O3
and D3 sources, with $Q'_i$ equal to 1 for a D3 brane.  We assume that
all O3 planes are standard O3${}^-$ planes in the terminology of
Ref.~\Bergman, and work on the covering space of the orientifold, so
that there are $M$ D3 branes and $M$ $\IZ_2$ image branes.  The
orientifold operation is $\O(-1)^{F_l}\CI_6$, where $I_6$ inverts
$X'_6$.  For $X'_6 = T^6$, the O3 planes are located at the $2^6$
fixed points on the base where each $x^a$ is equal to 0 or $1/2$, and
each O3 plane has charge $Q'_i=-1/2$.

In the absence of three-form flux, the theory preserves 4D $\CN'=4$
supersymmetry.  The 4D moduli are the zero-modes on $X'_6$ of
\eqn\Dthreemoduli{\tdil' = C'_{(0)} + i/g'_s,\quad g'_{6\,ab},\quad
C'_{(2)ab},\quad {\rm and}\quad C'_{(4)abcd},}
together with the D3 worldvolume scalars $\Phi_I$, $I=1,\ldots,M$.
The massless 4D gauge bosons are the zero-modes of
\eqn\Dthreegb{B'_{(2)\m a}\quad{\rm and}\quad C'_{(2)\m a},}
together with the D3 worldvolume gauge bosons $A_{I\,\m}$.

In the presence of three-form flux, the supersymmetry and massless
field content is reduced \GKP.  For the case that $X'_6=T^6$, this
model has been analyzed in great detail \refs{\KST,\FP}.  The possible
choices of NS and RR three-form flux are $H'_{(3)},F'_{(3)}\in
(2\pi)^2\a' H^3(T^6,2\IZ)$.  That is,
\eqn\OthreeFlux{\eqalign{H'_{(3)}
&= (2\pi)^2\a' m_{[abc]}dx^a\w dx^b\w dx^c,
\quad m_{[abc]}\in 2\IZ,\cr
F'_{(3)} &= (2\pi)^2\a' n_{[abc]}dx^a\w dx^b\w dx^c,
\quad n_{[abc]}\in 2\IZ,\quad a,b,c = 1,\ldots,6.}}

Without imposing any supersymmetry conditions, the equations of motion
alone imply the ISD condition
\eqn\GthreepISD{*_6 G'_{(3)} = i G'_{(3)},}
where
\eqn\Gpdef{G'_{(3)} = F'_{(3)} - \tdil' H'_{(3)}.}
Since $H'_{(3)}$ and $F'_{(3)}$ are discrete, this is a constraint on
$\tdil'$ and metric moduli.

The condition for $\CN' \ge 1$ 4D (Poincar\'e) supersymmetry further
refines this to
\eqn\DthreesusyG{G'_{(3)}\quad\hbox{(2,1) and primitive.}}
The complex structure on $T^6$ can be
parametrized by a complex $3\times3$ period matrix $\t'^i{}_j$,
\eqn\Dthreez{z^i = x^i + \t'^i{}_j y^j,}
where $z^i\cong z^i+1 \cong z^i+\t'^i{}_j$, for $i,j = 1,2,3$.  The
holomorphic three-form is
\eqn\DthreeOmega{\O'\propto dz^1\w dz^2\w dz^3,}
which depends on the complex structure moduli up to cubic order in the
period matrix.  The K\"ahler form is
\eqn\DthreeKahler{J' = ig'_{i\jbar}dz^i\w d\bar z^j.}

The (2,1) supersymmetry condition is easily implemented by varying the
superpotential of Gukov, Vafa, and Witten \GVW
\eqn\WGVW{W_{\rm GVW} = \int_{X'_6} G'_{(3)}\w\O',}
with respect to $\tdil'$ and all complex structure moduli $\t'^i{}_j$,
and in addition imposing the condition $W_{\rm GVW}=0$.  (The latter
follows from $D_\r W_{\rm GVW} = 0$, where $\rho$ is the complexified
overall volume modulus \refs{\GKP,\KST}.)  For generic supersymmetric
vacua, this fixes $\tdil$ and all of the complex structure moduli, but
for nongeneric flux, some of these moduli are left unfixed.  The
primitivity condition then becomes a linear constraint on the
$g'_{i\jbar}$, with coefficients determined by the flux and
axion-dilaton.  One subtlety in this procedure for $X'_6=T^6$ is that
not all of the (18 real) $\t'^i{}_j$ and (9 real) $g'_{i\jbar}$
correspond to the (21 real) physical metric moduli.  See App.~E for
further discussion.  Due to the subtleties discussed in App.~E
concerning $W_{\rm GVW}$ in theories with extended supersymmetry, we
have chosen in this paper to avoid any reference to a superpotential
for the O5 orientifold, and instead to state the supersymmetry
conditions directly in terms of conditions on the fields.

In the O3 orientifold, the decomposition of RR potentials into a
background part $C'^\bg_{(p)}$ and deformation $c'_{(p)}$,
\eqn\RRdecompp{C'_{(p)} = C'^\bg_{(p)} + c'_{(p)},}
is cleaner than it was for the O5 orientifold.  First, set the gauge
fields to zero and focus on deformations corresponding to moduli.
Then,
\eqna\Cdecompp
$$C'_{(0)} = c'_{(0)},\quad C'_{(2)} = C'^\bg_{(2)},\eqno\Cdecompp a$$
and it is only in $C'_{(4)}$ that both contributions appear:
$$C'_{(4)} = C'^\bg_{(4)} + c'_{(4)}.\eqno\Cdecompp b$$
Here, $F'_{(3)}$, $H'_{(3)}$, and $\tilde F'_{(5)}$, as given by
Eqs.~\OthreeFlux\ and \Othreebg{d}, satisfy
\eqn\FbgCbgp{\eqalign{ F'_{(3)} &= dC'^\bg_{(2)}\cr
\tilde F'_{(5)} &= dC'^\bg_{(4)} - C'^\bg_{(2)}\w H'_{(3)}.}}

In addition to the deformations just discussed, there are other
deformations of the supergravity background which when promoted to 4D
fields become gauge bosons.  In the closed string sector, these
deformations are the zero modes on $X'_6$ of $B'_{(2)d\m}$ and
$C'_{(2)d\m}$, and will be denoted by the lowercase symbols
$b'_{(2)d\m}$ and $c'_{(2)d\m}$, respectively.  When the couplings to
gauge bosons are included and all deformations are promoted to 4D
fields, the kinetic term for $c'_{(4)abcd}$ is the square of \FP
\eqn\Ckintermp{\partial_\m c'_{(4)abcf}
+ 4F'_{(3)[abc|} b'_{(2)|f]\m} - 4H'_{(3)[abc|} c'_{(2)|f]\m}.}
If we define complex gauge bosons
\eqn\cpxgaugebos{d'_{a\m} = c'_{(2)a\m} - \tdil' b'_{(2)a\m},\quad
{\rm and}\quad \bar d'_{a\m} = c'_{(2)a\m} - \bar\t_{\rm dil}{}'
b'_{(2)a\m},}
then Eq.~\Ckintermp\ can also be written as
\eqn\cpxCkintermp{\partial_\m c'_{(4)abcf} - {1\over\Im\tdil'}\Im
\Bigl(4\bar G'_{(3)[abc|} d'_{|f]\m}\Bigr).}

\subsec{T-Duality Map}

In this subsection, as in the previous one, symbols with (without) a
prime denote quantities associated with the O3 (O5) orientifold.  Let
$X'_6 = T^6$ and $\CB=T^4$.  Then, the metric on $X'_6$ can be written
as a trivial $T^2$ fibration over $\CB$:
\eqn\dssixprime{ds'^2_6 = ds^2_\CB + g^{(\Tfib)\prime}_{\a\b}
\eta'^a\eta'^\b,\quad \eta'^\a = dx^\a+a'^\a,}
where trivial means that $\CF'^\a = da'^\a = 0$.  The use of a
lowercase $a'^\a$ indicates that the $a'^\a{}_{mn}$ are moduli rather
than a quantized background.  Until the very end of this subsection,
we set all gauge fields to zero.

For the O5 orientifold, we write NS $B$-field as $B_{(2)} =
B^\bg_{(2)} + b_{(2)}$, where, from Sec.~3.9,
\eqn\Bcomponents{\eqalign{B^\bg_{(2)} &= B^{\bg0}_{(2)} = \half
B_{(2)mn} dx^m\w dx^n,\cr 
b_{(2)} &= b^{\bsk1}_{(2)} = \eta^\a\w b_{(2)\a},\quad
{\rm with}\quad b_{(2)\a} = b_{(2)\a m}dx^m.}}
The NS $B$-field for the O3 orientifold is given by $B'_{(2)} =
B'^\bg_{(2)}$, where $B'^\bg_{(2)} = B'^{\bg\,0}_{(2)} +
B'^{\bg\,1}_{(2)}$, and
\eqn\Bpcomponents{\eqalign{B'^{\bg0}_{(2)} &= \half
B'^\bg_{(2)mn} dx^m\w dx^n,\cr 
B'^{\bg\,1}_{(2)} &= \eta'^\a\w B'^\bg_{(2)\a},\quad
{\rm with}\quad B'^\bg_{(2)\a} = B'^\bg_{(2)\a m}dx^m.}}

T-duality is an exact duality between string vacua that relates
conformal field theories order by order in string perturbation theory.
At the level of the conformal field theories, it is has the simple
interpretation as the sign reversal $X^\a_R(\bar z) \to -X^\a_R(\bar
z)$ of the right-moving worldsheet scalars in the duality directions.
However, in the low energy supergravity description, it involves an
intermediate operation known as smearing.  In the O3 orientifold, the
locations of the O3 planes and D3 branes spontaneously break the
translation isometries in each of the internal $T^6$ directions.  The
breaking is spontaneous in the sense that, on the $\IZ_2$ covering
space, the entire O3/D3 system can be translated by an arbitrary
amount along any of the $T^6$ circles, to produce an inequivalent
vacuum with identical relative O3/D3 positions and identical physics.
The space of vacua respects the isometry, but a particular vacuum does
not.  Smearing is the operation of averaging a supergravity background
over all vacua related by such translations in order to restore an
isometry.  To perform the T-duality between the O3 and O5 supergravity
backgrounds, we first smear in the 4 and 5 directions and then perform
the supergravity analog of the conformal field theory T-duality in
these directions.  The information about D3 brane positions lost in
the smearing is recovered in the Wilson lines of the of D5 worldvolume
gauge fields in the 4 and 5 directions.

The effect of the smearing is that the warp factors $Z$ and $Z'$ are
related via
\eqn\Zmap{Z = Z'_{\rm smeared}
= {1\over V'_\fib}\int_{\Tfib'}\Vol'_\fib Z',}
where
\eqn\Volprime{\Vol'_\fib = V'_\fib\eta'^4\w\eta'^5,\quad
{\rm and}\quad V'_\fib = \bigl(g^{(\Tfib)\prime}\bigr)^{1/2}
= (2\pi)^4\a'^2/V_\fib,}
which is easily seen to relate solutions of the 6D Poisson
equation \Othreebg{c}\ to solutions of the 4D Poisson equation
\Zpoisson.

The T-duality action on the dilaton is
\eqn\dilmap{e^\phi(2\pi)^2\a'/\bigl(Z^{-1}V_\fib\bigr) = e^{\phi'},}
or equivalently
\eqn\gsmap{g_s (2\pi)^2\a'/V_\fib = g'_s.}

The T-duality action on the NS $B$-field and metric interchanges the
geometrical $S^1$ fibration of connection $a'^\a$ with the formal
$S^1$ fibrations of connection $B'^\bg_{(2)\a}$.\foot{This point has
been emphasized recently, first in Ref.~\BEM\ (and subsequent work
\BEMmore) and then in Ref.~\MinMirror.}  The relations are \MinMirror
\eqna\BGrels
$$\eqalignno{b_{(2)\a} &= -a'^\a\,(2\pi)^2\a', & \BGrels a\cr
\CA^\a &= -B'^\bg_{(2)\a}/\bigl((2\pi)^2\a'\bigr), & \BGrels b\cr
B^\bg_{(2)} &= B'^{\bg\,0}_{(2)} + a'^\a\w B'^\bg_{(2)\a}, &\BGrels
c}$$ 
together with
$$\eqalignno{g^{(\Tfib)}_{\a\b} & = 
g^{(\Tfib)\prime\,\a\b}\,\bigl((2\pi)^2\a'\bigr)^2,
&\BGrels d\cr
g^{\vphantom{\prime}}_{\CB\msk mn} & = g'_{\CB\msk mn}. &\BGrels e}$$

Note that the T-duality map does not leave the purely base component
$B^{\bsk0}_{(2)}$ invariant.\foot{I am grateful to A.~Tomasiello for
correspondence regarding this point.}  Nevertheless, the correction
term in Eq.~\BGrels{c}\ has a straightforward interpretation.  Let
underscored indices $\ul\m,\ul\a,\ul m$ denote components in the
$dx^\m,dx^\a,dx^m$ basis as opposed to the fibration-adapted bases
involving $\eta^\a$ or $\eta'^\a$.  Then Eq.~\BGrels{c}\ is equivalent
to
\eqn\Bbasesame{B^\bg_{(2)mn} = B'^\bg_{(2)\ul m\ul n}.}

This illustrates an important rule.  Many (but not all) of the
T-duality relations connecting O3 and O5 orientifolds take the
simplest form when expressed in terms of the $dx^\a,\eta^\a,dx^m$
basis for the O5 orientifold and the $dx^\m,dx^\a,dx^m$ basis for the
O3 orientifold.  This is perhaps to be expected, since in the O3
orientifold there is nothing special about the $\a=4,5$
directions---any other pair of directions could have been used to
define the fiber of a flat $T^2$ fibration.  The three-form fluxes of
the O3 orientifold are of the moduli-independent form \OthreeFlux.
Thus, $H'_{(3)\ul\a\ul m\ul n}$ and $H'_{(3)\ul m\ul n\ul r}$ are
quantized and moduli-independent, whereas $H'_{(3)mnr}$ involves a
combination of quantized fluxes and the metric moduli $a'^\a{}_{mn}$.
From the T-duality map \BGrels{}, the relation between quantized NS
sector fluxes is
\eqn\fluxFmap{\CF^\a_{mn} = H'_{(3)\ul\a\ul m\ul n}
/\bigl((2\pi)^2\a'\bigr)\quad{\rm and}\quad
H^\bg_{(3)mnr} = H'_{(3)\ul m\ul n\ul r}.}
The relation between NS sector moduli has already been given
explicitly in Eqs.~\BGrels{a,d,e}.

The rule described in the previous paragraph is particularly
applicable to the T-duality action on RR fluxes and moduli.  The
T-duality relations between RR fluxes are
\refs{\Bergshoeff,\Myers,\Hassan}
\eqn\RRprime{\eqalign{\tilde F^{\bsk0}_{(n)} 
&= \half\e^{\a\b}\tilde F'^{\bsk2}_{(n+2)\a\b}
/\bigl((2\pi)^2\a'\bigr),\cr
\tilde F^{\bsk1}_{(n)\a} &= -\e^{\a\b}\tilde F'^{\bsk1}_{(n)\b},\cr
\tilde F^{\bsk2}_{(n)\a\b} &= -\e^{\a\b}\tilde F'^{\bsk0}_{(n-2)}
\,(2\pi)^2\a'.}}
Here $\e^{45} = - \e^{54} = 1$, and we have assumed that the ordering
of the T-dualities in going from the O3 theory to the O5 theory is
that we first T-dualize in the 4-direction and then in the
5-direction.

In Eq.~\RRprime, both sides involve a combination of moduli-dependent
and quantized contributions.  It is desirable to disentangle the two
types of contributions.  When the RHS is expanded in terms of $a'^\a$,
$c'_{(0)}$, and underscored flux components, the relations become
\eqna\FluxMap
$$\eqalignno{F_{(1)m} &= \bigl(F'_{(3)}
- c'_{(0)}H'_{(3)}\bigr)_{\ul4\ul5\ul m}/(2\pi)^2\a', &\FluxMap a\cr
F_{(3)\b mn} &= -\e^{\b\a}\bigl(F'_{(3)} -
c'_{(0)}H'_{(3)}\bigr)_{\ul\a\ul m\ul n} -
2a'^\a{}_{[\ul m|}\bigl(F'_{(3)} 
- c'_{(0)}H'_{(3)}\bigr)_{\ul4\ul5|\ul n]}, &\FluxMap b\cr
\tilde F_{(5)\,45mnr} &= -\Bigl(\bigl(F'_{(3)} -
c'_{(0)}H'_{(3)}\bigr)_{\ul m\ul n\ul r} - 3a'^\a{}_{[\ul m|}\bigl(
F'_{(3)} - c'_{(0)}H'_{(3)}\bigr)_{\ul\a |\ul n\ul r]} &\cr
&\qquad\quad + 6a'^4{}_{[m|}a'^5{}_{|n|}\bigl(F'_{(3)}
- c'_{(0)}H'_{(3)}\bigr)_{\ul4\ul5|\ul r]}\Bigr)(2\pi)^2\a'.
&\FluxMap c}$$
In addition, the map of odd fluxes is
$$\tilde F_{(3)mnr} =
\bigl(\tilde F'_{(5)\,{\rm smeared}}\bigr)_{45mnr}
/(2\pi)^2\a'. \eqno \FluxMap d$$
Here, the five-form flux in the smeared supergravity backround is
\eqn\Ffivesmeared{\tilde F'_{(5)\,{\rm smeared}} = 
{1\over g'_s} \Vol'_\fib *_\CB dZ'_{\rm smeared}.}
By Eqs.~\Zmap\ and \gsmap, Eq.~\FluxMap{d}\ agrees with
Eq.~\BPScond{b}.  In the remaining relations \FluxMap{a,b,c}, we can
drop the components $H'_{(3)\,\ul4\ul5\ul m}$, which necessarily
vanish for the T-duality map to exist (cf.~Footnotes~\nongeofootI\ and
\nongeofootII).  Upon setting these terms to zero and using the
map~\BGrels{a}, Eqs.~\FluxMap{c,d}\ can be identified, term for term,
with Eqs.~\CFdefs\ and \fluxdeform{a,b}.  The result is that the
T-duality map between quantized fluxes is
\eqna\FluxbgMap
$$\eqalignno{F_{(1)m} &= F'_{(3)\,\ul4\ul5\ul m}/(2\pi)^2\a',
&\FluxbgMap a\cr
F^\bg_{(3)\,4mn} &= -F'_{(3)\,\ul5\ul m\ul n}\quad{\rm and}\quad
F^\bg_{(3)\,5mn} = F'_{(3)\,\ul4\ul m\ul n}, &\FluxbgMap b\cr
F^\bg_{(5)\,45mnr} &= -F'_{(3)\,\ul m\ul n\ul r}\,(2\pi)^2\a',
&\FluxbgMap c}$$
and the axionic partner of the dilaton maps as
$$\tilde c_{(2)\,45} = -(2\pi)^2\a' c'_{(0)}. \eqno\FluxbgMap d$$
From $H'_{(3)\,\ul4\ul5\ul m}=0$ together with Eq.~\GthreepISD, the
flux components $F'_{(3)\ul m\ul n\ul r}$ vanish.  Therefore,
\eqn\SecNineResult{F^\bg_{(5)\,45mnr}=0,} 
which is a result that we stated without proof in Sec.~3.9\null.
Eqs.~\FluxbgMap{d}\ and \gsmap\ together give
\eqn\tdilmap{\tdil = \tdil',}
where $\tdil$ is defined in Eq.~\newtdil\ and $\tdil'$ in
Eq.~\Dthreemoduli.  See App.~F for a discussion of the T-duality map
between RR potentials.

The T-duality map between supersymmetry parameters is \Hassan
\eqn\eeprime{\eqalign{\e_R &= \e'_R,\cr
\e_L &= - \bigl(g^{(\Tfib)}\bigr)^{-1/2} Z^{-1/2}
\G^{(10)}\G'_{[5}\G^{(10)}\G'_{4]}\e'_L = i\G_\fib\e'_L,}}
in agreement with the results of Sec.~4.1\null.  Here, as explained in
Ref.~\Hassan, given a vielbein $e'^A{}_M$ such that $g'_{MN} =
\eta_{AB} e'^A{}_M e'^B{}_N$ in the O3 theory, one obtains two
different vielbeine $(e_{L,R})^A{}_M$ in the dual O5 theory, and
correspondingly two different Dirac matrix representations
$(\G_{L,R})_M = (e_{L,R})^A{}_M\G_A$, for fixed choice of
representation of $\G_A$ satisfying the algebra $\{\G_A,\G_B\} =
2\eta_{AB}$.  The two vielbeine are related by a local Lorentz
transformation.  In writing Eq.~\eeprime, an arbitrary choice has been
made to define
\eqn\GisGR{\G_M = (\G_R)_M = (e_R)^A{}_M\G_A,}
and to re-express $(e_L)^A{}_M$ in terms of $(e_R)^A{}_M$ whereever it
appears.  The net effect of this is to supplement T-duality with a
local Lorentz transformation in the left moving worldsheet sector,
which leads to nontrivial T-duality map of $\e_L$.

Finally, when the deformations of the supergravity background
corresponding to 4D gauge bosons are reintroduced, the T-duality map
between these deformations is
\eqn\gbmap{\eqalign{& v^\a_\m = -b'_{(2)\a\m}/\bigl((2\pi)^2\a'\bigr),
\quad b_{(2)m\m} = b'_{(2)m\m},\cr
& c_{(2)\a\m} = -e^{\a\b}c'_{(2)\b\m},\quad
\tilde c_{(4)\,45m\m} = -c'_{(2)m\m}\,(2\pi)^2\a'.}}


\newsec{$\CN=2$ Examples}

In this section, we present three $\CN=2$ examples in succession.  The
first two examples contain $F_{(3)}$ flux only, while the third
example also contains $F_{(1)}$ and $H_{(3)}$ flux.  The three
examples are T-dual to the same O3 background, given in Sec.~6.4, via
three different choices of the cycles on which to T-dualize.

\subsec{Example 1: $\CN_+=2$, $\CN_-=0$}

Consider the choice of fibration curvature and background flux
\eqna\choiceIi
$$\eqalignno{&\CF^4 = 2n\bigl(dx^6\w dx^8+dx^7\w dx^9\bigr),
\quad\CF^5 = 0, &\choiceIi a\cr
\noalign{\vskip5pt}
&F^{\bg\,1}_{(3)}/\bigl((2\pi)^2\a'\bigr) = -2m\eta^4\w
\bigl(dx^6\w dx^8+dx^7\w dx^9\bigr), &\choiceIi b}
$$
with $F_{(1)}=H_{(3)}=0$.  This choice gives a contribution to the
Gauss's law constraint,
\eqn\NfluxI{N_\flux = 8mn,}
so that Eq.~\Gauss\ becomes
\eqn\GaussI{2M+8mn=32.}
As discussed in Sec.~3.2, $N_\flux$ is nonnegative.  So, $m$ and $n$
have the same sign, and the number of D5 branes is $2M=0$, 8, 16, 24,
or 32, depending on the choice of integers $m$ and $n$.  We assume
that $m,n\ne 0$, so that $2M$ is strictly less than 32.

The complex three-form flux is
\eqn\GIi{{1\over(2\pi)^2\a'}G_{(3)} = 2m\bigl(\eta^4+(n/m)\tdil\eta^5\bigr)
\w\bigl(dx^6\w dx^8+dx^7\w dx^9\bigr).}

If we define an almost complex structure via (cf.~Eq.~\etaz)
\eqn\cpxstrI{\eqalign{\eta^{z^1} &= \eta^4+\t_1\eta^5,\cr
dz^i &= dx^{2i+2}+\t^i{}_j dx^{2j+3},\quad i,j=2,3,}}
then, using the decomposition \OJpm, the type (2,1)
condition on the flux is
\eqn\twooneIi{\t_1 = (n/m)\tdil
\quad{\rm and}\quad\det_{2\times2}\t = -1.}
Note that the first condition implies that only for large fiber
complex structure $\Im\t_1\gg1$ can we simultaneously have $g_s\ll1$
and $V_\fib\gg\a'$.  Eq.~\twooneIi\ is equivalent to demanding that
the two factors appearing in $G_{(3)}$ be of definite Hodge type:
\eqn\alttwooneIi{\eqalign{\eta^4-(n/m)\tdil\eta^5
&\quad\hbox{of type (1,0),}\cr
dx^6\w dx^8+dx^7\w dx^9 &\quad\hbox{of type (1,1).}}}  
Thus, from Eq.~\choiceIi{a}, the complex fibration curvature
$\CF^{z^1} = \CF^4+\t_1\CF^5$ is also of type (1,1).  This provides a
check of our formal result that the complex structure is integrable
in the case of RR three-form flux only.  The vanishing of the (0,2)
component of $\CF^{z^1}$ is all that is required for the fibration to
be holomorphic; that condition is satisfied.

Finally, the primitivity condition is
\eqn\primI{J^\CB\w\bigl(dx^6\w dx^8+dx^7\w dx^9\bigr) = 0.}

Not all of the $\t^i{}_j$ and $g_{\CB\ssk i\jbar}$ left unfixed by
these constraints correspond to independent physical moduli.  (This is
the $T^4$ version of the ambiguity discussed in App.~E.)  We can fix
the redundancy at the cost of breaking manifest $SU(2)$ covariance in
the parametrization of the metric and complex structure on the base
$T^4$.  We write the $T^4$ metric as a flat $T^2_{\{67\}}$ fibration
over $T^2_{\{89\}}$,
\eqn\flatfibIi{ds^2_\CB = {V_2\over|\Im\t_2|}
\bigl|\tilde\eta^6+\t_2\tilde\eta^7\bigr|^2 +
{V_3\over|\Im\t_3|}\bigl|dx^8+\t_3 dx^9\bigr|^2,}
where
\eqn\etatildeIi{\eqalign{\tilde\eta^6 
&= \tilde a^6{}_8 dx^8 + \tilde a^6{}_9 dx^9,\cr
\tilde\eta^7 &= \tilde a^7{}_8 dx^8 + \tilde a^7{}_9 dx^9,}}
with $\tilde\CF^m = d\tilde a^m = 0$, $m=6,7$.  We take the
(1,0)-forms on $X_6$ to be
\eqn\newcpxstrI{\eta^{z^1} = \eta^4+\t_1\eta^5,\quad
\tilde\eta^{z^2} = \tilde\eta^6+\t_2 \tilde\eta^7,
\quad{\rm and}\quad
dz^3 = dx^8+\t_3 dx^9,}
instead of those given in Eq.~\cpxstrI.  Then,
\eqn\newOJBI{\eqalign{\O^\CB &=
\biggl({V_2V_3\over|\Im\t_2||\Im\t_3|}\biggr)^{1/2}
\bigl(\tilde\eta^6+\t_2\tilde\eta^7\bigr)
\w\bigl(dx^8+\t_3 dx^9\bigr),\cr
J^\CB &= (\sign\t_2)V_2 \tilde\eta^6\w\tilde\eta^7
+ (\sign\t_3)V_3 dx^8\w dx^9.}}
The (2,1) condition becomes
\eqna\newtwooneIi
$$\eqalignno{\t_1 &= (n/m)\tdil, &\newtwooneIi a\cr
\t_2\t_3 &= -1, &\newtwooneIi b}$$
and the primitivity condition becomes
\eqn\newprimI{\tilde a^6{}_9 = \tilde a^7{}_8.}

The constraints on NS $B$-field moduli follow from
Eqs.~\FthreeBconstr{a,b}.  These equations become
\eqna\NSBIi
$$0 = h_{(3)} = b_{(2)\,4m}\CF^4\w dx^m,\eqno\NSBIi a$$
and
$$0 = -b_{(2)\,5m} dx^m\w F^\bg_{(3)\,4},\eqno\NSBIi b$$
respectively.  The first equation eliminates $b_{(2)\,4m}$, and the second
$b_{(2)\,5m}$.  So, there are no NS $B$-field moduli.

The massless RR moduli are those that do not couple to the gauge
fields.  (Those that do are eaten via the supersymmetric Higgs
mechanism). From the kinetic terms \gaugecouplings{}, they are
\eqn\RRmoduliIi{\tilde c_{(2)\,89},\ \tilde c_{(2)\,78},
\ \tilde c_{(2)\,69},\ \tilde c_{(2)\,67},\ \tilde c_{(2)\,68}
- \tilde c_{(2)\,79},\quad{\rm and}\quad
\tilde c_{(2)\,\m\n}\leftrightarrow\tilde c_{(6)\,456789}.}
Similarly, the massless gauge bosons are those that do not couple to
RR scalars.  From \gaugecouplings{}, they are the three linear
combinations of $v^4{}_\m$, $v^5{}_\m$, $c_{(2)\,4\m}$, and
$c_{(2)\,5\m}$ orthogonal to
\eqn\gaugebosIi{(2\pi)^2\a' (m/n) v^4{}_\m + c_{(2)\,4\m},}
where orthogonality is defined by the metric appearing in the gauge
kinetic terms.  (See Sec.~6.4 for further discussion in the context of
the dual O3 orientifold).

Finally, there are $6M$ massless scalars $A_{I\bsk\a}$, $\Phi_I{}^m$
and $M$ massless gauge bosons $A_{I\bsk\m}$ from the D5 branes.

In summary, the massless bosonic fields are one graviton, $3+M$
vectors, and $16+6M$ moduli.  The moduli consist of
$$\displaylines{\hbox{$V_\fib$, $V_2$, $V_3$, 2 indep $\t$,
3 indep $\tilde a^m{}_n$},\cr
\hbox{6 $\tilde c_{(2)}$ scalars, and $6M$ D5 scalars.}}$$
These fields combine to form one 4D $\CN=2$ gravity multiplet, $2+M$
vector multiplets, and $3+M$ hypermultiplets.

The amount of unbroken supersymmetry is
\eqn\NsusyIi{\CN_+ = 2,\quad\CN_-=0,}
in the notation of Sec.~4.2.  To verify this, first note that since
$V_\fib,g_s>0$, we have $\Im\tdil>0$.  Then, from the moduli
constraints, $\Im\t_1>0$. So, $J^\fib = +\Vol_\fib$, and we conclude
that $\chi=\chi_+$ and $\CN_-=0$.  Next, observe that there is exactly
one antiholomorphic involution of $X_6$ compatible with the
constraints on complex structure moduli:
\eqn\involIi{\t_1,\t^i{}_j\to\t_1,\bar\t^i{}_j}
(or equivalently, $\t_1,\t_2,\t_3\to\t_1,\bar\t_2,\bar\t_3$ in the
complex structure \newcpxstrI).  Therefore, there are two independent
6D Killing spinors.  The spinors are of the form
\eqn\KillingspinorIi{\chi_{1+} = \z^\fib_+\otimes\z^\CB_+,\quad
\chi_{2+} = \z^\fib_+\otimes{\z^\CB_+}^*,}
for some $\z^\fib_+$ and $\z^\CB_+$, and are related by complex
conjugation of the base.

\subsec{Example 2: $\CN_+ = \CN_- = 1$}

Choose fibration curvature and background flux
\eqna\choiceIii
$$\eqalignno{&\CF^4 = 2ndx^6\w dx^8,\quad \CF^5 = 2ndx^7\w dx^8,
&\choiceIii a\cr
\noalign{\vskip5pt}
&F^{\bg\,1}_{(3)}/\bigl((2\pi)^2\a'\bigr) = -2m\bigl(\eta^4\w dx^7
-\eta^5\w dx^6\bigr)\w dx^9. &\choiceIii b}$$
Again, $m$, $n$, and $M$ are constrained by Eq.~\GaussI.  The
complex three-form flux is
\eqn\GIii{{1\over(2\pi)^2\a'}G_{(3)} = 2n\tdil
\bigl(\eta^4\w dx^7 - \eta^5\w dx^6\bigr)\w
\bigl(dx^8+(m/n)(-1/\tdil)dx^9\bigr).}
Define the complex structure as in Eq.~\cpxstrI.  Then, the type (2,1)
condition gives
\eqna\twooneIii
$$\eqalignno{\t_{2\times2} &= \diag(\t_2,\t_3), &\twooneIii a\cr
\t_1 &= \t_2, &\twooneIii b\cr
\t_3 &= (m/n)(-1/\tdil). &\twooneIii c}$$
So, the base factorizes into $T^2_{\{x^6,x^7\}}\times
T^2_{\{x^8,x^9\}}$ with respect to complex structure.  Primitivity
implies that $g_{\CB\msk i\jbar}$ factorizes in the same way.
Therefore, the base metric takes the form
\eqn\dsBIii{ds^2_\CB = {V_2\over|\Im\t_2|}\bigl|dz^2\bigr|^2
+ {V_3\over\Im\t_3}\bigl|dz^3\bigr|^2.}
Here, we have removed the absolute value bars from $\Im\t_3$ since
the nonnegativity of $\Im\tdil$ and $mn$ implies nonnegativity of
$\Im(-1/\t_3)$ and $\Im\t_3$.

From Eqs.~\FthreeBconstr{a,b}, the constraints on NS $B$-field moduli
are
\eqna\NSBIii
$$\eqalignno{0 &= h_{(3)} = \bigl(b_{(2)\,4m}\CF^4
+ b_{(2)\,5m}\CF^5\bigr)\w dx^m,&\cr
& = -2n\bigl(b_{(2)\,4m}dx^6 + b_{(2)\,5m}dx^7\bigr)\w dx^8\w dx^m
& \NSBIii a}$$
and
$$\eqalignno{0 &= b_{(2)\,4m} dx^m\w F^\bg_{(3)\,5}
- b_{(2)\,5m} dx^m\w F^\bg_{(3)\,4},\cr
& = -2m dx^m\w\bigl(b_{(2)\,4m}dx^6 + b_{(2)\,5m}dx^7\bigr)
\w dx^9. & \NSBIii b}$$
These equations are equivalent to
\eqna\NSBIiiequiv
$$\eqalignno{b_{(2)\,49} = b_{(2)\,59} = 0,
&\quad b_{(2)\,56} = b_{(2)\,47}, &\NSBIiiequiv a\cr
b_{(2)\,48} = b_{(2)\,58} = 0,
&\quad b_{(2)\,56} = b_{(2)\,47}, &\NSBIiiequiv b}$$
respectively.  So, there are three unconstrained NS $B$-field moduli,
\eqn\NSBmoduliIii{b_{(2)\,46},\quad b_{(2)\,57},\quad
{\rm and}\quad b_{(2)\,56}+b_{(2)\,47}.}

In the RR sector, the uneaten moduli that follow from the kinetic
terms \gaugecouplings{}\ are the uncharged scalars
\eqn\RRmoduliIii{\tilde c_{(2)\,67},\ \tilde c_{(2)\,89},
\ \tilde c_{(4)\,4689},\ \tilde c_{(4)\,5789},\ \tilde c_{(4)\,4789} 
+ \tilde c_{(4)\,5689},\quad{\rm and}\quad
\tilde c_{(2)\m\n}\leftrightarrow\tilde c_{(6)\,456789}.}
The massless gauge bosons are those that do not couple to RR scalars.
From \gaugecouplings{}, they are the three linear combinations of
$b_{(2)\,8\m}$, $b_{(2)\,9\m}$, $\tilde c_{(4)\,458\m}$, and
$\tilde c_{(4)\,459\m}$ orthogonal to
\eqn\gaugebosIii{(2\pi)^2\a' (m/n) b_{(2)\,8\m} - \tilde
c_{(4)\,459\m},}
where orthogonality is defined by the metric appearing in the gauge
kinetic terms.  (See Sec.~6.4 for further discussion in the context of
the dual O3 orientifold).

Finally, there are $6M$ massless scalars $A_{I\bsk\a}$, $\Phi_I{}^m$
and $M$ massless gauge bosons $A_{I\bsk\m}$ from the D5 branes.

In summary, as in Ex.~1, the massless bosonic fields are one graviton,
$3+M$ vectors, and $16+6M$ moduli.  However, the moduli now consist of
$$\displaylines{\hbox{$V_\fib$, $V_2$, $V_3$, 2 indep $\t$, 3 indep
$b_{(2)\a m}$,}\cr
\hbox{3 $\tilde c_{(2)}$ scalars, 3 $\tilde c_{(4)}$ scalars, 
and $6M$ D5 scalars.}}$$
These fields combine to form one 4D $\CN=2$ gravity multiplet, $2+M$
vector multiplets, and $3+M$ hypermultiplets.

The amount of unbroken supersymmetry is
\eqn\NsusyIii{\CN_+ = 1,\quad\CN_-=1,}
in the notation of Sec.~4.2.  Due to the factorization $\CB =
T^2_{\{z^2\}}\times T^2_{\{z^3\}}$ and the earlier observation that
$\Im\t_3>0$, we can write
\eqn\zetaBIii{\chi_\pm = \z^\fib_\pm\otimes\z^\CB_\pm,\quad 
{\rm where}\quad\z^\CB_\pm=\z^{(2)}_\pm\otimes\z^{(3)},}
for some $\z^\fib_\pm$, $\z^{(2)}_\pm$ and
$\z^{(3)}_{\vphantom{\pm}}$.  Here, $\z^{(i)}$ is a spinor on
$T^2_{\{z^i\}}$, and the chirality of $\z^{(3)}$ is fixed by the
condition $\Im \t_3>0$.  The two 6D Killing spinors $\chi_+$ and
$\chi_-$ are related by complex conjugation of $T^2_\fib$ and
$T^2_{\{z^2\}}$,
\eqn\involIii{\t_1,\t_2,\t_3\to \bar\t_1,\bar\t_2,\t_3,}
that is,
\eqn\KillingspinorIii{\z^\fib_+ = {\z^\fib_-}^*,\quad
\z^{(2)}_+ =  {\z^{(2)}_-}^*.}

\subsec{Example 3: More General Flux}

Choose fibration curvature and background flux
\eqna\choiceIiii
$$\eqalignno{\CF^5 &= -2ndx^6\w dx^9 &\choiceIiii a\cr
F_{(1)} &= 2m dx^6 &\choiceIiii b\cr
F^{\bg\,1}_{(3)}/\bigl((2\pi)^2\a'\bigr) 
&= -2m\eta^5\w dx^7\w dx^8 &\choiceIiii c\cr
H^\bg_{(3)}/\bigl((2\pi)^2\a'\bigr)
&= 2n dx^7\w dx^8\w dx^9 &\choiceIiii d}$$
Again, $m$, $n$, and $M$ are constrained by Eq.~\GaussI.  The flux
$G_{(3)}$ is
\eqn\GIiii{{1\over(2\pi)^2\a'}G_{(3)} = -2m\eta^5\w dx^7\w dx^8
-2n\tdil dx^4\w dx^6\w dx^9 + b_{(2)}\w 2m dx^6.}
From the moduli constraints, it is possible to show that $b_{(2)}$,
$J$, and the complex structure all decompose as $(49)\oplus(5678)$.
We omit the proof here, since it is tedious and uninstructive;
however, this factorization is dual to an analogous factorization for
the O3 orientifold discussed in the next subsection.  Due to the
factorization, we have
\eqn\Iiiimetric{ds^2_{X_6} = ds^2_{5678} + ds^2_{49},}
where the most general metric in the 5678 directions can be written as
\eqna\fiveeightm
$$ds^2_{5678} = {V_1\over|\Im\t_1|}
\bigl|\eta^5+\t_1Z^{1/2}\tilde\eta^8\bigr|^2
+Z{V_2\over|\Im\t_2|}\bigl|dx^6+\t_2 dx^7\bigr|^2,
\eqno\fiveeightm a$$
with
$$\tilde\eta^8 = dx^8+\tilde a^8{}_m dx^m,\quad m=6,7,
\eqno\fiveeightm c$$
parametrizing a flat $S^1$ fibration, and the most general metric in
the 49 directions can be written
\eqn\fourninem{ds^2_{49} = {V_3\over|\Im\t_3|}
\bigl|dx^4+\t_3Z^{1/2} dx^9\bigr|^2.}
Here $\t_1$ and $\t_3$ are pure imaginary, but $\t_2$ can have both
real and imaginary parts.

In this parametrization, the (3,0) form and K\"ahler form are
\eqna\IiiiJO
$$\O = \biggl({V_1V_2V_3\over
|\Im\t_1||\Im\t_2||\Im\t_3|}\biggr)^{1/2}
\bigl(\eta^5+\t_1Z^{1/2}\tilde\eta^8\bigr)\w
Z^{1/2}\bigl(dx^6+\t_2 dx^7\bigr)\w
\bigl(dx^4+\t_3 Z^{1/2}dx^9\bigr),
\eqno\IiiiJO a$$
and
$$J = Z^{1/2}v_1\eta^5\w\tilde\eta^8 + Z v_2 dx^6\w dx^7
+ Z^{1/2}v_3 dx^4\w dx^9, \eqno\IiiiJO b$$
where the lowercase $v_i$ are signed volumes,
$$v_i = V_i \sign(\Im\t_i),\quad i = 1,2,3\quad\hbox{(no sum).}
\eqno\IiiiJO c$$
By expressing $H_{(3)}$ in terms of
$\eta^5,\tilde\eta^8,dx^6,dx^7,dx^4,dx^9$, and using the factorization
of $b_{(2)}$, we find that
\eqnn\JJVH
$$\eqalignno{&\half(Z J\w J)\lrcorner\,(\Vol_\fib\w H_{(3)})
/\bigl((2\pi)^2\a'\bigr)&\cr
&\quad= 2n{v_\fib\over v_1 v_3}\Biggl(
dx^7+{b_{(2)\,58}\over(2\pi)^2\a'}dx^6
- Z^{-1/2}{v_1\over v_2}\biggl({b_{(2)\,57}\over(2\pi)^2\a'}
- {b_{(2)\,58}\over(2\pi)^2\a'}\tilde a^8{}_7
+ \tilde a^8{}_6\biggr)\eta^5\Biggr). &\cr
&&\JJVH}$$
From Eq.~\Gtildeconds{c}, the complex one-form flux is then
\eqnn\GoneIiii
$$\eqalignno{G_{(1)} &= -2ni{(2\pi)^2\a' v_\fib\over v_1 v_3 g_s}
\Biggl(dx^7+{1\over(2\pi)^2\a'}\biggl(b_{(2)\,58}
+i{m\over n}{v_1v_3 g_s\over v_\fib}\biggr)dx^6 &\cr
&\quad - Z^{-1/2}{v_1\over v_2}
\biggl({b_{(2)\,57}\over(2\pi)^2\a'}
-{b_{(2)\,58}\over(2\pi)^2\a'}\tilde a^8{}_7
+\tilde a^8{}_6\biggr)\eta^5\Biggr). &\GoneIiii}$$
The supersymmetry condition that $G_{(1)}$ be of type (0,1) implies that
\eqna\GoneResultIiii
$$\eqalignno{{1\over\t_2} &= {1\over(2\pi)^2\a'}
\biggl(b_{(2)\,58} - i{m\over n}{v_1 v_3 g_s\over v_\fib}\biggr),
&\GoneResultIiii a\cr
\tilde a^8{}_6 &= -{1\over(2\pi)^2\a'}\bigl(b_{(2)\,57}
- b_{(2)\,58}\tilde a^8{}_7\bigr). &\cr}$$
The second equation is equivalent to
$$\tilde a^8{}_6 = -b_{(2)5\tilde7}/\bigl((2\pi)^2\a'\bigr),
\eqno\GoneResultIiii b$$
where the indices $\tilde6,\tilde7,\tilde8$ refer to components in the
$dx^6,dx^7,\tilde\eta^8$ basis (cf.~Eq.~\fiveeightm{c}).

When these constraints are satisfied,
\eqn\Goneonshell{G_{(1)} = -{2ni\over\bar\t_2} {(2\pi)^2\a'v_\fib\over
v_1 v_3 g_s} d\bar z^2,\quad{\rm where}\quad
dz^2 = dx^6+\t_2 dx^7.}

This takes care of the condition on $G_{(1)}$.  The remaining
supersymmetry condition \Gtildeconds{b}\ on $\tilde G_{(3)}$ is most
easily imposed by first demanding the weaker condition
\eqn\weakerIiii{Z^{1/2}J\lrcorner\,G_{(3)} = iG_{(1)}}
(cf.~Eq.~\Gcontreqs{}).
From the above expressions for $J$ and $G_{(3)}$, we obtain
\eqn\JGIiii{Z^{1/2}J\lrcorner\,G_{(3)} = 2m{(2\pi)^2\a'\over
v_1} \Biggl(dx^7+\biggl({v_1\over v_3}\biggl({n\over m}
\t_{\rm dil} + {b_{(2)\,49}\over(2\pi)^2\a'}\biggr) +
{b_{(2)\,58}\over(2\pi)^2\a'} \biggr)dx^6\Biggr).}
Identifying this with $iG_{(1)}$ gives the constraint
\eqna\weakersimpIiii
$$(n/m)\tdil = (-b_{(2)\,49}+ iv_3)/\bigl((2\pi)^2\a'\bigr).
\eqno\weakersimpIiii a$$
This constraint allows us to simplify Eq.~\GoneResultIiii{a} to
$$-1/\t_2 = (-b_{(2)\,58}+iv_1)/\bigl((2\pi)^2\a').
\eqno\weakersimpIiii b$$
The tilded three-form flux is then
\eqn\tildedtff{\tilde G_{(3)} = \bigl((2\pi)^2\a' m/(v_1\t_2)\bigr)
\bigl(v_1\eta^5\w\tilde\eta^8 - v_3 dx^4\w dx^9\bigr)\w
(dx^6+\t_2 dx^7).}
The first factor is (1,1) and primitive, and the second is $dz^2$.  So,
$\tilde G_{(3)}$ is (2,1) and primitive, and no further constraints
arise from condition \Gtildeconds{b}.

From the kinetic terms \gaugecouplings{}, the massless RR scalars are
the uncharged fields
\eqn\RRmoduliIiii{\tilde c_{(2)\,\tilde79},\quad \tilde c_{(2)\,89},
\quad\tilde c_{(4)\,5789}+(2\pi)^2\a'\tilde c_{(2)\,\tilde69},\quad
\tilde c_{(4)\,5\tilde6\tilde79},\quad \tilde c_{(4)\,5689},\quad
{\rm and}\quad \tilde c_{(4)4678}.}
The massless gauge bosons are those that do not couple to RR scalars.
From \gaugecouplings{}, they are the three linear combinations of
$v^4{}_\m$, $b_{(2)\,9\m}$, $c_{(2)\,5\m}$, and $\tilde
c_{(4)\,459\m}$ orthogonal to
\eqn\gaugebosIiii{(m/n) b_{(2)\,9\m} - c_{(2)\,5\m},}
where orthogonality is defined by the metric appearing in the gauge
kinetic terms.  (See Sec.~6.4 for further discussion in the context of
the dual O3 orientifold).

Finally, there are $6M$ massless scalars $A_{I\bsk\a}$, $\Phi_I{}^m$
and $M$ massless gauge bosons $A_{I\bsk\m}$ from the D5 branes.

In summary, as in the previous two examples, the massless bosonic
fields are one graviton, $3+M$ vectors, and $16+6M$ moduli.  However,
in this case the moduli can be taken to be the independent fields
$$\displaylines{\hbox{$\t_1$, $\t_2$, $\t_3$, $v_2$, $\tdil$,
$\tilde a^8{}_6$, $\tilde a^8{}_7$, $b_{(2)\,5\tilde6}$,}\cr
\hbox{6 $\tilde c_{(p)}$ scalars, and $6M$ D5 scalars,}}$$
with $\t_1$ and $\t_3$ pure imaginary.  These fields combine to form
one 4D $\CN=2$ gravity multiplet, $2+M$ vector multiplets, and $3+M$
hypermultiplets.

The amount of unbroken supersymmetry is
\eqn\NsusyIii{\CN=2,}
due to the existence of an antiholomorphic involution on $\t_2$
compatible with the moduli constraints,
\eqn\involIiii{\t_2,\rho_{1\bar1}\to\bar\t_2,(\rho_{1\bar1})^*,
\quad{\rm where}\quad \rho_{1\bar1} 
= (-b_{(2)\,58}+iv_1)/\bigl((2\pi)^2\a'\bigr).}
In contrast to the previous two examples, there is no decomposition of
the $\CN=2$ algebra into $\CN_+$ and $\CN_-$ algebras generated by 6D
spinors of definite fiber chirality.  For any $\CN=1$ subalgebra of
the $\CN=2$, the corresponding 6D spinor $\hat\chi$ contains
components of both positive and negative fiber chirality (cf.~Sec.~4.4).

\nonnumberedsubsec{Nonintegrability of the Almost Complex Structure}

In the example just discussed, the ACS selected by the supersymmetry
conditions is nonintegrable.  The simplest way to verify this is via
the torsion classes $\CW_i$ of Eq.~\JOW.  This is the one place in the
paper in which we will find it useful to compute any of the $\CW_i$.
The criterion that we will use is \refs{\Gray,\Louis,\CardTor}:
$$\hbox{The ACS is integrable if and only if }\CW_1 = \CW_2 = 0.$$

In the example, the ACS is such that the three (1,0)-forms of $X_6$
are the three factors appearing in $\O$ of Eq.~\IiiiJO{a}.
Consequently, the three terms in $J$ of Eq.~\IiiiJO{b} are each of
type (1,1), and we have the following Hodge decomposition of $dJ$:
\eqna\dJdecompIiii
$$\eqalignno{dJ_{(2,1)+(1,2)} 
&= dZ\w \bigl(\half Z^{-1/2} v_1\eta^5\w\tilde\eta^8 
+ v_2 dx^6\w dx^7+\half Z^{-1/2}v_3 dx^4\w dx^9\bigr),\quad
&\dJdecompIiii a\cr
dJ_{(3,0)+(0,3)} &= 2nZ^{1/2} v_1 dx^6\w
dx^9\w\tilde\eta^8.\quad &\dJdecompIiii b}$$
From the second equation, $\CW_1\ne0$.  It is straightfoward to
compute
\eqn\Wnonzero{\CW_1 = -{n\t_2\over3Z}\biggl({v_1/(v_2v_3)\over
\Im\t_1\Im\t_2\Im\t_3}\biggr)^{1/2}.}
Therefore, the ACS is nonintegrable.

Note, however, that there is no topological obstruction to definining
an integrable complex structure.  For example, the fibration
\choiceIiii{b} is compatible with the complex structure
\eqn\intcpxstrIiii{w^1 = x^4+ix^5,\quad  w^2 = x^6+ix^9,\quad{\rm and}
\quad w^3 = x^7+ix^8.}
In this complex structure, $\CF^{w^1} = -2n dw^2\w d\bar w^2$, with no
(0,2) component, so that the fibration is indeed holomorphic.  This
can also be seen from the fact that $X_6$ is just a warped version of
one of the complex nilmanifolds classified in Ref.~\Salamon.  On the
other hand, the complex structure \intcpxstrIiii\ is not the one
selected by the supersymmetry conditions, and is incompatible with the
physical metric \Iiiimetric\ except at certain points in moduli space.

\subsec{Dual O3 Orientifold}

Consider the O3 orientifold with internal $T^6$ and flux \KST
\eqna\choiceID
$$\eqalignno{F'_{(3)}/\bigl((2\pi)^2\a'\bigr) & = 2m
\bigl(dx^4\w dx^6 + dx^5\w dx^7\bigr)\w dx^9
&\choiceID a\cr
H'_{(3)}/\bigr((2\pi)^2\a'\bigr)  & = 2n 
\bigl(dx^4\w dx^6 + dx^5\w dx^7\bigr)\w dx^8.
&\choiceID b}$$
The Gauss's law constraint \Othreebg{e} is Eq.~\GaussI.  The
complex flux $G'_{(3)}$ is
\eqn\GID{{1\over (2\pi)^2\a'} G'_{(3)} = -2n\tdil'
\bigl(dx^4\w dx^6 + dx^5\w dx^7\bigr)\w
\bigl(dx^8-(m/n)(1/\tdil') dx^9\bigr).}

If we parametrize the complex structure as in Eq.~\Dthreez, then the
supersymmetry conditions imply a factorization $T^6\to
T^4_{\{4567\}}\times T^2_{\{89\}}$ with respect to both complex and
K\"ahler structure.  The condition that $G'_{(3)}$ be of type (2,1)
implies
\eqn\twooneID{\t' =\t'_{T^4}\oplus\t'_{T^2},\quad{\rm with}\quad
\det\t'_{T^4} = -1,\quad\tdil'\t'_{T^2} = -m/n,}
where $\t'_{T^4}$ is a complex $2\times2$ matrix and $\t'_{T^2}$ is a
complex number.  The condition that $G'_{(3)}$ be primitive then gives
$$J' = J'_{T^4} + J'_{T^2}.$$

Working in terms of $\t'^i{}_j$ and $g'_{i\jbar}$ has been convenient
thus far for deriving the $T^6\to T^4\times T^2$ factorization, but is
bad for describing moduli.  As discussed in Sec.~5.1 and in App.~E,
the parametrization is redundent.  For this reason, we will adopt a
different parametrization shortly.  First, however, note that the
supersymmetry constraints on the $T^4$ factor can be phrased in a
parametrization independent way as follows:
\eqn\paramindep{dx^4\w dx^6 + dx^5\w dx^7\quad
\hbox{(1,1) and primitive on $T^4$.}}
Starting from this condition, it is not hard to show that there is
${\cal N}=2$ supersymmetry.  For example, at $g'^{(T^4)}_{mn} =
\delta_{mn}$, there is an $S^2$ of complex structures such that the
condition is satisfied.  The corresponding K\"ahler forms and
holomorphic (2,0) forms are
\eqn\sphereJO{J'_{T^4} = n^A J^{(A)},\quad \O'_{T^4} = n^A\O^{(A)},}
where ${\bf n} = (n^1,n^2,n^3)$ is a unit vector on $S^2$, and
\eqnn\HKstructure
$$J^{(1)}= dx^4\w dx^5 + dx^6\w dx^7,\quad
J^{(2)} = dx^4\w dx^7 + dx^5\w dx^6,\quad
J^{(3)} = dx^4\w dx^6 - dx^5\w dx^7,$$
$$\O^{(1)} = J^{(2)}+iJ^{(3)},\quad\O^{(2)},
\O^{(3)} = \hbox{cyc.\ perms.} \eqno\HKstructure$$
The $S^2$ of complex structures defines a single hyperK\"ahler
structure, and corresponds to two 6D negative chirality Weyl spinors
$\hat\chi'_{1,2}$ via
\eqn\PauliSandwich{J'^{(A)}{}_a{}^b 
= i\hat\chi'^\dagger\s^{(A)}\g_a{}^b\hat\chi,\quad
\hat\chi' = \pmatrix{\hat\chi'_1\cr\hat\chi'_2}.}
Here, the primes indicates that we are discussing the O3 orientifold,
and the hats indicate that $\chi'$ is a normalized spinor, rescaled
relative to the $\chi$ appearing in Eq.~\OthreeBecker.  One can also
see the ${\cal N}=2$ supersymmetry from the fact that there exists
exactly one antiholomorphic involution of $X'_6$ that preserves the
moduli constraints on $\t'^i{}_j$ and $\tdil'$,
\eqn\involID{\t'_{T^4},\t'_{T^2} \to \bar\t'_{T^4},\t'_{T^2}.}
This involution implies that for every Killing spinor such that
$\hat\chi'_1=\chi'_{T^4}\otimes\chi'_{T^2}$, there is an independent
Killing spinor such that $\chi'_2={\chi'_{T^4}}^*\otimes\chi'_{T^2}$.
The condition $\Im\t'_{T^2} = (m/n)\Im(-1/\tdil')>0$, fixes the
chirality of $\chi'_{T^2}$ to be positive.  Since $\chi'$ has negative
chirality by definition, this means that $\chi'_{T^4}$ also has
negative chirality.  On $T^4$, there are exactly two negative
chirality spinors, complex conjugate to one another up to an overall
phase that can be absorbed into the definition of the spinors.
Therefore, once we know that this example preserves any supersymmetry,
we know that it preserves $\CN=2$.

A nonredundent parametrization of the metric moduli is obtained by
writing the $T^4$ as a flat $T^2_{\{45\}}$ fibration over
$T^2_{\{67\}}$:
\eqna\flatfibID
$$\eqalignno{ds^2_{T^4} & = {V'_1\over\Im\t'_1}
|\eta'^4+\t'_1\eta'^5|^2
+{V'_2\over\Im\t'_2} |dx^6+\t'_2 dx^7|^2,
&\flatfibID a\cr
ds^2_{T^2} & = {V'_3\over\Im\t'_3}
|dx^8+\t'_3 dx^9|^2,
&\flatfibID b}$$
where
\eqna\apfib
$$\eqalignno{\eta'^4 &= dx^4 + a'^4{}_6 dx^6 + a'^4{}_7 dx^7,
&\apfib a\cr
\eta'^5 &= dx^5 + a'^5{}_6 dx^6 + a'^5{}_7 dx^7, &\apfib b}$$
with $a'^m{}_n$ constant on $T^6$.  That is, $a'^m$ defines a trivial
fibration $\CF'^m =da'^m=0$ over the base $T^2_{\{6,7\}}$.  The
natural holomorphic one-forms associated with this parametrization are
\eqn\newcpxstrID{\eta'^{z^1} = \eta'^4+\t'_1\eta'^5,\quad
dz^2 = dx^6+\t'_2 dx^7,\quad{\rm and}\quad
dz^3 = dx^8+\t'_3 dx^9.}

Therefore, the (2,1) condition on $G_{(3)}$ becomes
\eqna\newtwooneID
$$\eqalignno{\t'_1\t'_2 &=-1, &\newtwooneID a\cr
\t'_3\tdil' &= -m/n, &\newtwooneID b}$$
and the primitivity condition becomes
\eqn\apconstraint{a'^4{}_7 = a'^5{}_6.}

From the kinetic terms \Ckintermp, the massless $c_{(4)abcd}$ scalars
are the uncharged scalars
\eqn\RRmoduliID{c'_{(4)\,6789},\quad c'_{(4)\,5689},\quad
c'_{(4)\,4789},\quad c'_{(4)\,4589},\quad c'_{(4)\,4689} - c'_{(4)\,5789},
\quad{\rm and}\quad c'_{(4)\,4567}.}
The massless gauge bosons are those that do not couple to RR scalars.
From \Ckintermp, they are the three linear combinations of
$b'_{(2)\,8\m}$, $b'_{(2)\,9\m}$, $c'_{(2)\,8\m}$, and $c'_{(2)\,9\m}$
orthogonal to
\eqn\gaugebosID{(m/n) b'_{(2)\,8\m} + c'_{(2)\,9\m}.}
Here, orthogonality is defined by the metric appearing in the gauge
kinetic terms.   In term of the complex gauge bosons \cpxgaugebos, the
metric on the 89 subspace is proportional to
\eqnn\gaugemetricID
$$\ha\pmatrix{0&1\cr1&0}\matrix{{\scriptstyle d'}\cr
{\scriptstyle\bar d'}}\quad\otimes\quad\ha\pmatrix{0&1\cr1&0}
\matrix{{\scriptstyle z^3}\cr{\scriptstyle\bar z^3}}.
\eqno\gaugemetricID$$
The massive linear combination \gaugebosID\ is proportional to
$\Re\bigl((1/\bar\t')d'_{z^3\m}\bigr)$.\foot{Note that from $dz^3 =
dx^8 + \t' dx^9$, we have $d'_{z^3\m} = (d'_{9\m} -
\bar\t' d_{8\m})/(\t'-\bar\t')$.}  Therefore, the 3D space
of massless gauge bosons is spanned by
\eqn\masslessgbID{\Im\bigl((1/\bar\t')d'_{z^3\m}\bigr),\quad
\Re\bigl(d'_{\bar z^3\m}\bigr),\quad{\rm and}\quad
\Im\bigl(d'_{\bar z^3\m}\bigr).}

Finally, there are $6M$ massless scalars $\Phi'_I{}^m$ and $M$
massless gauge bosons $A'_{I\bsk\mu}$ from the D3 branes.

In summary, the massless bosonic fields are one graviton, $3+M$
vectors, and $16+M$ moduli.  The moduli consist of
$$\displaylines{%
\hbox{$V'_1,\ V'_2,\ V'_3$,\ 2 independent $\t'$,
\ 3 independent $a'^m{}_n$,}\cr
\hbox{6 $c'_{(4)}$ scalars, and $6M$ D3 scalars.}}$$

From the supersymmetry-breaking mass spectra given in
Ref.~\SUSYhiggs, this corresponds to ${\cal N}=4$ $\to$ ${\cal
N}=2$ with $m_1 = m_2$ for the two massive gravitini.  (This is a
third way to see that the background preserves $\CN=2$ supersymmetry).
The moduli space is $\CM = \CM_H\times\CM_V$, where in the
approximation that the warp factor is set to unity,
\eqna\ModSpaceI
$$\eqalignno{\CM_H &= {SO(4,3+M)\over SO(4)\times SO(3+M)},
&\ModSpaceI a\cr
\CM_V &= {SO(2,1+M)\over SO(2)\times SO(1+M)}
\times{SU(1,1)\over U(1)}, &\ModSpaceI b}$$
up to discrete identifications.  See App.~G for a discussion of the
metric on moduli space.

\subsec{T-Duality Map}

The T-duality map relating the O3 orientifold of Sec.~6.4 to the O5
orientifolds of Secs.~6.1--6.3 is exactly as described in Sec.~5.2,
once we perform the following relabelings of coordinates in the O3
orientifold:
\eqna\relabelingsI
$$\eqalignno{
\hbox{Sec.~6.1:} &\quad
\bigl(x^4,x^5,x^6,x^7,x^8,x^9\bigr)_{\hbox{\sevenrm Sec.~6.4}}
= \bigl(x^6,x^7,x^8,x^9,x^4,x^5\bigr)_{\rm new},
&\relabelingsI a\cr
\hbox{Sec.~6.2:} &\quad
\hbox{no relabeling,}
&\relabelingsI b\cr
\hbox{Sec.~6.3:} &\quad
\bigl(x^4,x^5,x^6,x^7,x^8,x^9\bigr)_{\hbox{\sevenrm Sec.~6.4}}
= \bigl(x^5,x^8,x^6,x^7,-x^9,x^4\bigr)_{\rm new}.
&\relabelingsI c\cr}$$
Note that in all three cases, the relabeling is a permutation of
positive Jacobian.  Therefore, the image of the ISD flux \choiceID{}\
is again ISD, rather than IASD.  In terms of the $T^6\to T^4\times
T^2$ factorization of Sec.~6.4, the three relabelings correspond to
T-dualizing along the $T^2$ factor, along a $T^2$ in the $T^4$ factor,
and along $S^1\subset T^4$ times $S^1\subset T^2$, respectively.

It is straightforward to follow all moduli and gauge fields through
the T-duality map, using the relations in Sec.~5.2.  The constraints
\newtwooneID{a,b}\ and \apconstraint\ map to \newtwooneIi{b,a}\ and
\newprimI\ in Sec.~6.1, to \twooneIii{a,b}\ and $b_{(2)\,56} =
b_{(2)\,47}$ in Sec.~6.2, and to \weakersimpIiii{b,a}\ and
\GoneResultIiii{b} in Sec.~6.3.  The volumes $V'_1,V'_2,V'_3$ are
equal to $V_2,V_3,(2\pi)^4\a'^2/V_\fib$ in Sec.~6.1, to
$(2\pi)^4\a'^2/V_\fib,V_2,V_3$ in Sec.~6.2, and to
$(2\pi)^2\a'|\t_1|,V_2,(2\pi)^2\a'|\t_3|$ in Sec.~6.3.  The RR moduli
\RRmoduliID\ map to \RRmoduliIi, \RRmoduliIii, and \RRmoduliIiii.  The
massive gauge bosons \gaugebosID\ map to \gaugebosIi, \gaugebosIii,
and
\gaugebosIiii.


\newsec{$\CN=3$ Example: The Warped Iwasawa Manifold}

\subsec{Example 4: $\CN+ = 1$, $\CN_- = 2$}

Consider the choice of fibration curvature and background flux
\eqna\choiceII
$$\eqalignno{&\CF^4 = 2h\bigl(dx^6\w dx^8 -m dx^6\w dx^9 -m dx^7\w
dx^8 + (m^2-n) dx^7\w dy^9\bigr), &\choiceII b\cr 
\noalign{\vskip5pt}
&\CF^5 = -2h\bigl(dx^6\w dx^9 + dx^7\w dx^8 - m dx^7\w
dx^9\bigr), &\choiceII a\cr
\noalign{\vskip5pt}
&F^{\bg\,1}_{(3)}/\bigl((2\pi)^2\a'\bigr) = 2f\Bigl(\eta^4\w\bigl(
dx^6\w dx^8 - ndx^7\w dx^9\bigr) &\cr
&\qquad\qquad + \eta^5\w\bigl(-ndx^6\w dx^9 -ndx^7\w dx^8 + mndx^7\w
dx^9\bigr)\Bigr), 
&\choiceII c}
$$
with $F_{(1)}=H_{(3)}=0$.  Here, $f,h,n=1,2$ and $m=0,1$.  This choice
gives a contribution to the Gauss's law constraint,
\eqn\NfluxI{N_\flux = 4fh\bigl(4n-m^2\bigr),}
so that Eq.~\Gauss\ becomes
\eqn\GaussI{2M+4fh\bigl(4n-m^2\bigr)=32.}
The condition $0\le N_\flux\le 32$ puts restrictions on the allowed
combinations of $f,h,m,n$.  The discussion of the dual O3 theory in
Sec.~7.2 shows why this is in some sense a natural class of
backgrounds to consider.

For the choice \choiceII{}\ to preserve $\CN=3$ supersymmetry, we
require that the complex flux \cpxfluxII{d}\ be (2,1) and primitive
with respect to three independent complex structures.  This is
equivalent to demanding that $G_{(3)}$ be of type (0,3) with respect a
single complex structure, which is then distinct from the previous
three \FP.  (See Sec.~7.2 for further discussion).

It is convenient to parametrize the metric and almost complex
structure as in Ex.~1:
\eqn\cpxstrII{\eqalign{\eta^{w^1} &= \eta^4+\t^{w^1}\eta^5,\cr
dw^i &= dx^{2i+2}+\t^{w^i}{}_j dx^{2j+3},\quad i,j=2,3.}}
Here, we have used the symbols $w^i$ instead of $z^i$, reserving the
latter for complex coordinates in which $G_{(3)}$ is of type (2,1) and
primitive.

The condition that $G_{(3)}$ be of type (0,3) results in the following
constraints:
\eqna\zerothreeconstrII
$$\eqalignno{&\t^{w^i}{}_j = \t\d^i{}_j,\quad{\rm where}\quad
-(h/f)\bar\t_\dil=\t^{w^1}=\t,\quad{\rm and}&\zerothreeconstrII a\cr
&\t^2 + m\t + n = 0,\quad\hbox{or equivalently,}\quad
\t = \half\bigl(-m+i\sqrt{4n-m^2}\bigr). &\zerothreeconstrII b}$$
Here, we have chosen the root $\t$ such that $\Im\t>0$, for agreement
of Eq.~\zerothreeconstrII{a}\ with $\Im\tdil = 1/g_s > 0$.  When
Eqs.~\zerothreeconstrII{a,b}\ are satisfied,
\eqn\GII{{1\over(2\pi)^2\a'}G_{(3)}
= 2f\,\eta^{\bar w^1}\w d\bar w^2\w d\bar w^3,}
which is indeed of type (0,3).  Since this statement is independent of
the K\"ahler moduli, the unfixed metric moduli are the five real
K\"ahler moduli $V_\fib$ and $g_{\CB\msk w^i\bar w^j}$ $(i,j=2,3)$,
with $g_{\CB\msk w^i\bar w^j} = \bigl(g_{\CB\msk w^j\bar
w^i}\bigr)^*$.  The constraints \zerothreeconstrII{a,b} also imply
that
\eqn\CFw{\CF^{w^1} = 2h d\bar w^2\w d\bar w^3,}
where $\CF^{w^1} = \CF^4+\t^{w^1}\CF^5$.

Now let us return to the supersymmetry conditions as originally
formulated.  The three independent complex structures satisfying
conditions \cpxfluxII{a,b}\ are related to the one just described via
\eqna\wzmap
$$\eqalignno{\hbox{complex structure 1:}
&\quad \bigl(\eta^{z^1},dz^2,dz^3\bigr)  = \bigl(\eta^{w^1},d\bar
w^2,d\bar w^3\bigr), & \wzmap a\cr
\hbox{complex structure 2:}
&\quad \bigl(\eta^{z^1},dz^2,dz^3\bigr)  = \bigl(\eta^{\bar
w^1},dw^2,d\bar w^3\bigr), & \wzmap b\cr
\hbox{complex structure 3:}
&\quad \bigl(\eta^{z^1},dz^2,dz^3\bigr)  = \bigl(\eta^{\bar w^1},d\bar
w^2,dw^3\bigr). & \wzmap c}$$
In these three complex structures,
\eqna\CFzs
$$\CF^{z^1} = 2h dz^2\w dz^3,\quad
\CF^{z^1} = 2h d\bar z^2\w dz^3,\quad{\rm and}\quad
\CF^{z^1} = 2h dz^2\w d\bar z^3,\eqno\CFzs{a,b,c}$$
respectively.  In all three cases, $\CF^{z^1}$ has no (0,2) component.
Therefore, the corresponding fibrations are holomorphic, and the complex
structures are integrable, in agreement with Sec.~4.2.  In constrast,
the almost complex structure of the $w$ coordinates is not integrable.

Specialize to complex structure 1.  Then we can write Eq.~\CFzs{a}\ as
\eqn\detazz{d\eta^{z^1} = N dz^2\w dz^3,\quad{\rm where}\quad N=2h.}
This is the familiar relation between the three left-invariant
one-forms
$$\eta^{z^1} = z^1 + N z^2 dz^3,\quad dz^2,\quad{\rm and}\quad dz^3,$$
on the Iwasawa manifold.  

The Iwasawa manifold can be defined as a coset of upper triangular
matrices as follows.\foot{This description is taken directly from the
twisted $T^3$ example in Sec.~2.2 of Ref.~\KSTT, with $\IR^3$ replaced
by $\IC^3$.}  Consider $\IC^3$, presented as the space of upper
triangular $3\times3$ matrices with ones along the diagonal,

\eqn\ICthree{g_N(z^1,z^2,z^3)
= \pmatrix{1&z^3&-z^1/N\cr 0&1&z^2\cr 0&0&1},
\quad{\rm where}\quad z^1,z^2,z^3\in\IC.}
Here, $N$ is a fixed positive integer. Let ${\cal G}^N_3(\IC)$ deonte
the corresponding group under matrix multiplication.  For any $N$,
this group is isomorphic to the complex three-dimensional Heisenberg
group ${\cal H}_3(\IC)$.  We can also define a discrete subgroup
${\cal G}^N_3(\L)$ containing the matrices

\eqn\ICthreet{g_N(a,b,c)
= \pmatrix{1&c&-a/N\cr 0&1&b\cr 0&0&1},
\quad{\rm where}\quad a,b,c\in\L,}
with $\L=\IZ+\t\IZ\subset\IC$ a lattice parametrized by the complex
modulus~$\t$.  This subgroup has a natural action on ${\cal G}^N_3
(\IC)$ by matrix multiplication.  Consider the right-coset $\CM^N =
{\cal G}^N_3(\IC)/{\cal G}^N_3(\L)$.  The resulting identifications
are
\eqn\ICresult{(z^1,z^2,z^3) \cong (z^1+a-Nbz^3,z^2+b,z^3+c).} 
This quotient defines the $N$th Iwasawa manifold.

The standard metric and K\"ahler form on the Iwasawa manifold are
\eqna\dsJIwasawa
$$\eqalignno{ds^2_{\rm Iwasawa} &= \bigl|\eta^{z^1}\bigr|^2
+ \bigl|dz^2\bigr|^2 + \bigl|dz^3\bigr|^2, &\dsJIwasawa a\cr
J_{\rm Iwasawa} &= {i\over2}\bigl(\eta^{z^1}\w\eta^{\bar z^1}
\mkern-3mu +dz^2\w d\bar z^2 + dz^3\w d\bar z^3\bigr). &\dsJIwasawa b}$$
However, in the orientifold example that we are considering, the
metric and K\"ahler form are given by\foot{In the conventions of this
paper, $ds^2 = g_{i\jbar} dz^id\bar z^j + g_{\ibar j}d\bar z^idz^j =
2g_{i\jbar} dz^id\bar z^j$ (cf.~App.~A).}
\eqna\dsJII
$$\eqalignno{ds_6{}^2 &= {V_\fib\over\Im\t}\bigl|\eta^{z^1}\bigr|^2
+2Z g_{\CB\msk z^i\bar z^j} dz^i d\bar z^j, &\dsJII a\cr J &=
{iV_\fib\over2\Im\t}\eta^{z^1}\w\eta^{\bar z^1} + iZg_{\CB\msk z^i\bar
z^j} dz^i\w d\bar z^j, &\dsJII b}$$
a generalization of \dsJIwasawa{a,b}\ that includes arbitrary fiber
and base K\"ahler structure, and also the warp factor $Z$.

Now let us turn to the NS $B$-field moduli.  The constraints that
follow from Eq.~\FthreeBconstr{a}\ are
$$\eqalign{& mb_{(2)\,46}+b_{(2)\,47}+b_{(2)\,56}=0,
\quad nb_{(2)\,46}-b_{(2)\,57} = 0,\cr
& mb_{(2)\,48}+b_{(2)\,49}+b_{(2)\,58}=0,
\quad nb_{(2)\,48}-b_{(2)\,59} = 0,}$$
or equivalently,
\eqn\NSBII{b_{(2)w^1w^i} = b_{(2)\bar w^1\bar w^i} = 0,\quad
i=2,3.}
Eq.~\FthreeBconstr{b}\ gives the same constraints.  The unlifted NS
$B$-field moduli are the orthogonal components
\eqn\NSBmoduliII{b_{(2)w^1\bar w^i},\quad b_{(2)\bar w^1 w^i},
\quad i=2,3.}

From the kinetic terms \gaugecouplings{a,b,c}, the massless RR sector
scalars are the nine uncharged fields
\eqn\RRmoduliII{\eqalign{&\tilde c_{(2)\,67},\quad \tilde c_{(2)\,89},
\quad\tilde c_{(2)\, 69} - \tilde c_{(2)\,78},
\quad\tilde c_{(2)\m\n}\leftrightarrow\tilde c_{(6)\,456789},\cr
& n\tilde c_{(2)\,68} +(m/2)\bigl(\tilde c_{(2)\,69} + \tilde
c_{(2)\,78}\bigr) + \tilde c_{(2)\,79},\cr
& \tilde c_{(4)\,5678} - \tilde c_{(4)\,4679},\quad
\tilde c_{(4)\,4789} - \tilde c_{(4)\,5689},\cr
& n\tilde c_{(4)\,4678} +(m/2)\bigl(\tilde c_{(4)\,4679} + \tilde
c_{(4)\,5678}\bigr) + \tilde c_{(4)\,5679},\cr
& {\rm and}\quad n\tilde c_{(4)\,4689} +(m/2)\bigl(\tilde c_{(4)\,4789}
+ \tilde c_{(4)\,5689}\bigr) + \tilde c_{(4)\,5789}.}}

The massive gauge bosons are those that couple to the other RR
scalars.  From \gaugecouplings{a,b,c}, they lie in the 6D space
spanned by
\eqn\gaugebosII{\eqalign{&(2\pi)^2\a'(f/h)v^4{}_\m
- c_{(2)\,4\m},\quad
(2\pi)^2\a'(f/h)v^5{}_\m - mc_{(2)\,4\m} - c_{(2)\,5\m},\cr
&(2\pi)^2\a'(f/h)b_{(2)\,7\m} - \tilde c_{(4)\,456\m},\quad
(2\pi)^2\a'(f/h)b_{(2)\,6\m} + m\tilde c_{(4)\,456\m}
+ \tilde c_{(4)\,457\m},\cr
&(2\pi)^2\a'(f/h)b_{(2)\,9\m} - \tilde c_{(4)\,458\m},\quad
(2\pi)^2\a'(f/h)b_{(2)\,8\m} + m\tilde c_{(4)\,458\m}
+ \tilde c_{(4)\,459\m}.}}
The massless gauge bosons span the orthogonal 6D space, where
orthogonality is defined with respect to the metric appearing in the
gauge kinetic terms.  (See Sec.~7.2 for further discussion in the dual
O3 orientifold).

Finally, there are $6M$ massless scalars $A_{I\,\a}$, $\Phi_I{}^m$,
and $M$ massless gauge bosons $A_{I\,\m}$ from the D5 branes.

In summary, the massless bosonic fields are one graviton, $6+M$
vectors, and $18+6M$ moduli.  The moduli consist of
$$\displaylines{\hbox{$V_\fib$, 4 $g_{\CB\msk z^i\bar z^j}$,
4 indep $b_{(2)\a m}$,}\cr
\hbox{9 $\tilde c_{(p)}$ scalars, and $6M$ D5 scalars.}}$$
These fields combine to form one 4D $\CN=3$ gravity multiplet, and
$3+M$ vector multiplets.

The amount of unbroken supersymmetry is
\eqn\NsusyII{\CN_+ = 1,\quad \CN_-=2,}
in the notation of Sec.~4.2.  The $\CN_+=1$ supersymmetry corresponds
to complex structure~1 \wzmap{a}, with the modulus $\t_1=\t$ of
$\Im\t_1>0$ appearing in $\eta^{z^1}$ (cf.~Eq.~\fibpm{a}).  The
$\CN_-=2$ supersymmetries correspond to complex structures~2 and~3
\wzmap{b,c}, with the modulus $\t_1 = \bar\t$ of $\Im\t_1<0$ appearing
in $\eta^{z^1}$.

\subsec{Dual O3 Orientifold}

For the O3 orientifold with internal $T^6$, choices of flux preserving
4D $\CN=3$ supersymmetry were first discussed in Ref.~\FP.  The
requirement for $\CN=3$ supersymmetry is
\eqn\Neqthreecond{G'_{(3)}\quad\hbox{(2,1) and primitive
w.r.t.\ to three independent complex structures.}}
In contrast, from the equations of motion alone, without imposing any
supersymmetry conditions, one obtains the condition that the flux be
ISD: $*_6 G'_{(3)} = iG'_{(3)}$.  As has already been mentioned in
Footnote~\ISDfoot, the space of ISD forms includes not only primitive (2,1)
forms, but also nonprimitive (1,2) forms $J\w\o^{(0,1)}$, where
$\o^{(0,1)}$ is a (0,1) form, as well as (0,3) forms.  The condition
\Neqthreecond\ is equivalent to~\FP
\eqn\altNeqthreecond{G'_{(3)}\quad\hbox{(0,3) for some choice of
complex structure.}}
The complex structure in this last condition is a fourth complex
structure, independent of the three complex structures of the previous
condition.\foot{In fact, we could have also rephrased the
supersymmetry conditions for the $\CN=2$ examples.  In each example of
Sec.~6, there exists a complex structure such that $G_{(3)}$, $\tilde
G_{(3)}$, or $G'_{(3)}$ (whichever is appropriate) is of the form
$J\w\o^{(0,1)}$.  However, unlike the $\CN=4\to\CN=3$ case, where the
conditions \Neqthreecond\ and \altNeqthreecond\ are truly equivalent,
in the $\CN=4\to\CN=2$ case, this rephrasing of the supersymmetry
conditions seems to involve the assumption that $m_1=m_2$ for the two
massive gravitini.}

To construct a large class of $\CN=3$ backgrounds, let us write
\eqn\GIID{{1\over(2\pi)^2\a'}G'_{(3)}
= 2\l d\bar w^1\w d\bar w^2\w d\bar w^3,}
where $\l$ is a positive real number, and where the $T^6$ factorizes
as $T^2\times T^2\times T^2$ with respect to complex structure in the
following way:
\eqna\cpxstrIID
$$\eqalignno{&w^i = x^i+\t' y^i,\quad i=1,2,3, &\cpxstrIID a\cr
&(h/f)\tdil' = -\bar\t',\quad f,h\in\IN. &\cpxstrIID b}$$
The coordinates $x^i,y^i$ are related to those used elsewhere in this
paper by
\eqn\SecviiNot{(x^1,y^1,x^2,y^2,x^3,y^3)_{\rm here}
= (-x^5,x^4,x^6,x^7,x^8,x^9)_{\hbox{\sevenrm rest of paper}}.}  
We adopt this notation here and similar notation in Sec.~8.2 since it
allows equations involving cyclic permutation of the $x^i,y^i$ to be
written succinctly.

When expanded in real coordinates, the complex flux becomes
\eqn\GcubictII{\eqalign{{1\over(2\pi)^2\a'}G'_{(3)}
&= 2\l\Bigl(dx^1\w dx^2\w dx^3 + \bar\t'\bigl(dx^1\w dx^2\w dy^3
+ \hbox{cyc.\ perms.\ of 123}\bigr)\cr
& + \bar\t'^2\bigl(dx^1\w dy^2\w dy^3
+ \hbox{cyc.\ perms.\ of 123}\bigr)
+ \bar\t'^3 dy^1\w dy^2\w dy^3\Bigr).}}
If, in addition, we assume as part of our ansatz that $\t'$ satisfies
the quadratic equation
\eqn\quadraticP{P(\t') = l\t'^2 + m\t' + n = 0,\quad
l,m,n\in\IZ,\quad l>0,}
then Eq.~\GcubictII\ reduces to an expression linear in $\bar\t'$.
From Eqs.~\cpxstrIID{a,b}, we can then read off the RR and NS fluxes.
It is possible to show from the quantization condition \OthreeFlux\
that by redefinitions of $l$, $m$, $n$, and $\l$, we can take $\l =
fl$.  Then,
\eqna\FandHIID
$$\eqalignno{&{1\over(2\pi)^2\a'}F'_{(3)}
= 2f\Bigl(l dx^1\w dx^2\w dx^3
- n\bigl(dx^1\w dy^2\w dy^3
+ \hbox{cyc.\ perms.\ of 123}\bigr) &\cr
&\qquad\qquad + (mn/l)dy^1\w dy^2\w dy^3\Bigr),
&\FandHIID a\cr
&{1\over(2\pi)^2\a'}H'_{(3)}
= 2h\Bigl(l \bigl(dx^1\w dx^2\w dy^3
+ \hbox{cyc.\ perms.\ of 123}\bigr) &\cr
& \quad\vphantom{\Big(\Bigr)} - m\bigl(dx^1\w dy^2\w dy^3
+ \hbox{cyc.\ perms.\ of 123}\bigr)
+ (m^2/l - n)dy^1\w dy^2\w dy^3\Bigr).
&\FandHIID b}$$
The contribution to the Gauss's law constraint \Gauss\ from this choice
of flux is
\eqn\GaussIID{N_\flux = 4fh(4ln-m^2).}
Not all choices of $(l,m,n)$ are inequivalent.  From change of lattice
basis on the $T^6$, there is an $SL(2,\IZ)_{\t'}$ identification of
the complex structure modulus $\t'$.  Modulo identifications, we can
assume that $\t'$ lies in the fundamental domain of $SL(2,\IC)$,
\eqn\FundDomain{\t'\in\CF_0 = \{\t'\in\IC \mid -\half\le\Re\t'<\half,
\ |\t'|\ge1\}.}
From Eq.~\quadraticP, we have
\eqn\tlmn{\t' = {1\over2l}\bigl(-m+i\sqrt{4ln-m^2}\bigr),
\quad |\t'| = \sqrt{n/l},}
so the condition \FundDomain\ becomes
\eqn\tconditions{n>l\ge |m|\quad{\rm and}\quad 4ln>m^2.}

The possible values of $(f,h,l,m,n)$ are strongly constrained by $0\le
N_\flux\le 32$ and the fact that $m^2 \equiv 0,1 \pmod{4}$. For $(f,h)
= (1,1)$, in addition to the $\CN=4$ solution without flux, there are
just four choices of $(l,m,n)$ satisfying the inequalities
\tconditions:
\eqna\possiblelmnII
$$\matrix{(l,m,n) = (1,1,1), \quad\hfill & N_\flux = 12,
\quad\hfill & 2M = 20,\quad\hfill & \t' = e^{2\pi i/3},\quad\hfill\cr
\noalign{\vskip5pt}
(l,m,n) = (1,0,1),\quad\hfill & N_\flux = 16, \quad\hfill & 2M = 16,
\quad\hfill & \t' = i,\quad\hfill\cr
\noalign{\vskip5pt}
(l,m,n) = (1,1,2),\quad\hfill & N_\flux = 28, \quad\hfill & 2M = 4,
\quad\hfill & \t' = \half(-1+i\sqrt7),\quad\hfill\cr
\noalign{\vskip5pt}
(l,m,n) = (1,0,2),\quad\hfill & N_\flux = 32, \quad\hfill & 2M = 0,
\quad\hfill & \t' = i\sqrt2.\quad\hfill} \eqno\possiblelmnII a$$ 
Allowing arbitrary values of $(f,h)$, there are four additional
choices.  For $(f,h) = (2,1)$ or (1,2), we can take
$$\matrix{(l,m,n) = (1,1,1), \quad\hfill & N_\flux = 24,
\quad\hfill & 2M = 8,\quad\hfill & \t' = e^{2\pi i/3},\quad\hfill\cr
\noalign{\vskip5pt}
(l,m,n) = (1,0,1),\quad\hfill & N_\flux = 32, \quad\hfill & 2M = 0,
\quad\hfill & \t' = i.\quad\hfill} \eqno\possiblelmnII b$$
Note that the flux \FandHIID{a,b}\ is correctly quantized with
$F'_{(3)},H'_{(3)}\in (2\pi)^2\a' H^3(T^6,2\IZ)$ in all of four cases.

If the complex structure and axion-dilaton are deformed from $\t^i{}_j
= \t'\d^i{}_j$ and $(h/f)\tdil' =\t'$, with $\t'$ given by the values
in \possiblelmnII{a,b}, then the complex flux is no longer of type
(0,3).  On the other hand, the flux is still (0,3) for arbitrary
choice of K\"ahler moduli.  So, the metric moduli are the nine real
degrees of freedom $g'_{w^i\bar w^j}$.

From the kinetic terms \Ckintermp, the space of massless RR scalars is
spanned by the nine uncharged scalars
\eqn\RRmoduliIID{\eqalign{&\quad nc'_{(4)x^1y^1x^2x^3} +
c'_{(4)x^1y^1y^2y^3} + (m/2) 
\bigl(c'_{(4)x^1y^1x^2y^3}+c'_{(4)x^1y^1y^2x^3}\bigr),\cr
& c'_{(4)x^1y^1x^2y^2},\quad c'_{(4)x^1y^1x^2y^3} -
c'_{(4)x^1y^1y^2x^3}, \quad{\rm and}\quad\hbox{cyc.\ perms.\ of
123}.}} 
This is equivalent to the space spanned by the RR scalars of Hodge
type (2,2),
$$c'_{(4)w^iw^j\bar w^k\bar w^l},\quad i,j,k,l = 1,2,3.$$

The massive gauge bosons are those that couple to the other RR
scalars.  From \Ckintermp, they lie in the 6D space spanned by
\eqn\gaugebosIID{(f/h)nb'_{(2)x^i\m}-mc'_{(2)x^i\m}-c'_{(2)y^i\m},
\quad{\rm and}\quad (f/h)b'_{(2)y^i\m} + c'_{(2)x^i\m},
\quad i=1,2,3.}
In terms of the complex gauge bosons \cpxgaugebos, this is the space
spanned by $d'_{w^i\m}$, for $i=1,2,3$.  In fact, we can see directly
from \cpxCkintermp\ that $c'_{(4)w^1w^2w^3\bar w^i}$ is the axion
eaten by $d'_{w^i\m}$ \FP.  The massless bosons are the linear
combinations orthogonal to this space, where orthogonality is defined
with respect to the metric appearing in the gauge kinetic terms.  In
the $\bigl(d'_{w^i\m},d'_{\bar w^i\m},\bar d'_{w^i\m},\bar d'_{\bar
w^i\m}\bigr)$ basis of complex gauge bosons \cpxgaugebos, the latter
is proportional to
\eqn\gaugemetricIID{\ha\pmatrix{0&1\cr1&0}
\matrix{{\scriptstyle d'}\cr
{\scriptstyle\bar d'}}\quad\otimes\quad
\pmatrix{0&g'^{w^i\bar w^j}_6\cr g'^{\bar w^i w^j}_6&0}
\matrix{{\scriptstyle w^i}\cr{\scriptstyle\bar w^i}}.}
Therefore, the six massless gauge bosons are (the real and imaginary
parts of) $d'_{\bar w^i\m}$, for $i = 1,2,3$ \FP.

Finally, there are $6M$ massless scalars $\Phi'_I{}^m$, and $M$
massless gauge bosons $A'_{I\,\m}$ from the D3 branes.

In summary, the massless fields are one graviton, $6+M$ vectors, and
$18+6M$ moduli.  The moduli consist of
$$\hbox{9 $g_{\CB\msk w^i\bar w^j}$,\quad
9 $\tilde c_{(p)}$ scalars,\quad and $6M$ D3 scalars.}$$
These fields combine to form one 4D $\CN=3$ gravity multiplet, and
$3+M$ vector multiplets.

From Ref.~\Castellani, the moduli space of the $\CN=3$ theory is
completely determined by the number $n_V$ of vector multiplets.  It is
the coset
\eqn\NthreeModspace{\CM = {U(3,n_V)\over U(3) \times U(n_V),}}
up to discrete identifications, where, in this example, $n_V = 3+M$.
For the parametrization of the coset in terms of the above moduli, see
Ref.~\FP.

\subsec{T-Duality Map}

For the O3 orientifold of Sec.~7.2, the metric is
\eqn\gwwherm{ds'^2_6 = 2g_{w^i\bar w^j}dw^i d\bar w^j,}
with $g_{w^i\bar w^j}$ constrained only to be Hermitian.  We can
alternatively write this metric in the fibration form \dssixprime,
\eqn\gwwfib{ds'^2_6 = {V'_\fib\over\Im\t'}\bigl|dw^1+a'^{w^1}\bigr|^2
+2g_{\CB\msk w^i\bar w^j}dw^i d\bar w^j,}
where the components of $a'^\a$ are holomorphically constrained,
\eqna\hermvsfib
$$a'^{w^1} = a'^{w^1}{}_{w^i}dw^i,\quad{\rm and}\quad a'^{\bar w^1} =
a'^{\bar w^1}{}_{\bar w^i}d\bar w^i,\quad i=2,3, \eqno\hermvsfib a$$
and where
$$\eqalignno{V'_\fib &= 2 g_{w^1\bar w^1}\Im\t', &\hermvsfib b\cr
g_{\CB\msk w^i\bar w^j} &= g_{w^i\bar w^j} - g_{w^1\bar w^1}
a'^{w^1}{}_{w^i}a'^{\bar w^1}{}_{\bar w^j},\quad i,j =2,3. &\hermvsfib
c}$$
Here, from Eqs.~\cpxstrIID{a}\ and \SecviiNot, the complex coordinates
in the O3 orientifold are
\eqn\wtprime{w^1 = -x^5 + \t'x^4\quad{\rm and}\quad
w^i = x^{2i+2} + \t'x^{2i+3},\quad i=2,3.}
Applying the T-duality map of Sec.~5.2, we find agreement between
Secs.~7.1 and 7.2, provided that we identify $\t=\t'$.  For example,
the map between complex components of $a'^\a$ and $b_{(2)\a}$ is
\eqn\bwawmap{b_{(2)w^1} = a'^{\bar w^1}/(\t'-\bar\t')\quad{\rm
and}\quad b_{(2)\bar w^1} = -a'^{w^1}/(\t'-\bar\t'),}
so, the constraints \NSBII\ and \hermvsfib{a}\ agree.  Here,
$\eta^{w^1} = \eta^4+\t\eta^5$ for the O5 orientifold, from which the
relation between $b_{(2)w^1},b_{(2)\bar w^1}$ and
$b_{(2)\,4},b_{(2)\,5}$ is
\eqn\bcpxreal{b_{(2)w^1} = {1\over\t-\bar\t}(b_{(2)\,5}-\bar\t
b_{(2)\,4})\quad{\rm and}\quad b_{(2)w^1} = -{1\over\t-\bar\t}
(b_{(2)\,5}-\t b_{(2)\,4}).}
Similarly, using the underscore notation of Sec.~5.2,
\eqn\CFvsGprime{\eqalign{\CF^{w^1}_{mn} &= \CF^4_{mn}+\t\CF^5_{mn} =
H'_{(3)\,\ul{4mn}}+\t'H'_{(3)\,\ul{5mn}}\cr
&= -(\t'-\bar\t')H'_{(3)\ul{\bar w^1 mn}} = (h/f)G'_{(3)\ul{\bar w^1
mn}},}}
from which Eq.~\CFw\ follows.


\newsec{$\CN=1$ Example}

\subsec{Example 5: $\CN_+ = 1$, $\CN_- = 0$}

Consider the choice of fibration curvature and background flux
\eqna\choiceIII
$$\eqalignno{&\phantom{-}\CF^4 = 2h\bigl(dx^6\w dx^8 - dx^6\w dx^9 -
dx^7\w dx^8 + dx^7\w dx^9\bigr),&\choiceIII a\cr
&{-}\CF^5 = 2h\bigl(dx^6\w dx^8 - dx^6\w dx^9 - dx^7\w dx^8 +
2 dx^7\w dx^9\bigr), &\choiceIII b\cr
&{1\over(2\pi)^2\a'}F^{\bg\,1}_{(3)} = 2f\eta^5\w dx^6\w
dx^8 -2f \eta^4\w dx^7\w dx^9, &\choiceIII c}$$
with $F_{(1)} = H_{(3)}= 0$.  This choice gives a contribution to the
Gauss's law constraint
\eqn\NfluxIII{N_\flux = 12fh,}
so that Eq.~\Gauss\ becomes
\eqn\GaussIII{2M + 12fh=32.}
As discussed in Sec.~3.2, $N_\flux$ is nonnegative.  So, $f$ and $h$
have the same sign, and the possible values of $(f,h)$ are (1,1),
(2,1), and (1,2), corresponding to $2M =$ 20, 8, and 8 D5 branes,
respectively.

Define an almost complex structure as in Eq.~\cpxstrI.  Then, using
the decomposition \OJpm, it is possible to show that the base $T^4$
decomposes as $T^4\to T^2\times T^2$ with respect to complex
structure:
\eqn\cpxstrIII{\t^i{}_j = \t\d^i{}_j,\quad i,j = 2,3,}
where
\eqn\twooneIII{-1/\t_1 = \t = (h/f)\tdil = e^{2\pi i/3}.}
When these constraints are satisfied, the complex three-form flux
\cpxfluxII{d}\ is
\eqn\GIII{G_{(3)} = -\bigl(2if/\sqrt{3}\bigr)
\bigl(\eta^{z^1}\w dz^2\w d\bar z^3 + \eta^{z^1}\w d\bar z^2\w dz^3
+ \eta^{\bar z^1}\w dz^2\w dz^3\bigr).}
The primitivity condition \cpxfluxII{b}\ then constrains $J$ to be of
the form
\eqn\JIII{J = {i\over2\Im\t_1}V_\fib\eta^{z^1}\w\eta^{\bar z^1}
+ {1\over2\Im\t}\Bigl(iV_2 dz^2\w d\bar z^2 + iV_3 dz^3\w d\bar z^3
-2V_{(23)}\Im\bigl(dz^2\w d\bar z^3\bigr)\Bigr),}
corresponding to the base metric
\eqn\basemIII{ds^2_\CB = {1\over|\Im\t|}\Bigl(V_2 dz^2 d\bar z^2
+ V_3 dz^3 d\bar z^3 + V_{(23)}\Re\bigl(dz^2 d\bar z^3\bigr)\Bigr).}

From Eqs.~\FthreeBconstr{a,b}, the constraints on NS $B$-field moduli
are 
\eqna\NSBIII
$$\eqalignno{0 &= 2h(b_{(2)\,4m} - b_{(2)\,5m}) dx^m\w
(dx^6\w dx^8 - dx^6\w dx^9 - dx^7\w dx^8 + dx^7\w dx^9) &\cr
&\qquad -2h b_{(2)\,5m}dx^m\w dx^7\w dx^9, &\NSBIII a\cr
0 &= -2f b_{(2)\,4m}dx^m\w dx^7\w dx^9
+ 2f b_{(2)\,5m}\w dx^m\w dx^6\w dx^8. &\NSBIII b}$$
Together, these constraints imply
\eqna\NSBequivIII
$$\eqalignno{&b_{(2)\,47} = b_{(2)\,49} = b_{(2)\,56} = b_{(2)\,58} =
0, &\NSBequivIII a\cr
& b_{(2)\,46} = b_{(2)\,57},\quad{\rm and}\quad
b_{(2)\,48} = b_{(2)\,59}. &\NSBequivIII b}$$
So, there are two unconstrained NS $B$-field moduli,
\eqn\NSBmoduliIII{b_{(2)\,46} + b_{(2)\,57}\quad{\rm and}\quad
b_{(2)\,48} + b_{(2)\,59}.}

From the kinetic terms \gaugecouplings{a,b,c}, the massless RR sector
scalars are the uncharged fields
\eqn\RRmoduliIII{\eqalign{&\tilde c_{(2)\,67},\quad\tilde c_{(2)\,89},
\quad \tilde c_{(2)\,69} - \tilde c_{(2)\,78},
\quad\tilde c_{(2)\m\n}\leftrightarrow\tilde c_{(6)\,456789},\cr
& \tilde c_{(4)\,4678} + \tilde c_{(4)\,5679},\quad{\rm and}\quad
\tilde c_{(4)\,5789} + \tilde c_{(4)\,4689}.}}

The massive gauge bosons are those that couple to the other RR
scalars.  From \gaugecouplings{a,b,c}, they lie in the 9D space
spanned by
\eqn\gaugebosIII{\eqalign{&(2\pi)^2\a'(f/h)v^4{}_\m - c_{(2)\,5\m},
\quad v^5{}_\m,\quad c_{(2)\,4\m}-c_{(2)\,5\m},\cr
&(2\pi)^2\a' (f/h)b_{(2)\,6\m}-\tilde c_{(4)\,456\m},\quad
b_{(2)\,7\m},\quad \tilde c_{(4)\,456\m}+\tilde c_{(4)\,457\m},\cr
&(2\pi)^2\a' (f/h)b_{(2)\,8\m}-\tilde c_{(4)\,458\m},\quad
b_{(2)\,9\m},\quad \tilde c_{(4)\,458\m}+\tilde c_{(4)\,459\m}.}}
The massless bosons are the three linear combinations orthogonal to
this space, where orthogonality is defined with respect to the metric
appearing in the gauge kinetic terms.  (See Sec.~8.2 for further
discussion in the dual O3 orientifold).

Finally, there are $6M$ massless scalars $A_{I\,\a}$, $\Phi_I{}^m$,
and $M$ massless gauge bosons $A_{I\,\m}$ from the D5 branes.

In summary, the massless fields are one graviton, $3+M$ vectors, and
$6+6M$ moduli.  The moduli consist of
$$\displaylines{\hbox{$V_\fib$, $V_2$, $V_3$, $V_{(23)}$, 2 indep
$b_{(2)\a m}$,}\cr
\hbox{6 $\tilde c_{(p)}$ scalars, and $6M$ D5 scalars.}}$$
These fields combine to form one 4D $\CN=1$ gravity multiplet, $3+3M$
chiral multiplets, and $3+M$ vector multiplets.

The amount of unbroken supersymmetry is
\eqn\NsusyIII{\CN_+ = 1,\quad \CN_- = 0,}
in the notation of Sec.~4.2.  To verify this, first note that the
complex structure modulus of the fiber is $\t_1$, with $\Im\t_1>0$
from \twooneIII.  Therefore, the supersymmetry is of $+$ type.  Since
the base factorizes as $T^2_{\{67\}}\times T^2_{\{89\}}$ with respect
to complex structure, we can define chirality operators
$\tilde\g_{(2)} = \tilde\g_2{}^2$ on $T^2_{\{67\}}$ and
$\tilde\g_{(3)} = -\tilde\g_3{}^3$ on $T^2_{\{89\}}$, where the
indices 2 and 3 are holomorphic $z^2$ and $z^3$ indices.\foot{Here,
the relative minus sign is necessary for chiralities to be
multiplicative, that is, for $\tilde\g_\CB =
\tilde\g_{(2)}\tilde\g_{(3)}$.  In the conventions of this paper,
$\tilde\g_\CB = g_\CB^{1/2}\tilde\g^{6789}$ (cf.~Eq.~\pregtprod),
which equivalent to $\tilde\g_\CB = -\tilde\g_2{}^2{}_3{}^3$ when
expressed in terms of holomorphic coordinates.}  Then, since complex
structure moduli of $T^2_{\{67\}}$ and $T^2_{\{89\}}$ are $\t$, with
$\Im\t>0$ from \twooneIII, there is one independent negative chirality
6D Killing spinor $\chi$.  It is of the form
\eqn\KillingspinorIII{\chi =
\z^\fib_+\otimes\z^{(2)}\otimes\z^{(3)},}
where $\tilde\g_{(2)}\z^{(2)} = \z^{(2)}$ and $\tilde\g_{(3)}\z^{(3)}
= -\z^{(2)}$. The three factors in Eq.~\KillingspinorIII\ are on equal
footing in the sense that $\tilde\g_i{}^i \hbox{ (no sum) }= +1$ in
all three cases.

\subsec{Dual O3 Orientifold}

In this subsection, it is convenient to use the notation
\eqn\SecviiiNot{(x^1,y^1,x^2,y^2,x^3,y^3)_{\rm here}
= (x^4,x^5,x^6,x^7,x^8,x^9)_{\hbox{\sevenrm rest of paper}}.}  
In this notation, consider the O3 orientifold with internal $T^6$ and
flux
\eqna\choiceIIID
$$\eqalignno{{1\over(2\pi)^2\a'}F'_{(3)} & = a^0 dx^1\w dx^2\w dx^3
+ a\bigl(dx^1\w dx^2\w dy^3 + \hbox{cyc.\ perms.\ of 123}\bigr)&\cr
&\hskip-2em -b\bigl(dx^1\w dy^2\w dy^3 + \hbox{cyc.\ perms.\ of 123}\bigr)
+ b_0 dy^1\w dy^2\w dy^3,&\choiceIIID a\cr
{1\over(2\pi)^2\a'}H'_{(3)} & = c^0 dx^1\w dx^2\w dx^3
+ c\bigl(dx^1\w dx^2\w dy^3 + \hbox{cyc.\ perms.\ of 123}\bigr)&\cr
&\hskip-2em -d\bigr(dx^1\w dy^2\w dy^3 + \hbox{cyc.\ perms.\ of 123}\bigr)
+ d_0 dy^1\w dy^2\w dy^3.&\choiceIIID b}$$
This class of fluxes was analyzed in Sec.~4.1 of Ref.~\KST.  Let us
quickly review that analysis.  The (2,1) constraint on the complex
flux $G'_{(3)}$ implies that
\eqn\tauIIIDdiag{\t'^i{}_j = \t'\d^i{}_j,}
that is, $T^6\to T^2\times T^2\times T^2$ with respect to complex
structure, where $\t'$ and $\tdil'$ satisfy
\eqna\twooneIIID
$$\eqalignno{P_1(\t') &\equiv a^0\t'^3-3a\t'^2-3b\t'-b_0 = 0,
&\twooneIIID a\cr
P_2(\t') &\equiv c^0\t'^3-3c\t'^2-3d\t'-d_0 = 0,
&\twooneIIID b}$$
and
$$\bigl(a^0-\tdil' c^0\bigr)\t'^2 - 2(a-\tdil' c)\t'
-(b-\tdil' d) = 0. \eqno\twooneIIID c$$

A solution exists only if
\eqn\commonfactor{P_1(\t') = 2(f\t'+g)P(\t'),\quad
P_2(\t') = 2(h\t'+k)P(\t'),}
where $P(\t')$ is a quadratic polynomial of the form \quadraticP, and
$f,g,h,k\in\IZ$.  In Eq.~\commonfactor, we have added factors of 2
compared to the corresponding equation in Ref.~\KST, in order to
automatically take into account the $2\IZ$ quantization condition on
the fluxes.  The relations
\eqn\threeabcd{\eqalign{2(fm+gl)&=-3a,\cr 2(fn+gm)&=-3b,}\quad
\eqalign{2(hm+kl)&=-3c,\cr 2(hn+km)&=-3d,}}
give consistency conditions modulo 3, and the flux contribution to the
Gauss's law constraint \Othreebg{e} is
\eqn\NfluxIIID{N_\flux = {4\over3}(fk-gh)(m^2-4ln).}
It can be shown from Eq.~\threeabcd\ that $N_\flux$ is divisible by
12.  

Ref.~\KST\ went on to consider the particular example
\eqna\partEx
$$\bigl(a^0,a,b,b_0\bigr) = (2,0,0,2),\quad
\bigl(c^0,c,d,d_0\bigr) = (2,-2,-2,-4),\eqno\partEx a$$
$$(f,g,h,k,l,m,n) = (1,-1,1,2,1,1,1).\eqno\partEx b$$
In App.~H, we prove that this is the unique choice of supersymmetric
flux in the class \choiceIIID{a,b}\ with the minimum value $N_\flux =
12$ (and 2M = 20 D3 branes), modulo $SL(2,\IZ)_{\t'}
\times SL(2,\IZ)_{\tdil'}$ equivalences.\foot{This was suggested, but
not proven in Ref.~\KST.}  This choice gives
$$\t' = \tdil' = e^{2\pi i/3}.\eqno\partEx c$$
The only other value of $N_\flux\le32$ such that 12 divides $N_\flux$
is $N_\flux = 24$ (with 2M = 8 D3 branes).  By arguments analogous to
those in App.~H, one can show that in this case there are two distinct
choices of flux modulo equivalences: one choice differs from
Eqs.~\partEx{a,b,c}\ by $f,g,\tdil'\to 2f,2g,2\tdil'$, the other
differs by $h,k,\tdil'\to 2h,2k,\tdil'/2$.  So, in the entire class
\choiceIIID{a,b}, there are just three inequivalent choices of
supersymmetric flux.

The three possibilities are summarized by
\eqna\indepchoiceIIID
$$\eqalignno{{1\over(2\pi)^2\a'}F'_{(3)} &= 2f\bigl(dx^1\w dx^2\w dx^3
+ dy^1\w dy^2\w dy^3\bigr),&\indepchoiceIIID a\cr
{1\over(2\pi)^2\a'}H'_{(3)} &= 2h dx^1\w dx^2\w dx^3 - 2h
\bigl(dx^1\w dx^2\w dy^3 + \hbox{cyc.\ perms.\ of 123}\bigr)&\cr
&\hskip-2em +2h\bigl(dx^1\w dy^2\w dy^3 
+ \hbox{cyc.\ perms.\ of 123}\bigr)
-4h dy^1\w dy^2\w dy^3.&\indepchoiceIIID b}$$
$$N_\flux = 12fh,\qquad \t' = (h/f)\tdil' = e^{2\pi i/3},
\eqno\indepchoiceIIID c$$
where $(f,h) = (1,1)$, (2,1), or (1,2).  

The complex three-form flux is
\eqn\GIIID{G'_{(3)} = \bigl(2if e^{-2\pi i/3}/\sqrt{3}\bigr)
\bigl(dz^1\w dz^2\w d\bar z^3 + dz^1\w d\bar z^2\w dz^3
+ d\bar z^1\w dz^2\w dz^3\bigr).}
The primitivity condition \DthreesusyG\ implies that $J'$ is of the
form
\eqn\JIIID{\eqalign{J' &= {1\over2|\Im\t'|}\bigl(iV'_1 dz^1\w d\bar z^1
+ iV'_2 dz^2\w d\bar z^2 + iV'_3 dz^3\w d\bar z^3\cr
&\quad -2V'_{(12)}\Im(dz^1\w d\bar z^2)-2V'_{(23)}\Im(dz^2\w d\bar z^3)
- 2V'_{(31)}\Im(dz^3\w d\bar z^1)\bigr),}}
where $z^i = x^i+\t y^i$.  In real coordinates, this becomes
\eqn\JrealIIID{\eqalign{J' &= v'_1 dx^1\w dy^1 + v'_2 dx^2\w dy^2
+ v'_3 dx^3\w dy^3\cr &\quad + v'_{(12)}(dx^1\w dy^2 - dy^1\w dx^2)
+ (12\to23) + (12\to31),}}
where $v'_i=V'_i\sign\t'$ and $v'_{(ij)} = V'_{(ij)}\sign\t'$.  The
metric is
\eqn\metricIIID{ds_6{}^2 = {1\over|\Im\t'|}\Bigl(\sum_{i=1}^3 
V'_i dz^i d\bar z^i + \sum_{{{\scriptstyle i,j=1}\atop
{\scriptstyle i\ne j}}}^3 V'_{(ij)}
dz^i d\bar z^j\Bigr),}
where $V'_{(ij)}=V'_{(ji)}$.

From the kinetic terms \Ckintermp, the massless RR scalars are the
uncharged scalars
\eqn\RRmoduliIIID{\eqalign{&c'_{(4)x^2y^2x^3y^3},
\quad c'_{(4)x^3y^3x^1y^1}, \quad c'_{(4)x^1y^1x^2y^2},\quad
c'_{(4)x^1y^1x^2y^3} + c'_{(4)x^1y^1x^3y^2},\cr 
&\hskip2.5em c'_{(4)x^2y^2x^3y^1} + c'_{(4)x^2y^2x^1y^3},
\quad\hbox{and}\quad c'_{(4)x^3y^3x^1y^2} + c'_{(4)x^3y^3x^2y^1},}}
which are the axionic partners of $V'_1$, $V'_2$, $V'_3$, $V'_{(23)}$,
$V'_{(31)}$, and $V'_{(12)}$, respectively, in $\CN=1$ chiral
multiplets.

The massive gauge bosons are those that couple to the other RR
scalars.  From \Ckintermp, they lie in the 9D space spanned by
\eqn\gaugebosIIID{(f/h)b'_{(2)x^i\m}+ c'_{(2)x^i\m},\quad
b'_{(2)y^i\m},\quad{\rm and}\quad c'_{(2)x^i\m}+c'_{(2)y^i\m},
\quad i=1,2,3.}
In terms of the complex gauge bosons \cpxgaugebos, this is the space
spanned by
\eqn\cpxgaugebosIIID{\Im\bigl((1/\bar\t')d'_{z^i\m}\bigr),\quad
\Re\bigl(d'_{\bar z^i\m}\bigr),\quad{\rm and}\quad
\Im\bigl(d'_{\bar z^i\m}\bigr),\quad i=1,2,3.}
The massless bosons are the linear combinations orthogonal to this
space, where orthogonality is defined with respect to the metric
appearing in the gauge kinetic terms.  In the
$\bigl(d'_{z^i\m},d'_{\bar z^i\m},\bar d'_{z^i\m},\bar d'_{\bar
z^i\m}\bigr)$ basis, the latter is proportional to
\eqnn\gaugemetricIIID
$$\ha\pmatrix{0&1\cr1&0}\matrix{{\scriptstyle d'}\cr
{\scriptstyle\bar d'}}\quad\otimes\quad
\pmatrix{0&g'^{z^i\bar z^j}_6\cr g'^{\bar z^i z^j}_6&0}
\matrix{{\scriptstyle z^i}\cr{\scriptstyle\bar z^i}}.
\eqno\gaugemetricIIID$$
Therefore, the three massless gauge bosons are
\eqn\masslessgaugebosIIID{\Re\bigl((1/\bar\t')d'_{z^i\m}\bigr) \propto
(f/h) \bigl(b'_{(2)x^i\m}+ b'_{(2)y^i\m}\bigr) + c'_{(2)y^i\m},\quad
i=1,2,3.}

Finally, there are 6M moduli $\Phi_I$ and $M$ massless gauge bosons
$A_{I\,\m}$ from the D3 branes.

In summary, the massless fields are one graviton, $3+M$ vectors, and
$6+6M$ moduli.  The moduli consist of
$$\displaylines{\hbox{$V'_1$, $V'_2$, $V'_3$, $V'_{(23)}$,
$V'_{(31)}$, $V'_{(12)}$,}\cr
\hbox{6 $\tilde c_{(p)}$ scalars, and $6M$ D5 scalars.}}$$
These fields combine to form one 4D $\CN=1$ gravity multiplet, $3+3M$
chiral multiplets, and $3+M$ vector multiplets.

From the supersymmetry-breaking mass spectra given in
Ref.~\SUSYhiggs, this number of massless moduli and gauge
bosons corresponds to $\CN=4\to\CN=1$ with $m_1=m_2=m_3$ for the three
massive gravitini.  There are $3+3M$ massless chiral multiplets and
$3+M$ massless vector multiplets.  The metric on moduli space is
discussed in App.~G.  Since the complex structure modulus for each
$T^2$ factor satisfies $\Im\t'>0$, the 6D spinor $\chi'$ generating
the supersymmetries can be written as the product of three 2D spinors,
each of definite chirality along a $T^2$ (cf.~the discussion at the
end of Sec.~8.1).

\subsec{T-Duality Map}

If we return from the notation of Sec.~8.2 to the conventional
labeling of coordinates via \SecviiiNot, then it is straightforward to
show that the metric \metricIIID\ can be written as a flat $T^2$
fibration \dssixprime, with
\eqna\flatfibIIID
$$\eqalignno{V'_\fib &= V'_1, &\flatfibIIID a\cr
a'^4 &= V'^{-1}_1\bigl(V'_{(12)}dx^6 +
V'_{(31)}dx^8\bigr), &\flatfibIIID b\cr
a'^5 &= V'^{-1}_1\bigl(V'_{(12)}dx^7 +
V'_{(31)}dx^9\bigr), &\flatfibIIID c\cr
ds^2_\CB &= {1\over|\Im\t'|}\Bigl(\bigl(V'_2 - (V'_{(12)})^2/V_1\bigr)
dz^2 d\bar z^2 + \bigl(V'_3 - (V'_{(31)})^2/V_1\bigr) dz^3 d\bar z^3 &\cr
&\qquad + \bigl(V'_{23} - V'_{(12)}V'_{(31)}/V_1\bigr) 
\bigl(dz^2 d\bar z^3 + dz^3 d\bar z^2\bigr)\Bigr). &\flatfibIIID d}$$
The base metrics \basemIII\ and \flatfibIIID{d}\ agree, provided that
we make the identifications
\eqnn\mapIII
$$\openup1\jot\displaylines{V_2 = V'_2 - (V'_{(12)})^2/V'_1,\quad
V_3 = V'_3 - (V'_{(31)})^2/V'_1,\cr
V_{(23)} = V'_{(23)} - V'_{(12)}V'_{(31)}/V'_1,\quad
\t = \t'.\hfil\llap{\mapIII}\hfilneg}$$
All other quantities map exactly as described in Sec.~5.2.


\newsec{Preview: $\CN=2$ Calabi-Yau Duals without Flux\foot{I am
grateful to P.~Berglund, V.~Braun, B.~Florea, C.~Johnson,
R.~Reinbacher, N.~Warner, and C.~Vafa for comments and suggestions
related to this section.  In particular, P.~Berglund and N.~Warner
first suggested studying the lift to M theory.}}

So far, we have said nothing about $g_s$ or $\a'$ corrections.  We
know that there are at least some such corrections, due to the
following pathology in the leading $g_s,\a'$ supergravity description
of the orientifold.\foot{Here, leading order in $g_s$ means
``$0+\half$'' loop order, that is, tree level in the closed string
sector plus tree level in the open string sector.}  As $x\to x'$, the
Green's function
\Green\ behaves like a Green's function on $\IR^4$:
\eqn\Gnearsource{G_\CB(x,x')\sim {1\over8\pi^2}\biggl({1\over|x-x'|^2}
+ c\biggr),}
where $c$ is a constant.  So, near an O5 plane (with $Q_{\rm O5} =
-2$),
\eqn\ZnearOplane{Z(x)\sim 1-g_s\a'\biggl({1\over
|x-x^{\phantom{\prime}}_{\rm O5}|^2} + c\biggr).}
Consequently, as we let $x$ approach $x^{\phantom{\prime}}_{\rm O5}$,
we encounter a locus $Z=0$ where the metric is singular and $e^\phi =
g_sZ^{-1/2}$ diverges.  Beyond this locus, we enter the region $Z<0$,
where the metric and $e^\phi$ are imaginary.  A similar pathology
afflicts the leading supergravity description of all orientifolds.  On
the other hand, orientifold planes are not singular objects in string
theory.  Therefore, the 10D effective field theory descending from
string theory must get corrected in a way that resolves the pathology.
In the $\CN=2$ and $\CN=4$ cases, it is easy to understand this
resolution.

In the $\CN=2$ example of Sec.~6.2, there is an isometry in the $x^9$
direction (spontantanously broken by the positions of the eight O5
planes and $2M$ D5 branes).  If we T-dualize in this direction, we
obtain an O6 orientifold, with O6 planes and D6 branes wrapping the
$x^4,x^5,x^9$ directions.  Since there was no NS flux before the
T-duality, there is no new $S^1_{\{x^9\}}$ fibration introduced by the
T-duality.  The only flux after the T-duality is $F_{(2)}$, which is
(the pullback to 6D of) a two-form in the $T^3_{\{x^6,x^7,x^8\}}$
directions.  As noted in Ref.~\KSTT, this IIA background lifts to a
purely geometrical M theory background.  The $F_{(2)}$ flux gives the
fibration of the M theory circle $S^1_{\{x^{10}\}}$ over the IIA
geometry.  The dilaton gives the size of the fiber.  Finally, the D6
branes and O6 planes, the only objects that are singular in the IIA
supergravity description, lift locally to smooth Taub-NUT and smooth
Atiyah-Hitchin spaces times $\IR^{6,1}$, respectively \Sen.  (The
Atiyah-Hitchin space is the $\IZ_2$ quotient of negative mass Taub-NUT
space in which the region too close to the center has been smoothly
excised.  The excised region corresponds roughly to the unphysical
region $Z<0$ of the orientifold.)  Thus, the IIA orientifold lifts to
$$\hbox{M theory on }Y_7,\quad\hbox{ where }\quad Y_7 =
S^1_{\{x^9\}}\times\hbox{ smooth }Y_6.$$
There is no flux in M theory after the lift.  The manifold $Y_7$ is
truly a product, with no discrete identifications mixing the two
factors.  The orientifold $\IZ_2$ operation lifts to a geometrical
$\IZ_2$ that inverts $x^6,x^7,x^8,x^{10}$.  So, it goes into the
definition of $Y_6$, but does not act on $S^1_{\{x^9\}}$.  Since the M
theory compactification preserves 4D $\CN = 2$ supersymmetry, it
follows that
$$Y_6\hbox{ is a smooth Calabi-Yau threefold.}$$
Compactifying on $S^1_{\{x^9\}}$ then gives a standard $\CN=2$
Calabi-Yau compactification of IIA.  Compactifying on
$S^1_{\{x^{10}\}} \subset Y_6$ (with $Y_6$ viewed as a fibration)
gives the O6 orientifold.  Alternatively, F theory on $Y_6\times T^2$
gives the original O5 orientifold of Sec.~6.2, provided that we
identify the F theory torus with $S^1_{\{x^{10}\}} \subset Y_6$ times
$S^1\subset T^2$.  Since this F theory $T^2$~fibration has no global
section, there is automatically $F_{(3)}$ flux in the IIB orientifold
\WittenNPSup.

Since the geometry $\IR^{3,1}\times S^1\times Y_6$, with no flux,
solves the equations of motion of 11D supergravity, the only essential
ingredient that is missing in the leading IIA description but present
in the M theory dual description is the dependence of the 11D
background on the $x^{10}$ direction.  The 10D IIA supergravity theory
is the dimensional reduction of 11D supergravity truncated to lowest
Fourier modes along the $x^{10}$ circle.  The full Kaluza-Klein
reduction of 11D supergravity includes not only the IIA supergravity
fields, but also the entire tower of their D0-charged massive cousins
from the bound states of $N$ D0 branes, for all $N\in\IN$.  Away from
the O6 planes, if $g_s$ is tuned to be small, the D0 bound states are
heavy, and it is a good approximation to ignore the massive tower.  By
tuning $g_s$ smaller and smaller, this approximation becomes valid
closer and closer to the O6 planes.  However, $e^\phi$ always diverges
at some locus near the O6 planes.  As we approach this locus, the D0
bound states become massless and cannot be neglected.

It is natural to ask what class of Calabi-Yau threefolds arises from
the duality just described.  Immediately, we know at least one piece
of topological data.  In Sec.~6.2 we determined that there were
$n_V=2+M$ vector multiplets and $n_H=3+M$ hypermultiplets in addition
to the gravity multiplet in the massless spectrum.  On the other hand,
for Calabi-Yau compactifications of type IIA, $n_V = h^{1,1}$ and $n_H
= h^{2,1}+1$.  Therefore,
\eqn\hodgenos{h^{1,1}(Y_6) = h^{2,1}(Y_6) = 2+M,}
where the possible values of $M$ are $M=0$, 4, 8, and 12.  In the case
that $M=16$, there is $\CN=4$ supersymmetry and $Y_6$ is
$T^2_{\{x^4,x^5\}} \times K3_{\{x^6,x^7,x^8,x^10\}}$ \Kthreelift.

Beyond this, it is possible to determine the intersection numbers from
the special K\"ahler geometry of the moduli space \ModSpaceI{b}.
Furthermore, for each $M$, there is not one Calabi-Yau, but
topologically distinct geometries distinguished by the integers $m,n$
of Sec.~6 such that $4mn = 16-M$:
\eqn\hodgemn{\matrix{M & \qquad & (h^{1,1},h^{2,1})\hfill
& \qquad &\hfill(m,n)\phantom{,}\hfill\cr
\noalign{\vskip-5pt}
\hbox to 15pt {\hrulefill} &&\hbox to 50pt {\hrulefill}
&&\hbox to 95pt {\hrulefill}\cr
\noalign{\vskip3pt}
\hfill0\hfill && (2,2)\hfill && (4,1),\quad (2,2),\quad (1,4)\hfill\cr
\hfill4\hfill && (6,6)\hfill && (3,1),\quad (1,3)\hfill\cr
\hfill8\hfill && (10,10)\hfill && (2,1),\quad (1,2)\hfill\cr
\hfill12\hfill && (14,14)\hfill && (1,1)\hfill}}
The integer $n$ tell us about $H_1(Y_6,\IZ)$, which is pure torsion
(and subsequently about $\pi_1(Y_6)$, whose abelianization is
$H_1(Y_6,\IZ)$).  In the dual O3 orientifold of Sec.~6.4, S-duality
interchanges the integers $m$ and $n$.\foot{To be precise, $m$ and $n$
are interchanged under S-duality followed by the coordinate relabeling
$(x^8,x^9)_{\rm new} = (x^9,-x^8)_{\rm old}$.}  Therefore,
compactification of type IIA on $Y_6(m,n)$ and $Y_6(n,m)$ gives the
same low energy effective field theory.  Since both compactifications
are of type IIA, the relation between $Y_6(m,n)$ and $Y_6(n,m)$ is
something different from Calabi-Yau mirror symmetry.

The details of this duality will appear in a separate paper \CYduals.
There, we will derive the further topological data mentioned in the
previous paragraph.  In addition, from the standard identifications
between 11D supergravity and 10D type IIA supergravity, we will obtain
an approximate metric for $Y_6$, together with its corresponding
closed K\"ahler form and closed (3,0) form.\foot{The approximate
metric is Calabi-Yau on a noncompact space that excludes the singular
region $Z''\le0$, where $Z''$ is the warp factor of the $O6$
orientifold.  This metric is the lowest Fourier mode of the exact
Calabi-Yau metric on the compact manifold $Y_6$, in a Fourier
expansion along the $x^{10}$ direction.  It can in principle be
systematically extended to the exact Calabi-Yau metric by including
all higher Fourier modes, that is, by solving the 6D Einstein
equations order by order in an $x^{10}$ Fourier expansion.}

An added incentive for studying this duality is that it could help us
to compute other $g_s$ and $\a'$ corrections in the original
orientifold.  For corrections that map to worldsheet and D~instantons
in the Calabi-Yau duals, we can hope to take advantage of the existing
results for Calabi-Yau compactifications and then map the results back
to the orientifold.


\newsec{Conclusions and Outlook}

Type IIB orientifolds with wrapped O5 planes and internal flux yield a
class of 4D vacua that remains largely unexplored compared to its O3
and O7 counterparts.  We have studied the simplest such vacua here.
We have seen that these vacua, in which O5 planes and D5 branes wrap
the $T^2$ fiber over a $T^4$ base, provide a computable toy model for
studying moduli stabilization in a torsionful compactification.  The
supersymmetry conditions can be expressed in terms of a complex
three-form flux $\tilde G_{(3)}$ just as for D3/D7 or heterotic
compactifications, and in addition a complex one-form flux $G_{(1)}$.
This was illustrated in Secs.~6--8 through five different examples
preserving 4D $\CN=1$, 2, and 3 supersymmetry.  As a check, we have
also seen that the results of each example agree with the T-dual
analysis in the corresponding $T^6/\IZ_2$ orientifold.  By providing
an intrinsic description in the O5 orientifold, we have generalized
the results of Ref.~\KSTT, which relied solely on the T-duality map at
a special locus in moduli space.

One might have hoped that by giving a description of the O5
orientifold not grounded in T-duality, we would have found new
solutions that do not T-dualize to the $T^6/\IZ_2$ O3 orientifold.
However, as observed in Sec.~3.1, the condition for the T-duality to
fail to exist is that $H_{(3)\a\b m}$, with two fiber indices, is
nonzero, and this is exactly the case in which the analysis of the O5
orientifold qualitatively changes: the pseudo-BPS conditions of
Sec.~3.3 no longer hold; moreover, this component of flux can be
thought of as a transversely varying NS $B$-field oriented parallel to
the D5 brane worldvolumes, which gives rise to noncommutativity.  It
would be interesting to prove that $\CN>1$ supersymmetry demands that
$H_{(3)\a\b m} =0$.

There are a number of other questions left unanswered by this paper.
In Sec.~3.9, we stated the flux quantization conditions in the O5
orientifold, but offered no proof.  It is a significant omission that
we were not able to derive these conditions directly in the O5
orientifold.  The quantization conditions agree with those of the dual
O3 orientifold via Eqs.~\FluxbgMap{}.  However, it should be possible,
and would be more satisfactory, to instead derive the conditions
directly from the appropriate twisted K-theory analysis.\foot{I am
indebted to G.~Moore for a helpful discussion on flux quantization.}

In our description of the massless spectra, we were more explicit than
in Ref.~\KSTT\ for the O5 orientifold or Ref.~\KST\ for the dual O3
orientifold.  As a result, one feature that we noticed is that in all
examples, the partial supersymmetry breaking from $\CN=4$ to $\CN=1$,
2, or 3 is of the ``nongeneric'' type in which the massive gravitini
have the same mass.  A natural question is whether this is required.
Do there exist supersymmetric $T^6/\IZ_2$ flux vacua, with quantized
flux satisfying the Gauss's law constraint, in which the superHiggs
mechanism gives different masses to the gravitini?  Also, our analysis
of the massless spectrum proceeded on a case by case basis.  Is there
there a more elegant way of computing the massless spectrum through
the cohomology of a suitable operator?

Other than in the introduction, we mentioned $SU(3)$ torsion classes
in this paper only in Sec.~6.3, where the criterion $\CW_1\ne0$ was
used to prove that the almost complex structure of Ex.~3 was
nonintegrable.  We did not employ $SU(2)$ torsion classes at all.  The
reason is that the torsion classes simply were not essential for our
purposes of analyzing the supersymmetry conditions and moduli
stabilization in this particular class of O5 vacua.  Nevertheless, in
recent months there has been considerable effort devoted to developing
a framework that can be used to describe the most general 4D Minkowski
vacua of type IIA and IIB string theory.  (In the case of noncompact
internal geometry this is closely related to studying type II or M
theory solutions with an $AdS_5$ factor \GauntAdS.)  The correct
framework for this analysis is in terms of $SU(2)$ structures and
torsion classes \refs{\DallAgataIIA,\GauntAdS,\DallAgataIIB}.  For the
example of Sec.~6.3, in which $\tilde F_{(3)}$, $F_{(1)}$, and
$H_{(3)}$ are all nonzero, we were able to avoid talking about $SU(2)$
structures by instead working in terms of the right-$SU(3)$
structures, and the volume form on the fiber.  However, this is the
one example of the paper in which supersymmetry does not provide us
with a single 6D spinor, but two linearly independent spinors $\chi_R$
and $\chi_L = \g_\CB\chi_R$.  The two spinors canonically determine an
$SU(2)$ structure rather than an $SU(3)$ structure.  Therefore, this
example should provide a useful testing ground for the IIB
supersymmetry conditions as formulated in Ref.~\DallAgataIIB\ in terms
of $SU(2)$ torsion classes.  One just needs to compute the torsion
classes corresponding to the example.  (In the other examples, the
$SU(2)$ torsion collapses to $SU(3)$ torsion.)  Working in terms of
these torsion classes might have the added benefit of proving that the
almost complex structure is always nonintegrable for the subclass of
O5 backgrounds of the type discussed here when there is NS flux, a
result that we proved only for the specific example of Sec.~6.3.

Finally, perhaps the most interesting direction to pursue is to make
more precise the duality of Sec.~9 relating $\CN=2$ orientifold vacua
to standard type IIA Calabi-Yau vacua.  We have described a number of
computable features of the dual Calabi-Yau theefolds and are
optimistic about the prospects for identifying these manifolds.  As
one application of the identification, note that although it is common
to speak of connected webs of $\CN=2$ string vacua, given a pair of
$\CN=2$ vacua, it is not necessarily known whether the moduli spaces
of the two are connected.  For the class of $\CN=2$ Calabi-Yau vacua,
what is known is that the subspace of smooth Calabi-Yau threefolds
that are hypersurfaces in toric varieties is connected.  This subspace
has trivial fundamental group, and has been completely described by
Kreuzer and Skarke, who tabulated all 473,800,776 reflexive polyhedra
in four dimensions \KreuzerSkarke.  In at least the case that $(M,m,n)
= (12,1,1)$ and $h^{1,1}=h^{2,1}=14$, the Calabi-Yau $Y_6$ of Sec.~9
is conceivably contained in this subspace, and the dual orientifold
contained in the corresponding web of known connected $\CN=2$ vacua.
This intriguing duality between $\CN=2$ orientifold vacua and standard
type IIA Calabi-Yau vacua is currently under investigation~\CYduals.


\bigbreak\bigskip\bigskip\centerline{{\bf Acknowledgements}}\nobreak
It is a pleasure to thank Allan Adams, Per Berglund, Andreas
Brandhuber, Ram Brustein, Richard Corrado, Jerome Gauntlett, Jaume
Gomis, Fawad Hassan, Anton Kapustin, Albion Lawrence, Christian
R\"omelsberger, and especially Andrew Frey for helpful discussions and
useful references.  In addition, I thank S.~Kachru, P.~Tripathy, and
S.~Trivedi for the enjoyable collaboration from which this
investigation is a continuous outgrowth.  Finally, I am grateful to
Stockholm University, the Aspen Center for Physics, the Kavli
Institute for Theoretical Physics, the University of Pennsylvania, and
Harvard University for hospitality during the course of this work.
This work was supported in part by the DOE under contract
DE-FG03-92-ER40701 and by the National Science Foundation under Grant
No.~PHY99-07949.


\appendix{A}{Conventions}

The conventions for coordinate indices are as follows:
\smallskip

{\settabs\+\hskip1em&$i,j,\ldots$ and $\ibar,\jbar,\ldots$\quad&denote 
complex 6D internal or 4D base indices.\cr 
\+&$M,N,\ldots$&denote 10D spacetime indices in the range
$0,\ldots,9$,\cr
\+&$\m,\n,\ldots$&denote 4D spacetime indices in the range
$0,\ldots,3$,\cr
\+&$a,b,\ldots$&denote 6D internal indices in the range
$4,\ldots,9$,\cr
\+&$\a,\b,\ldots$&denote $T^2$ fiber indices in the range
$4,5$,\cr
\+&$m,n,\ldots$&denote 4D base indices in the range
$6,\ldots,9$,\cr
\+&$i,j,\ldots$ and $\ibar,\jbar,\ldots$&denote complex 6D internal 
indices in the range $1,2,3$.\cr}
\smallskip

\noindent When 10D tensors (such as $R_{MN}$ or the fluxes) are
decomposed into $\m,\a,m$ components, the decomposition is with
respect to the basis $dx^\m,\eta^\a,dx^m$, not $dx^\m,dx^\a,dx^m$,
with one exception.  In the context of the O3 orientifold, underscored
indices $\ul\m,\ul\a,\ul m$ denote components in the basis
$dx^\m,dx^\a,dx^m$.

The squares and partially contracted squares of 10D tensors are
defined by
\eqn\squaretens{\eqalign{A^2_{(p)} &= A_{(p)M_1\ldots M_p}
A_{(p)}{}^{M_1\ldots M_p},\cr
A^2_{(p)MN} &= A_{(p)M Q_2\ldots Q_p}
A_{(p)N}{}^{Q_2\ldots Q_p},}}
with the metric \dsstring{}\ used for raising and lowering indices.
This metric is also used to define the square of the fibration
curvature,
\eqn\squareF{\CF^2 = g_{\a\b}g^{mp}g^{nq}\CF^\a_{mn}\CF^\b_{pq}.}

We follow the differential geometry conventions of Nakahara \Nakahara\
for normalization of differential forms, exterior differentiation,
Hodge duality, and curvature.  (In the terminology of Misner, Thorne,
and Wheeler \MTW, we employ $+++$ sign conventions.)  In particular,
on the 6D internal manifold,
\eqna\signconv
$$\eqalignno{&\o\w *_6 \o = {1\over p!} \o_{a_1\ldots a_p}
\o^{a_1\ldots a_p} \Vol_6, &\signconv a \cr
&{1\over3!} J\w J\w J = {i\over8}\O\w\bar\O = \Vol_6.
&\signconv b}$$
Compared to earlier work \refs{\GKP,\KST,\KSTT}, the Hodge star
operator maps the same forms to one another, but
$\bigl(\Vol_6\bigr)_{\rm here}$ is minus $\bigl(\Vol_6\bigr)_{\rm
there}$.  Therefore, the two possible selfduality conditions on middle
dimensional forms,
\eqn\SDconds{\eqalign{*_6 \o_{(3)} &= +i\o_{(3)}\quad
\hbox{imaginary-selfduality (ISD)},\cr
*_6 \o_{(3)} &= -i\o_{(3)}\quad
\hbox{imaginary-antiselfduality (IASD)},}}
are the same here as in the earlier work \refs{\GKP,\KST,\KSTT}.

Our normalization conventions for 6D Hermitian metrics in complex
coordinates $z^i,\bar z^j$ follow by compatibility with the expression
for the metric in arbitrary 6D coordinates $x^{a,b}$:
\eqn\hermgconv{ds^2 = g_{ab}dx^a dx^b = g_{i\jbar}dz^i d\bar z^j
+ g_{\ibar j}d\bar z^i dz^j = 2g_{i\jbar} dz^i dz^\jbar.}
Note the factor of 2 on the RHS of the last equality.

The notation $\o_{(p)} = \o^{\bsk0}_{(p)}+\o^{\bsk1}_{(p)}
+\o^{\bsk2}_{(p)}$ is defined in Sec.~3.2, and denotes the
decomposition of an internal 6D $p$-form $\o_{(p)}$ into components
$\o^{\bsk i}_{(p)}$ of rank $i$ on the $T^2$ fiber.

Slashes denote contraction of tensors with 10D Dirac matrices,
\eqn\slashdef{\Asl_{(p)M_1\ldots M_k} = {1\over(p-k)!} A_{(p)M_1\ldots
M_k} \G^{M_{k+1}}\ldots\G^{M_p}.}
Antisymmetrized products of Dirac matrices are
\eqn\antiDirac{\eqalign{\G^{M_1\ldots M_n} &=
\G^{[M_1}\G^{M_2}\ldots\G^{M_n]},\cr
\g^{a_1\ldots a_n} &= \g^{[a_1}\g^{a_2}\ldots\g^{a_n]}.}}
The antisymmetrized product of all ten $\G^M$ gives the 10D chirality
operator,
\eqn\tenDchi{\G^{(10)} = -g^{1/2}\G^{0123456789},}
where $g_{MN}$ is the 10D metric \dsstring{}.  Uppercase Dirac
matrices satisfy
\eqn\uDirac{\{\G_M,\G_N\} = 2g_{MN}.}
Lowercase Dirac matrices satisfy
\eqn\lDirac{\{\g_\m,\g_\n\} = 2\eta_{\m\n},\quad \{\g_a,\g_b\} =
2g_{ab},} 
where $\eta = \diag(-1,1,1,1)$ and $g_{ab}$ is the 6D internal metric
\sixDg.

The relation between uppercase and lowercase Dirac matrices is
\eqn\ulDirac{\G_\m = Z^{-1/4}\g_\m\otimes1,\quad
\G_a = Z^{-1/4}\g^{(4)}\otimes\g_a,}
where
\eqn\gfour{\g^{(4)} = i\g^{0123}.}
Here, aside from the overall warp factor (cf.~Eq.~\tensix), we follow
Candeles, Horowitz, Strominger, and Witten \Candelas.  We also follow
the latter in employing a Majorana representation. The $\G_M,\g_\m$
are real and Hermitian, except for $\G_0,\g_0$, which are real and
antiHermitian.  The $\g_a$ are imaginary and Hermitian.

In addition, we define
\eqn\chiG{\eqalign{\G^{(4)} &= i Z^{-1}\G^{0123},\cr
\G^{(6)} &= i Z^{1/2} g_6^{1/2}\G^{456789},\cr
\G_\fib &= i Z^{-1/2}
\bigl(g^{(\Tfib)}\bigr)^{1/2}\G^{45},\cr
\G_\CB &= Z g_\CB^{1/2}\G^{6789},}}
and similarly,
\eqn\chig{\eqalign{\g^{(6)} &= i g_6^{1/2}\g^{456789},\cr
\g_\fib &= i
\bigl(g^{(\Tfib)}\bigr)^{1/2}\g^{45},\cr
\g_\CB &= Z^2 g_\CB^{1/2}\g^{6789}.}}
Since $\bigl(\g^{(4)}\bigr)^2 = 1$, we then have the relations
\eqn\chiG{\eqalign{&\G^{(4)} = \g^{(4)}\otimes1,\quad
\G_\fib = 1\otimes\g_\fib,\quad
{\rm and}\quad\G_\CB = 1\otimes\g_\CB,\cr
&\G^{(6)} = \G_\fib\G_\CB = 1\otimes\g^{(6)},
\hskip1.4em \G^{(10)} = \G^{(4)}\G^{(6)} = \g^{(4)}\otimes\g^{(6)},}}
as well as
\eqn\evenoddprod{\G^{a_1\ldots a_{2n}} = Z^{n/2}\,1\otimes
\g^{a_1\ldots a_{2n}},\quad
\G^{a_1\ldots a_{2n+1}} = Z^{(2n+1)/4}\,\g^{(4)}\otimes
\g^{a_1\ldots a_{2n}}.}

In our representation conventions, $\g^{(4)}$, $\g^{(6)}$, and
$\g_\fib$ are imaginary and Hermitian, while $\g_\CB$ is real and
Hermitian.  Thus, complex conjugation $\chi\to\chi^*$ of a 6D internal
spinor reverses its $\g^{(6)}$ and $\g_\fib$ chirality, but leaves its
$\g_\CB$ chirality unchanged.  Complex conjugation $u\to u^*$ of a 4D
spacetime spinor $u$ reverses its $\g^{(4)}$ chirality.

We can further decompose the lowercase Dirac matrices as
\eqn\gtdecomp{\g_\a = \tilde\g_\a\otimes1,\quad
\g_m = Z^{1/2}\tilde\g_\fib\otimes\tilde\g_m,}
where
\eqn\gtprod{\tilde\g_\fib = i\bigl(g^{(\Tfib)}\bigr)^{1/2}
\tilde\g^{45},\quad
\tilde\g_\CB = g_\CB^{1/2}
\tilde\g^{6789},}
and
\eqn\gtcom{\{\tilde\g_\a,\tilde\g_\b\} = 2g^{(\Tfib)}_{\a\b},
\quad\{\tilde\g_m,\tilde\g_n\} = 2g_{\CB\ssk mn}.}
Since $\g_\fib{}^2 = 1$, we then have the relations
\eqn\chiGg{\g_\fib = \tilde\g_\fib\otimes1,
\quad\g_\CB = 1\otimes\tilde\g_\CB,}
and finally
\eqn\gtprod{\g_\a{}^\b = \tilde\g_\a{}^\b\otimes1,
\quad\g_m{}^n = 1\otimes\tilde\g_m{}^n,\quad
\g_{\a mn} = Z\tilde\g_\a\otimes\tilde\g_{mn},}
which are needed in Sec.~4.2.

The convention for $\pm$ subscripts on 6D, 4D, and 2D spinors is that
the subscripts always indicate the fiber chirality and/or minus the
base chirality.  Therefore, the only 6D spinors that we refer to
directly are of negative $\g^{(6)}$ chirality.


\appendix{B}{Coordinate Identifications}

The fiber coordinates are periodically identified via
\eqn\fiberid{x^\a\cong x^\a+1
\quad\hbox{at fixed }\{x^p,x^\b\}_{\b\ne\a}.}
In the case that the base $\CB$ is a torus, Eq.~\Chern\ implies that
\eqn\fibgauge{\CA^\a = \half \CF^\a_{mn} x^mdx^n,\quad
\CF^\a_{mn} = {\rm const} \in \IZ,}
up to a fiber coordinate redefinition
\eqn\fibtransf{ x^\a_{\rm new} = x^\a+\L^\a,\quad
\CA^\a_{\rm new} = \CA^\a-d\L^\a,\quad\hbox{where }\L^\a =
\L^\a\bigl(\{x^m\}\bigr).}
In the gauge \fibgauge, the base coordinates have a periodic
identification that also acts on the fiber coordinates:
\eqn\baseid{(x^m,x^\a)\equiv
(x^m+1,x^\a-\half\CF^\a_{mn}x^n)\quad
\hbox{at fixed }\{x^p,x^\b\}_{(p,\b)\ne(m,\a)}.} 
This twisted identification is the unique coordinate identification
that both projects to $x^m\equiv x^m+1$ on the base and is compatible
with
\eqn\etatwoways{\eta^\a = (dx^\a+\CA^\a)\big|_{x^m}
= (dx^\a+\CA^\a)|_{x^m+1}\quad\hbox{at fixed }
\{x^p,x^\b\}_{(p,\b)\ne(n,\a)}.}
Other gauge choices lead to similar coordinate identifications.


\appendix{C}{Derivation of Pseudo-BPS Constraints}

Let
\eqn\tris{\eqalign{\Delta_1 &= - g^{\m\n} \bigl(R_{\m\n}+
2\nabla_\m\nabla_\n \phi\bigr),\cr
\Delta_2 &= - g^{\a\b} \bigl(R_{\a\b}+
2\nabla_\a\nabla_\b \phi\bigr) + \quarter \CF^2,\cr
\Delta_3 &= - \nabla^2\phi + 2(\nabla\phi)^2,}}
and define
\eqn\sumtri{\Delta = \Delta_1 + \Delta_2 + \Delta_3.}
The warp factor $Z$ and dilaton $\phi$ are assumed to depend only on
the base coordinates.  For such a function $f$, we have
\eqn\deldel{\eqalign{\nabla_\m\nabla_\n f
& = Z^{-1}g_\CB^{mm'}(\log Z),_{m'} \eta_{\m\n}\cr
\nabla_\a\nabla_\b f
& = Z^{-1}g_\CB^{mm'}(\log Z),_{m'} \d_{\a\b}\cr
\nabla_m\nabla_n f &= \nabla_{\mkern-2mu \CB\ssk m}
\nabla_{\mkern-2mu \CB\ssk n} -\quarter Z^{-1}
\Bigl(-Z,_{k'} g_\CB^{k'k} g_{\CB\ssk mn}
+ Z,_m \d^k_n + Z,_n\d^k_m\Bigr) f,_k.}}
Using Eqs.~\Ricci\ and \deldel, it is straightfoward to show that
\eqn\triZ{\eqalign{\nabla_\CB\cdot
\Bigl(\bigl(Z^{-1}e^{-2\phi}\bigr)
\nabla_\CB \bigl(Z^{-3/2}e^{-\phi}\bigr)\Bigr)
&=Z^{1/2} \bigl(Z^{-1}e^{-2\phi}\bigr)\bigl(Z^{-3/2}e^{-\phi}\bigr)
\Delta\cr
&\mkern-20mu +\bigl(Z^{-1}e^{-2\phi}\bigr)\bigl(Z^{3/2}e^\phi\bigr)
\Bigl(\nabla_\CB\bigl(Z^{-3/2}e^{-\phi}\bigr)\Bigr)^2,}}
where contractions are performed using base metric $g_{\CB\ssk
mn}$.  On the other hand, from Eqs.~\trrev\ and \stressT, we
have\foot{Here, $\tilde F_{(5)} = (1-*)\tilde F^{\rm int}_{(5)}$,
where $\tilde F^{\rm int}_{(5)} = \sum \tilde F^{\bsk i}_{(5)}$
denotes the purely internal part of $\tilde F_{(5)}$, with no 4D
spacetime indices.  Note that this implies a relative factor of 2 in
$\half\smallfrac{1}{5!} \bigl(\tilde F^{(5)}\bigr)^2 =
\smallfrac{1}{5!} \bigl(\tilde F^{\rm int}_{(5)}\bigr)^2$.}
\def\triStrut{{\vphantom{\bigl(\bigl)}}}
\eqn\triF{\eqalign{\Delta_1 &= e^{2\phi}
\Bigl(F_{(1)}{}^2 + \smallfrac{1}{3!}\tilde F_{(3)}{}^2
+\half\smallfrac{1}{5!}\tilde F_{(5)}{}^2\Bigr)
+ (2\pi)^2\a'e^\phi\sum Q_i
{\d^4(x-x_i)\over Z\sqrt{g_\CB}}, \cr
\Delta_2 - \half\Delta_1 & =  
-\half e^{2\phi}
\Bigl(\smallfrac{1}{3!}(\tilde F_{(3)}^{\bsk1})\triStrut^2
+ \smallfrac{1}{5!}(\tilde F_{(5)}^{\bsk1})\triStrut^2
+ \smallfrac{2}{5!}(\tilde F_{(5)}^{\bsk2})\triStrut^2\Bigr)
+ \quarter \CF\,\triStrut^2, \cr
-\Delta_3 &= e^{2\phi}
\Bigl(F_{(1)}{}^2 + \half\smallfrac{1}{3!}\tilde F_{(3)}{}^2
-\half\smallfrac{1}{3!}H_{(3)}{}^2\Bigr)
+ \half (2\pi)^2\a'e^\phi\sum Q_i{\d^4(x-x_i)\over Z\sqrt{g_\CB}}.}}
Therefore,
\eqn\triFsum{\eqalign{\Delta &= \half e^{2\phi}
\Bigl(F_{(1)}{}^2
+ \smallfrac{2}{3!}(\tilde F_{(3)}^{\bsk0})\triStrut^2
+ \smallfrac{1}{3!}(\tilde F_{(3)}^{\bsk1})\triStrut^2
+ \smallfrac{2}{5!}(\tilde F_{(5)}^{\bsk1})\triStrut^2
+ \smallfrac{1}{5!}(\tilde F_{(5)}^{\bsk2})\triStrut^2\Bigr)\cr
&\quad+\half\smallfrac{1}{3!}(H_{(3)}^{\bsk0})\triStrut^2
+ \half\smallfrac{1}{2!}\CF\,\triStrut^2
+ (2\pi)^2\a'e^\phi\sum Q_i{\d^4(x-x_i)\over Z\sqrt{g_\CB}}.}}
By substituting Eq.~\triFsum\ into Eq.~\triZ\ and adding to the result
$Z^{-3}e^{-2\phi}$ times Eq.~\newBianchi{a}, we obtain the desired
pseudo-BPS condition Eq.~\BPSlike.

In writing Eqs.~\triF\ and \triFsum, we have assumed that
$H_{(3)}^{\bsk2} = 0$.  If this is not the case, then there is an
additional term
\eqn\dtritwo{\d\bigl(\Delta_2-\half\Delta_1\bigr) 
= -e^{2\phi}\smallfrac{1}{3!} (\tilde F_{(3)}^{\bsk2})\triStrut^2
-\smallfrac{1}{5!} (H_{(3)}^{\bsk2})\triStrut^2}
on the RHS of the second equation \triF, and additional terms
\eqn\dtri{\d\Delta = \smallfrac{3}{2} e^{2\phi}
\smallfrac{1}{3!}(\tilde F_{(3)}^{\bsk2})\triStrut^2
-\half\smallfrac{1}{3!}(H_{(3)}^{\bsk2})\triStrut^2}
on the RHS of Eq.~\triFsum.  

Here $\tilde F_{(3)}^{\bsk2} = -C_{0}H_{(3)}^{\bsk2}$, where $C_{(0)}$
is a local section of a $U(1)$ bundle (such that $F_{(1)}=dC_{(0)}$)
and cannot be written as a single-valued function.  Since $\tilde
F_{(3)}^{\bsk2}$ and $H_{(3)}^{\bsk2}$ are gauge-invariant
globally-defined three-forms, this implies that $F_{(1)}$ and
$H_{(3)}^{\bsk2}$ cannot simultaneously be nonzero. Therefore,
$F_{(3)}^{\bsk2} = 0$.  However, even when $F_{(1)} = 0$ and
$H_{(3)}^{\bsk2}\ne0$, the second term in Eq.~\dtri\ still presents a
problem; since it is negative, we obtain an additional negative term
on the RHS of Eq.~\BPSlike, and can no longer deduce the Hodge duality
relations \BPScond{}.  Therefore, we restrict to the case
$H_{(3)}^{\bsk2}=0$, in which we know how to proceed.


\appendix{D}{IIB Fermion Variations}

The string frame fermion variations to linear order in the spinors are
\eqn\IIBpmfermvar{\eqalign{%
\d_{L,R}\l_{L,R} & = \half\Bigl(\dsl\phi\mp
\half\Hsl_{(3)}\Bigr)\e_{L,R},\cr
\d_{R,L}\l_{L,R} & = \half e^\phi 
\Bigl(\pm\Fsl_{(1)} + \half
\tilde \Fsl_{(3)}\Bigr)\e_{R,L},\cr
\d_{L,R}\psi_{L,R\mkern3mu M} & = \Bigl(\nabla_M 
\mp\smallfrac{1}{4}\Hsl_{(3)M}\Bigr)\e_{L,R},\cr
\d_{R,L}\psi_{L,R\mkern3mu M} & = \eighth e^\phi\Bigl(\!\!
\mp\Fsl_{(1)} - \tilde\Fsl_{(3)}
\mp\half\tilde\Fsl_{(5)}\Bigr)\G_M\e_{R,L}.}}
Here, subscripts to the left (right) of a comma are correlated with
upper (lower) sign choices.  The subscripts $L$ and $R$ indicate
whether the spin content of a fermion comes from the left-moving or
right-moving sector of the worldsheet conformal field theory.  In this
appendix, we follow Hassan \Hassan, except for a sign difference in
the chirality of all of the 10D spinors, due to opposite conventions
for $\G^{(10)}$.  We take
\eqn\spinorchi{\G^{(10)}\e_{L,R} = -\e_{L,R},\quad
\G^{(10)}\l = +\l,\quad{\rm and}\quad \G^{(10)}\psi_{L,R\mkern3mu M} =
-\psi_{L,R\mkern3mu M},}
where $\G^{(10)}$ is given by Eq.~\tenDchi.  Slashes denote
contraction with Dirac matrices as defined in Eq.~\slashdef.  Also,
\eqn\Dspinor{\nabla_M \e_{L,R} = \bigl(\partial_M 
+ \smallfrac{1}{4}w_{Mab}\G^{ab}\bigr)\e_{L,R},}
where $w_{Mab}$ is the spin connection, and
\eqn\RRpot{\tilde F_{(n)M_1\ldots M_n}
= n\partial_{[M_1} C_{(n-1)M_2\ldots M_n]}
-{n!\over3!(n-3)!}H_{(3)[M_1M_2M_3} C_{(n-3)M_4\ldots M_n]}.}
The last equation means that
\eqn\dFRR{\eqalign{\tilde F_{(n)}
& = F_{(n)} - H_{(3)}\w C_{(n-3)},\cr
d\tilde F_{(n)} & = H_{(3)}\w F_{(n-2)} + \hbox{local sources}.}}

By defining the complex quantities
\eqn\complexquantities{\d = \d_L + \d_R,\quad
\e = \e_L+i\e_R,\quad \l = \l_L+i\l_R,\quad{\rm and}\quad
\psi_M = \bigl(\psi_L+i\psi_R\bigr)_M,}
we can equivalently write the fermion variations as
\eqn\IIBfermvar{\eqalign{\d\l
& = \smallfrac{1}{2}\bigl(i\dsl\phi 
- e^\phi \Fsl_{(1)}\bigr)\e^* + 
\smallfrac{1}{4}\bigl(e^\phi \tilde\Fsl_{(3)}
-i \Hsl_{(3)}\bigr)\e,\cr
\d\psi_M & = \Bigl(\nabla_M + \smallfrac{i}{8}e^\phi
\bigl(\Fsl_{(1)}\G_M+\half\Fsl_{(5)}\G_M\bigr)\Bigr)\e
-\smallfrac{i}{8}\Bigl(e^\phi\tilde\Fsl_{(3)}\G_M 
-2i\Hsl_{(3)M}\Bigr)\e^*.}}
In terms of the axion-dilaton and complex flux,\foot{Here, we have
given the axion-dilaton for type IIB supergravity with no orientifold
projection.  This is also the axion-dilaton for orientifolds with O3
or O7 planes and Becker-type spinor constraints.  However, as
discussed in Sec.~4.2, a different quantity plays the role of the
axion-dilaton for O5 orientifolds, since the zero mode of $C_{(0)}$ is
projected out.  For the type I orientifold with O9 planes, $\tdil = a
+ ie^{-\phi}$, where $a$ is the scalar dual to $C_{(2)\m\n}$.}
\eqn\tauG{\eqalign{\tdil & = C_{(0)}+ie^{-\phi},\cr
G_{(3)} & = \tilde F_{(3)} - ie^{-\phi}H_{(3)} 
= F_{(3)}-\tau H_{(3)},}}
this becomes
\eqn\IIBfermvarsimp{\eqalign{e^{-\phi}\d\l
& = -\smallfrac{1}{2}\bigl(\dsl\tdil\bigr)\e^*
+ \smallfrac{1}{4} \Gsl_{(3)}\e\cr
e^{-\phi}(\d\psi_M -\smallfrac{i}{4}\G_M\d\l^*)
& = \Bigl(e^{-\phi}\bigl(\nabla_M
-\eighth \G_M\dsl\phi\bigr)
+ \smallfrac{i}{4}F_{(1)M} 
+ \smallfrac{i}{16}\tilde\Fsl_{(5)}\G_M\Bigr)\e\cr
& \qquad - i\Bigl(\eighth \Gsl_{(3)}\G_M 
+ \smallfrac{1}{16}\G_M \Gsl_{(3)}\Bigr)\e^*.}}
Here, we have assumed a real representation of the Dirac matrices.  We
have also used the relation
\eqn\HHM{\{\Hsl_{(3)},\G_M\} = 2\Hsl_{(3)M}} 
in deriving \IIBfermvarsimp.

When expressed in terms of the complex quantities $\tdil$ and
$G_{(3)}$, the equations simplify in Einstein frame.  Using a prime to
indicate the latter, the relations between string frame and Einstein
frame are
\eqn\StoE{\eqalign{g'_{MN} & = e^{-\phi/2}g_{MN},\cr
\e' &= e^{\phi/8}\e,}\quad
\eqalign{\G'_M &= e^{-\phi/4}\G_M,\cr
\l' &= e^{\phi/8}\l,}\quad
\eqalign{\nabla'_M &= \nabla_M - \eighth \G_M{}^N
\partial_N\phi,\cr
\psi'_M &= e^{-\phi/8}(\psi_M - \smallfrac{i}{4}\G_M\l^*).}}
Therefore, the Einstein frame analog of Eqs.~\IIBfermvarsimp\ is
\eqn\IIBfermvarE{\eqalign{\d\l' 
& = -\half e^{\phi}\bigl(\dsl'\tdil\bigr)\e'^*
+ \smallfrac{1}{4} e^{\phi/2}\Gsl'_{(3)}\e',\cr
\d\psi'_M & = \Bigl(\nabla'_M
+ \smallfrac{i}{4}e^{\phi}F_{(1)M} 
+ \smallfrac{i}{16}\tilde\Fsl'_{(5)}\G'_M\Bigr)\e'
- ie^{\phi/2}\Bigl(\eighth \Gsl'_{(3)}\G'_M 
+ \smallfrac{1}{16}\G'_M \Gsl'_{(3)}\Bigr)\e'^*,}}
where a slash and prime denotes the analog of Eq.~\slashdef\ with
$\G_M\to\G'_M$.

Eqs.~\IIBfermvarE\ give the Einstein frame fermion variations in the
usual string theory conventions, in which $F_{(p)}$, $H_{(3)}$, and
$G_{(3)}$ are the same in string frame and Einstein frame.  In the
supergravity literature, the quantity $G_{(3)}$ is given a slightly
different definition.  The relation between that definition and the
one given in Eqs.~\tauG\ has been discussed elsewhere
\refs{\GP,\Hassan}.


\appendix{E}{The Meaning of $W_{\rm GVW}$ in the O3 Orientifold with
Internal $T^6$}

As mentioned in Sec.~5.1, one subtlety in our description of metric
moduli stabilization for the O3 orientifold based on internal $T^6$ is
that not all of the (18 real) $\t^i{}_j$ and (9 real) $g_{i\jbar}$
correspond to the (21 real) physical metric moduli.

In contrast to the case for a proper Calabi-Yau threefold, there is
not a one-to-one correspondence
\eqn\cpxvsg{\eqalign{&\d\t^i{}_j D_{\t^i{}_j}\O = \o^{(1,2)}
\quad\hbox{(sum on $i,j$),}\cr
&\d g_{kl} = \O^{\vphantom{()}}_{(k|}{}^{\bar k\bar l}
\o^{(1,2)}_{\bar k\bar l|l)},}}
between deformations of complex structure ((1,2) forms or
$T^{(0,1)}$-valued (1,0) forms) and deformations of $g_{ij}$.  The
reason is that for a manifold such as $T^6$ with nontrivial
$H^{(0,1)}$, there exist nonprimitive (1,2) forms
\eqn\nptwoone{\o^{(1,2)} = J\w\o^{(0,1)},}
that generate vanishing metric deformations.  For a $T^6$, the three
such forms are $J\w d\bar z^i$.  Correspondingly, there are three
complex unphysical degrees of freedom in $\t^i{}_j$ that do not
correspond to metric degrees of freedom.

This can be understood in the supergravity theory as follows.  In the
absence of flux, the theory has $\CN=4$ supersymmetry.  The fluxes
break this to $\CN<4$ supersymmetry, but the breaking should still be
described within the formalism of $\CN=4$ gauged supergravity.  Unlike
global $\CN=4$ supersymmetry, $\CN\ge1$ supergravity cannot be cleanly
described in $\CN=1$ language, due to the lack of a simple off-shell
formulation for $\CN=1$ gravitino multiplets with a single auxiliary
field analogous to the $D$ or $F$ fields for vector or chiral
multiplets.\foot{Here, we have in mind an off-shell multiplet whose
on-shell dynamical degrees of freedom are a spin $3/2$ field and a
spin 1 field.  For work on off-shell gravitino multiplets, see
Ref.~\Gates.}  Nevertheless, upon choosing an $\CN=1$ subalgebra of
$\CN=4$, one can still organize all fields into $\CN=1$ multiplets and
decompose the bosonic parts of all supersymmetry variations into
$D$-terms, $F$-terms, and ``other-terms.'' In this sense, we can still
define an $\CN=1$ superpotential, whose variation gives the $F$-terms.
However, this superpotential is part of a family of superpotentials
parametrized by a choice of the $\CN=1$ subalgebra.  Equivalently, it
is parametrized by a choice of a point in the space
\eqn\Yspace{Y = {SU(4)_\CR\over U(1)_\CR\times SU(3)},\quad
{\rm dim}_\IC Y = 3,}
of embeddings of the $U(1)$ $\CR$-symmetry of $\CN=1$ in the $SU(4)$
$\CR$-symmetry of $\CN=4$.  Here, the $SU(3)$ is the commutant of
$U(1)_\CR$ in $SU(4)_\CR$.  For $T^6$, $W_{\rm GVW}$ is exactly this
type of family of superpotentials, parametrized by the three complex
unphysical degrees of freedom in $\t^i{}_j$.  To obtain the $F$-terms,
$W_{\rm GVW}$ should only be varied with respect to the physical
degrees of freedom.  The $F$-terms impose the conditions
\eqn\physFterms{G_{(3)} = \hbox{(1,2)-nonprimitive} + (2,1),}
where the (1,2)-nonprimitive component is of the form \nptwoone.

On the other hand, in an operational sense, varying $W_{\rm GVW}$ with
respect to $\tdil$ and all of the $\t^i{}_j$, including unphysical
degrees of freedom, is a convenient thing to do: it is more easily
implemented in practice than varying with respect to only physical
deformations, and this seeming unphysical procedure imposes exactly
the (2,1) condition on $G_{(3)}$, which is still a subset of the
supersymmetry conditions.  The philosophy of Ref.~\KST\ was simply to
use $W_{\rm GVW}$ in this seemingly unphysical way, as a convenient
tool for imposing the (2,1) condition.

One disadvantage of this approach is that it is not always clear how
many physical moduli there are among the redundent set of all
$\t^i{}_j$ and $g_{i\jbar}$ unfixed by the supersymmetry conditions.
For the examples that we present, this is an issue only in the $\CN=2$
case.  (For our $\CN = 1$ and $\CN=3$ examples, $\tdil$ and all
complex structure moduli are fixed, and all of the unfixed K\"ahler
moduli are physical.)  We avoid the problem by choosing a noncanonical
decomposition of unfixed metric moduli into K\"ahler and complex
structure moduli in Sec.~6.4.  The choice is natural from the point of
view of the dual O5 orientifold of Sec.~6.2.  The choice corresponds
exactly to the choice of $\chi_+=0$ or $\chi_-=0$ in Sec~4.2, which
leads to an integrable complex structure possessing a fiber $\oplus$
base decomposition as in Eq.~\OJpm.


\appendix{F}{T-Duality Map for RR Potentials}

The T-duality map for the RR potentials $C_{(p)}$ is analogous to
Eq.~\RRprime.  The relations between components in the $\eta^\a,dx^m$
basis of the O5 orientifold and the $\eta'^\a,dx^m$ basis of the O3
orientifold are
\eqn\RRcmap{\eqalign{C^{\bsk0}_{(p)} &=
\half\e^{\a\b}C'^{\bsk2}_{(p+2)\a\b}/\bigl((2\pi)^2\a'\bigr),\cr
C^{\bsk1}_{(p)} &= -\e^{\a\b}C'^{\bsk1}_{(p)\b},\cr
C^{\bsk2}_{(p)\a\b} &= -\e^{\a\b}C'^{\bsk0}_{(p-2)} (2\pi)^2\a'.}}
We assume that $H'_{(3)\ul{\a\d m}}=0$ so that the T-duality exists,
but retain $F'_{(3)\ul{mnr}}$.  (The equations of motion ultimately
require that $F'_{(3)\ul{mnr}}=0$.)

In the O3 orientifold, a choice of gauge for the potentials leading to
the quantized flux \OthreeFlux\ is
\eqn\BCpgauge{\eqalign{B'^\bg_{(2)} &= \half H'_{(3)\ul{\a nr}} x^r
dx^\a\w dx^n  + \smallfrac{1}{3!} H'_{(3)\ul{mnr}} x^r dx^m\w dx^n,\cr
C'^\bg_{(2)} &= \half F'_{(3)\ul{\a\d r}}x^r
dx^\a\w dx^\d + \half F'_{(3)\ul{\a nr}} x^r dx^\a\w dx^n +
\smallfrac{1}{3!} F'_{(3)\ul{mnr}} x^r dx^m\w dx^n.\cr}}
In the $\eta'^\a,dx^m$ basis (where $\eta'^\a = dx^\a+a'^\a{}_m$),
this becomes
\eqn\BCpgaugeeta{\eqalign{B'^\bg_{(2)} &= \half H'_{(3)\ul{\a nr}} x^r
\eta'^\a\w dx^n  + \bigl(\smallfrac{1}{3!} H'_{(3)\ul{mnr}} -\half
H'_{(3)\ul{\a n r}}a'^\a{}_m\bigr) x^r dx^m\w dx^n,\cr
C'^\bg_{(2)} &= \half F'_{(3)\ul{\a\d r}}x^r
\eta'^\a\w\eta'^\d + \bigl(\half F'_{(3)\ul{\a nr}} -F'_{(3)\ul{\a\d
r}}a'^\d{}_n\bigr) x^r \eta'^\a\w dx^n\cr
& + \bigl(\smallfrac{1}{3!} F'_{(3)\ul{mnr}} -\half F'_{(3)\ul{\a nr}}
a'^\a{}_m + \half F'_{(3)\ul{\a\d r}}\a'^\a{}_m a'^\d{}_n\bigr)
x^r dx^m\w dx^n.\cr}}
In addition, there are background components of $C^\bg_{(4)}$ with one
and two $\eta'^\a$ indices, such that
\eqn\Cfourbg{\tilde F'_{(5)\,\rm{smeared}} = dC'^\bg_{(4)} -
C'^\bg_{(2)}\w H'_{(3)},}
with $\tilde F'_{(5)}$ given by Eq.~\Ffivesmeared.

The moduli are $c'_{(0)}$ and $c'_{(4)} = C'^{\vphantom{\bg}}_{(4)} -
C'^\bg_{(4)}$.  The latter are given by
\eqn\Cfourmoduli{\eqalign{c'_{(4)} &=\smallfrac{1}{2!2!}
c'_{(4)\ul{\a\d rs}} dx^\a\w dx^\d\w dx^r\w dx^s
+\smallfrac{1}{3!}c'_{(4)\ul{\a nrs}} dx^\a\w dx^n\w dx^r\w dx^s\cr
&+\smallfrac{1}{4!}c'_{(4)mnrs}dx^m \w dx^n\w dx^r\w dx^s.}}
In the $\eta'^\a,dx^m$ basis, this becomes
\eqn\Cfourmodulieta{\eqalign{c'_{(4)} &= \smallfrac{1}{2!2!}
c'_{(4)\ul{\a\d rs}} \eta'^\a\w \eta'^\d\w dx^r\w dx^s\cr
& +\smallfrac{1}{3!}\bigl(c'_{(4)\ul{\a nrs}} -\half c'_{(4)\ul{\a\d
rs}}a'^\d{}_n\bigr) \eta'^\a\w dx^n\w dx^r\w dx^s\cr
&+ \bigl(\smallfrac{1}{4!}c'_{(4)mnrs}
-\smallfrac{1}{3!}c'_{(4)\ul{\a nrs}}a'^\a{}_n
+ \smallfrac{1}{2!2!}c'_{(4)\ul{\a\d rs}}a'^\a{}_m a'^\b{}_n\bigr)
dx^m \w dx^n\w dx^r\w dx^s.}}
Applying the T-duality map \RRcmap, replacing $a'_{(2)}$ with
$-b_{(2)}$ via Eq.~\BGrels{a}, and at the same time using the flux
relations \FluxbgMap{}, we obtain
\eqn\Cfinal{\eqalign{C_{(0)} &= F_{(1)m} x^m,\cr
C_{(2)\,45} &= \tilde c_{(2)\,45},\cr
C_{(2)\a n} &= \half F^\bg_{(3)\a nr}x^r + b_{\a n} C_{(0)},\cr
C_{(2)mn} &= C^\bg_{(2)mn} + \tilde c_{(2)mn},\cr
C_{(4)\a nrs} &= C^\bg_{(4)\a nrs} + \tilde c_{(4)\a nrs} -
\bigl(b_{(2)}\w \tilde c_{(2)}\bigr)_{\a nrs},\cr
C_{(4)\,45mn} &= F^\bg_{(5)\,45mnr}x^r + \bigl(b_{(2)}\w\half
F_{(3)\,5qp}x^p dx^q\bigr)_{45mn} + \half\bigl(b_{(2)}\w 
b_{(2)}\,C_{(0)}\bigr)_{45mn},\cr
C_{(6)\,45mnrs} &= \tilde c_{(6)\,45mnrs} 
+ \bigl(b_{(2)}\w\tilde c_{(4)}\bigr)_{45mnrs} 
+ \half\bigl(b_{(2)}\w b_{(2)}\w\tilde c_{(2)}\bigr)_{45mnrs},}}
where
\eqn\Cbgfinal{C^\bg_{(2)rs} = C'^\bg_{(4)\,45rs},\quad
C^\bg_{(4)\a nrs} = -\e^{\a\b}C'^\bg_{(4)\a nrs},}
and where the T-duality map of RR moduli is
\eqn\ctildefinal{\eqalign{& \tilde c_{(2)\,\ul{45}}
= - c'_{(0)},\quad \tilde c_{(2)rs} = c'_{(4)\,\ul{45rs}},\cr
&\tilde c_{(4)\a nrs} = -\e^{\a\b}c'_{(4)\ul{\b nrs}},\quad
\tilde c_{(6)\,456789} = c'_{(4)\,\ul{6789}}.}}
Eq.~\Cfinal\ is of the same form as Eqs.~\CFdefs\ and \ctilde{},
with the gauge choice
\eqn\Cgauge{C_{(0)} = F_{(1)m}x^m,\quad
C^\bg_{(2)\a n} = \half F^\bg_{(3)\a nr}x^r, \quad
C^\bg_{(4)\,45mn} = F^\bg_{(5)\,45mnr}x^r.}
Similarly, from Eqs.~\BGrels{b,c}, the gauge choices for $\CA^\a$ and
$B^\bg_{(2)}$ resulting from Eq.~\BCpgauge\ and the T-duality map are
\eqn\NSgauge{\CA^\a = \half\CF^\a_{mn} x^m dx^n,\quad
B^\bg_{(2)} = \smallfrac{1}{3!} H_{(3)mnr} x^m dx^n\w dx^r.}
One can check that the potentials \Cfinal\ do indeed give rise to the
fluxes \BPScond{a}, \fluxdeform{}, and \DiracCohomB{}.


\appendix{G}{Moduli Space Metrics}

In this appendix, we discuss the kinetic terms for the moduli of the
$T^6/\IZ_2$ O3 orientifold.  We restrict to the $\CN=2$ flux
\choiceID{}\ of Sec.~6.4 and the $\CN=1$ flux \indepchoiceIIID{}\ of
Sec.~8.2, and work to leading order in $Z-1$.\foot{Properly treating
the warp factor in the 4D kinetic terms for the 6D metric moduli is a
problem that we do not attempt to address here.  It was partially
studied in Ref.~\DeWolfe, however, the treatment there neglected terms
involving $\partial_\m Z = (\partial Z/\partial g_{ab}) \partial_\m
g_{ab}$.  It is is currently under investigation \GM.}  For the
$\CN=3$ case, the kinetic terms were derived by Frey and Polchinski
\FP, and shown to describe a sigma model with target \NthreeModspace,
as required by Ref.~\Castellani.  Here, we follow Frey and Polchinski,
except for a minor difference in conventions,\foot{We take $x^a\cong
x^a+1$ on the $T^6$, whereas Frey and Polchinski take $x^a\cong x^a +
2\pi$.  So, some coefficients differ by powers of $2\pi$.  In
addition, we do not absorb factors $g_s = e^{\langle\phi\rangle}$ into
the normalizations of the 4D fields, since, in the $\CN=2$ case,
$\phi$ is an unlifted modulus whose vev is not a priori determined.}
and the inclusion of the axion-dilaton modulus.

The 10D string frame action for the metric, dilaton, and axion
$C_{(0)}$ is
\eqn\actiontenS{S^{\rm10S}_{G,\phi,C_{(0)}} = {1\over 2\cdot
2\k_{10}^2}\int d^{10}x\sqrt{-G_S}
\biggl(e^{-2\phi}\bigl(R_{\rm10S}+4(\partial\phi)^2\bigr) 
-\ha \bigl(\partial C_{(0)}\bigr)^2\biggr),}
where $2\k_{10}^2 = (2\pi)^6\a'^4$ and $(G_{\rm S})_{MN}$ denotes the
metric \Othreebg{a}.  The integral runs over the $\IZ_2$ covering
space $\IR^{3,1}\times T^6$, so there is an overall factor of $1/2$
multiplying the action.  In the 10D Einstein frame,
\eqn\StoE{(G_{\rm E})_{MN} = e^{-\phi/2}(G_{\rm S})_{MN},}
the action becomes
\eqn\actiontenE{S^{\rm10E}_{G,\tdil} = {1\over4\k_{10}^2}\int
d^{10}x\sqrt{-G_E}\biggl(R_{\rm10E}
-{\partial_M\tdil\partial^M\bar\t_\dil\over2\bigl(\Im\tdil\bigr)^2}
\biggr),}
where now the contractions are performed using the new metric $G_{\rm
E}$.  The dimensional reduction down to 4D is
\eqn\actionfour{S^4_{G,\tdil} = {1\over2\pi\a'}\int
d^4x\sqrt{-G_4}\D\biggl(R_4
-{\partial_\m\tdil\partial^\m\bar\t_\dil\over2\bigl(\Im\tdil\bigr)^2}
-{\partial_\m\D\partial^\m\D\over
\vphantom{\bigl(}2\D^2\vphantom{\bigr)^2}}-{1\over4}G^{ac}G^{bd}
\partial_\m G_{ab}\partial^\m G_{cd}\biggr).}
Here, $(2\pi)^6\a'^3\D = (\det G_6)^{1/2}/2$, with $G_4$ and $G_6$ the
restrictions of $G_{\rm E}$ to 4D and 6D, respectively.  Finally, in
terms of the 4D Einstein frame metric $(g_{\rm E})_{\m\n} =
(G_4)_{\m\n}\D$ and rescaled 6D metric $\g_{ab} =
(G_6)_{ab}/\bigl((2\pi)^2\a'\D\bigr)$, this becomes
\eqn\actionfourE{S^{\rm4E}_{g,\tdil,\g} = {1\over2\pi\a'}\int
d^4x\sqrt{-g_{\rm E}}\biggl(R_{\rm E}
-{\partial_\m\tdil\partial^\m\bar\t_\dil\over2\bigl(\Im\tdil\bigr)^2}
-{1\over4}\g^{ac}\g^{bd}
\partial_\m\g_{ab}\partial^\m\g_{cd}\biggr).}
As in Ref.~\FP, once the moduli $c_{(4)abcd}$ and $\Phi_I{}^a$ are
included, the total 4D Einstein-Hilbert plus moduli action is
\eqna\Scomponents
$$S^{\rm4E}_{g,\tdil,\a,\b,\g} = S^{\rm4E}_g + S^{\rm4E}_\tdil +
S^{\rm4E}_{\a,\b,\g},\quad{\rm where} \eqno\Scomponents a$$
$$\eqalignno{S^{\rm4E}_g &= {1\over2\pi\a'}\int d^4x
\sqrt{-g_{\rm E}}\,R_{\rm E}, &\Scomponents b\cr
S^{\rm4E}_\tdil &= -{1\over2\pi\a'}\int d^4x\sqrt{-g_{\rm E}}
\,{\partial_\m\tdil\partial^\m\bar\t_\dil\over2\bigl(\Im\tdil
\bigr)^2}, &\Scomponents c\cr
S^{\rm4E}_{\a,\b,\g} &= -{1\over2\pi\a'}\int d^4x\sqrt{-g_{\rm E}}
\biggl(\g_{ab}\partial_\m\a_I{}^a\partial^\m\a_I{}^b
+ {1\over4}\g_{ac}\g_{bd}\bigl(\CD_\m\b^{ab}\CD^\m\b^{cd}
+\partial_\m\g^{ab}\partial^\m\g^{cd}\bigr)\biggr).&\cr
& &\Scomponents d}$$
In Eq.~\Scomponents{c},
\eqna\abg
$$\eqalignno{\a_I{}^a &= \Phi_I{}^a/(2\pi), &\abg a\cr c_{(4)abcd} &=
(2\pi)^4\a'^2\e_{abcdef}\b^{ef}, \quad {\rm where}\quad\e_{456789}=1,
&\abg b}$$
and
$$\CD_\m\b^{ab} = \partial_\m\b^{ab} +
\a_I{}^{[a|}\partial_\m\a_I{}^{|b]}.\eqno\abg c$$
The second term in $\CD_\m\b^{ab}$ is the analog of the gauge
Chern-Simons term in $F_{(3)}$ of type~I or in $H_{(3)}$ of the
heterotic theories.  It is well known that the sigma model
\Scomponents{d}\ with $a,b$ running over $D$ values and $I$ running
over $M$ values parametrizes the coset $\CT_{D,D+M}$, where
$\CT_{m,n}$ is the Grassmannian
\eqn\grassmannian{\CT_{m,n} = {SO(m,n)\over SO(m)\times SO(n)},}
up to discrete identifications \MaharanaSchwarz.

Thus, for $\CN=4$, with $2M=32$ D3 branes and no flux, the scalar
manifold of the low energy effective field theory is
\eqn\NfourModspace{\CM_{\CN=4} = {SU(1,1)\over U(1)}
\ \times\ {SO(6,22)\over SO(6)\times SO(22)},}
up to discrete identifications, where the first factor is from $\tdil$
and the second from $\a,\b,\g$.  When there is nonvanishing flux, the
supersymmetry is reduced, and the moduli constraints reduce this
moduli space to a proper submanifold that is also a product of
homogenous spaces.

For $\CN=3$ flux, the axion-dilaton and some $\b,\g$ moduli are fixed.
In the $w$-coordinates of Sec.~8.2, the unlifted components of
$\a,\b,\g$ are $\a_I{}^i$, $\a_I{}^\ibar$, $\b^{i\jbar}=-\b^{\jbar
i}$, $\g^{i\jbar}=\g^{\jbar i}$, and the above sigma model
parametrizes the coset $U(3,3+M)/\bigl(U(3)\times U(3+M)\bigr)$, as
discussed in Refs.~\refs{\Castellani,\FP}.

For the $\CN=1$ flux of Sec.~8.2, the moduli constraints are similar.
The axion-dilaton is again fixed.  In the $z$-coordinates of Sec.~8.2,
the nonvanishing components of $\a$, $\b$, $\g$ are as in the $\CN=3$
case in $w$-coordinates, but with the addition restriction that
$\b^{i\jbar}$ and $\g^{i\jbar}$ be symmetric (imaginary and real,
respectively).  That is, $\b^{i\jbar} = \b^{j\ibar}$ and $\g^{i\jbar}
= \g^{j\ibar}$.  It would be interesting to describe this $6+6M$
dimensional real hypersurface in $U(3,3+M)/\bigl(U(3)\times
U(3+M)\bigr)$ as a coset, perhaps $U(3,1+M)/\bigl(U(3)\times
U(1+M)\bigr)$.

Finally, for the $\CN=2$ flux of Sec.~6.4, the moduli constraints are
such that the scalar manifold factorizes into an $a=4,5,6,7$ part and
an $a=8,9$ part.  If there were no further constraints, the moduli
space would be $\bigl(SU(1,1)/U(1)\bigr) \times
\CT_{2,2+M}\times\CT_{4,4+M}$.  The first two factors form the manifold
$\CS\CT_{2,2+M}$ in the special K\"ahler series
\eqn\STseries{\CS\CT_{2,n} = {SU(1,1)\over U(1)}
\ \times\ {SO(2,n)\over SO(2)\times SO(n)}.}
The third factor is quaternionic.  Therefore, we would identify the
vector multiplet moduli space with $\CS\CT_{2,2+M}$ and the
hypermultiplet moduli space with $\CT_{4,4+M}$.  However, there are
additional moduli constraints that further reduce these special
K\"ahler and quaternionic manifolds.  First, there is the constraint
$(m/n)(-1/\tdil) = \t'_3$, where $\t'_3$ is the complex structure
modulus of $\g_{ab}$ in the $8,9$ directions.  Therefore,
$\CS\CT_{2,2+M}$ is further lifted to\foot{The only other homogeneous
symmetric space $\CG/\CH$ of dimension $2(2+M)$, with $M$ divisible by
four, that is special K\"ahler is $U(1,2+M) /\bigl(U(1)\times
U(2+M)\bigr)$.  However, this is not a hypersurface in
$\CS\CT_{2,2+M}$.  For a nice review of special K\"ahler and
hyperK\"ahler/quaternionic geometry in connection with both locally
and globally $\CN=2$ supersymmetric field theories, see Ref.~\Fre.}
\eqn\Ntwovec{\CM_V = \CS\CT_{2,1+M}.} 
If, in the $4,5,6,7$ directions, we write (cf.~Eqs.~\flatfibID{a,b})
\eqn\gammaparam{\g_{ab}dx^a dx^b = {\g_1\over\Im\t'_1}
\bigl|\eta'^4+\t'_1\eta'^5\bigr|^2
+ {\g_2\over\Im\t'_2} \bigl|dx^6+\t'_2 dx^7\bigr|^2,}
with $\eta'^4,\eta'^5$ given by Eqs.~\apfib{a,b}, then the remaining
moduli constraints are
\eqn\constrgamma{\t'_1\t'_2 = -1,\quad a'^4{}_7=a'^5{}_6,
\quad{\rm and}\quad \b^{46}= -\b^{57}.}
The hypermultiplet moduli space $\CM_H$ is the hypersurface
\constrgamma\ in $\CT_{4,4+M}$.  Since it must be quaternionic, we
conclude that\foot{Likewise, the only other homogeneous symmetric
quaternionic manifolds of dimension $4(3+M)$ are $Sp(2M+8)
/\bigl(Sp(2)\times Sp(2M+6)\bigr)$ and $U(2,3+M) /
\bigl(U(2)\times U(3+M)\bigr)$.  Neither of these is a
hypersurface in $\CT_{4,4+M}$.}
\eqn\appMH{\CM_H = \CT_{4,3+M}.}

As we have already observed in Secs.~6.4 and 8.2, for classes of flux
discussed in this paper, the partial breaking of $\CN=4$ to $\CN=1,2$
supersymmetry is such that the resulting massive gravitini all have
equal mass.  It is noteworthy that for the $T^6/\IZ_2$ O3 orientifold,
there does not exist an single proper example in the literature (that
is, a choice of flux consistent with Dirac quantization and the
Gauss's law constraint) in which the breaking involves massive
gravitini of unequal mass.  The ``generic'' case of partial
supersymmetry breaking with unequal gravitino mass is apparantly not
as generic as one might have thought.  It is not clear that such a
case is possible.  However, if it is, then the resulting mass spectra
and superHiggs mechanism would be governed by the results of
Ref.~\SUSYhiggs, in which the reduced moduli spaces for this case have
also been worked out.


\appendix{H}{Uniqueness of Flux in Sec.~8.2}

In this appendix we prove the claim that the flux \partEx{} is the
unique choice of supersymmetric flux in the class \choiceIIID{a,b}\
with the minimal value $N_\flux=12$, modulo $SL(2,\IZ)_{\t} \times
SL(2,\IZ)_{\tdil}$ equivalences.  (Here $SL(2,\IZ)_{\t}$ is the
diagonal subgroup of $SL(2,\IZ)^3 \subset SL(6,\IZ)$, where
$SL(6,\IZ)$ is the duality group of $T^6$, and $SL(2,\IZ)^3$ is the
product of the duality groups of the $T^2$ factors in the $T^6\to
T^2\times T^2\times T^2$ factorization of the complex structure for
this example.)  For notational simplicity, we drop the primes on $\t$
and $\tdil$ that were used in Sec.~8.2 solely to denote that O3 from
O5 quantities.  Instead, primes will indicate $SL(2,\IZ)$ transformed
quantities below.  It is convenient to rewrite Eq.~\quadraticP\ in
matrix form as
\eqn\Pmatrix{P(\t) = \ha 
\pmatrix{\t & 1}\hatP \pmatrix{\t\cr 1},
\quad{\rm where}\quad\hatP = \pmatrix{2l & m\cr m & 2n}.}
Then,
\eqn\detP{\det\hatP = 4ln-m^2,}
and under $SL(2,\IZ)_\t$ transformations,
\eqna\SLt
$$\eqalignno{\t\to\t', &\quad{\rm where}\quad \pmatrix{\t\cr1} 
= {1\over \g\t'+\d}M\pmatrix{\t'\cr1}, &\SLt a\cr
\hatP\to \hatP', &\quad{\rm where}\quad
\hatP' = M^T\hatP M, &\SLt b}$$
with
$$M = \pmatrix{\a&\b\cr\g&\d}\in SL(2\IZ)_\t.\eqno\SLt c$$

For $N_\flux = 12$, we have $(fk-gh)(4ln-m^2) = 9$.  Since
$4ln-m^2\equiv -1\pmod{4}$, the only possible factorization is
\eqn\factorization{fk-gh=3,\quad 4ln-m^2 =3.}
Here we have used the fact that $4ln-m^2 > 0$ in order for $P(\t)$ to
have two complex conjugate roots with nonzero imaginary part.  This is
required for the $T^6$ to be nondegenerate.

It is well known that $\t$ can also be mapped into the fundamental
domain \FundDomain\ by $SL(2,\IZ)_\t$ transformations.  For
$\t\in\CF_0$, the conditions \tconditions\ combined with the second
equality in Eq.~\factorization\ imply that $3 \ge 4l(n-l)$, with $n\ge
l > 0$.  Therefore, $l=n$.  From Eq.~\factorization, we then have $3 =
(2l+|m|)(2l-|m|)$.  So, $l = \pm m = n = 1$, corresponding to $\t =
\half\bigl(\mp1+i\sqrt{3}\bigr)$ from Eq.~\tconditions.  The two
possibilities are dual via $\t\to\t\pm1$.  (Only the $m=+1$ solution
is actually in the fundamental domain as defined in \FundDomain).
Thus,
\eqn\lmnResult{(l,m,n) = (1,1,1)\quad{\rm and}\quad
\t = e^{2\pi i/3},\quad\hbox{up to $SL(2,\IZ)_\t$ duality.}} 

Now consider the equivalences on $(f,g,h,k)$.  Under $SL(2,\IZ)_\tdil$
duality,
\eqn\SLtdil{A=\pmatrix{f&g\cr h&k}\to A'=NA,\quad{\rm where}
\quad N\in SL(2,\IZ)_\tdil.}
From $a = b = -\smallfrac{2}{3}(f+g)$, $c = d =
-\smallfrac{2}{3}(h+k)$, together with $fk-gh = 3$, we conclude that
$\gcd(f,h)=1$.  (Assume, to the contrary, that $\gcd{f,h}=3$.  Then,
demanding that $a,b,c,d\in2\IZ$ implies that $\gcd(g,k)=3$ and
subsequently that 9 divides $fk-gh$.  This contradicts $fk-gh=3$.)
Therefore, $\g f+\d h =1$ for some $\g,\d$, so that the
$SL(2,\IZ)_\tdil$ duality $A' =
\bigl({f'\atop h'}{\g'\atop k'}\bigr) = \bigl({h\atop \g}{-f\atop
 \d}\bigr)\bigl({f\atop h}{g\atop k}\bigr)$ gives $h'=1$.
Next, the duality $A'' = \bigl({f''\atop h''}{g''\atop k''}\bigr) =
\bigl({0\atop-1}{1\atop f'}\bigr)\bigl({f'\atop 1}{g'\atop
k'}\bigr)$ gives $h''=0$.  Finally, the $SL(2,\IZ)_\tdil$ duality
$A''' = \bigl({f'''\atop 0^{\phantom{\prime\prime\prime}}}{g'''\atop
k'''}\bigr) = \bigl({1\atop 0}{\b\atop 1}\bigr)\bigl({f''\atop
0^{\phantom{\prime\prime}}}{g''\atop k''}\bigr)$ allows us to replace
$g''$ with any $g'''\equiv g''\pmod{k''}$.  From $\det A''' = \det A''
= \det A' = \det A = 3$, we have $f'''k'''=3$.  We can assume that
$f''',k'''>0$ (via $SL(2,\IZ)$ duality by $\bigl({-1\atop
0}{0\atop-1}\bigr)$, if necessary).  There are two possible
factorizations: $f'''=1$ and $k'''=3$, or $f'''=3$ and $k'''=1$.  The
second case is excluded by $c=d = -2k'''/3\in\IZ$.  In the first case,
the congruence $-3a = 2(f'''+g''')\equiv 0\pmod{3}$ gives
$g'''\equiv-1\pmod{3} \equiv-1\pmod{k'''}$.  So, $g''' = -1$ and
$\bigl({f\atop h}{g\atop k}\bigr) =
\bigl({1\atop0}{-1\atop\phantom{-1}3}\bigr)$ up to $SL(2,\IZ)_\tdil$
duality.  This is equivalent to
\eqn\fghkResult{\pmatrix{f&g\cr h&k} = \pmatrix{1&-1\cr1&2}
\quad\hbox{up to $SL(2,\IZ)_\tdil$ duality.}}
Together, Eqs.~\lmnResult\ and \fghkResult\ prove the desired
uniqueness of the choice \partEx{b}.

\listrefs
\bye